\documentclass[a4paper,11pt]{article}
\usepackage{jheppub}
\usepackage{amssymb}
\usepackage{mathtools}

\usepackage{epsfig}
\usepackage{epstopdf}
\usepackage{latexsym}
\usepackage{enumerate} 
\usepackage{graphicx}
\usepackage{placeins}
\usepackage{floatrow}
\usepackage{subfig}
\usepackage{caption}
\usepackage{booktabs}
\usepackage{bbm}
\usepackage{xcolor}
\usepackage{physics}
\usepackage{tensor}
\usepackage{tikz}
\usetikzlibrary{matrix}
\usetikzlibrary{decorations.markings,calc,shapes,decorations.pathmorphing,patterns,decorations.pathreplacing,arrows.meta}
\usetikzlibrary{positioning}
\usepackage{hyperref}

\tikzset{
    fermion/.style={ultra thick},
    fcut/.style={ultra thick,dashed},
    photon/.style={decorate, decoration={snake, segment length=2.8 mm, amplitude=.5 mm}},
    scalar/.style={dotted, thick},
}

\DeclareMathAlphabet{\mathcal}{OMS}{cmsy}{m}{n}
\title{Classical Observables from Causal Response Functions}
 \author{Shovon Biswas$^a$,}
 \affiliation{$^a$Department of Physics and Astronomy, University of British Columbia,\\
 6224 Agricultural Road, Vancouver, BC, Canada V6T 1Z1}

\emailAdd{shovon@phas.ubc.ca}
 \author{Julio Parra-Martinez$^b$}
 \affiliation{$^b$Institut des Hautes Études Scientifiques, 
 91440 Bures-sur-Yvette, France}
 \emailAdd{julio@ihes.fr}
 
\abstract{We revisit the calculation of classical observables from causal response functions, following up on recent work by Caron-Huot \emph{et al.} \cite{CaronHuot2024}. We derive a formula to compute asymptotic in-in observables from a particular soft limit of five-point amputated response functions. Using such formula, we re-derive the formulas by Kosower, Maybee and O'Connell (KMOC) for the linear impulse and radiated linear momentum of particles undergoing scattering, and we present an unambiguous calculation of the radiated angular momentum at leading order. Then, we explore the consequences of manifestly causal Feynman rules in the calculation of classical observables by employing the causal (Keldysh) basis in the in-in formalism. We compute the linear impulse, radiated waveform and its variance at leading and/or next-to-leading order in the causal basis, and find that all terms singular in the $\hbar \to 0$ limit cancel manifestly at the integrand level. We also find that the calculations simplify considerably and classical properties such as factorization of six-point amplitudes are more transparent in the causal basis.}
\usepackage{amsfonts,amsmath,bm,physics,slashed}
\makeatletter
\gdef\@fpheader{}
\makeatother

\def\a{\alpha}
\def\b{\beta}
\def\g{\gamma}

\def\d{\delta}
\def\D{\Delta}
\def\e{\epsilon}
\def\ve{\varepsilon}

\def\th{\theta}

\def\k{\kappa}

\def\m{\mu}
\def\n{\nu}

\def\r{\rho}

\def\s{\sigma}

\def\f{\phi}
\def\F{\Phi}
\def\vf{\varphi}
\def\ve{\varepsilon}

\def\Ps{\Psi}
\def\o{\omega}
\def\O{\Omega}
\def\I{\text{I}}
\def\II{\text{II}}
\def\pd{\partial}

\def\Op{\mathbb{O}}

\def\k{\bm{k}}

\def\in{\text{in}}
\def\out{\text{out}}

\definecolor{darkbrown}{rgb}{0.4, 0.26, 0.13}
\begin{document}
\maketitle
\flushbottom

\section{Introduction}
\label{sec:into}
How is classical physics precisely encoded in a perturbative theory of quantum fields? 
The naive textbook answer is that it is encoded in tree-level diagrams, but the answer in the presence of heavy particles is not so simple \cite{Holstein:2004dn}.
This question has become relevant in recent years following the detection of gravitational waves by LIGO \cite{LIGOScientific:2016aoc}, as an approach to the classical relativistic two-body problem from the classical limit of quantum field theory amplitudes has proven very effective \cite{Bern:2019nnu,Bern:2019crd, Bern:2021dqo,Bern:2021yeh,Herrmann:2021lqe,Herrmann2021,DiVecchia:2021ndb, DiVecchia:2021bdo, DiVecchia:2022nna, DiVecchia:2022piu,Brandhuber:2023hhy,Herderschee:2023fxh,Elkhidir:2023dco, Georgoudis:2023lgf, Georgoudis:2023eke, Georgoudis:2024pdz, Elkhidir:2023dco, Georgoudis:2023lgf, Bini:2024rsy, Ivanov:2024sds} and complemented classical field theory methods \cite{Buonanno:1998gg,Porto:2016pyg,Blanchet:2013haa}. In a seminal paper \cite{Kosower2019}, Kosower, Maybee and O’Connell (KMOC) developed a formalism for computing classical observables from quantum scattering amplitudes in the $\hbar\rightarrow 0$ limit. The main advantage of the scattering-amplitude-based method is that it can borrow various powerful tools like the double-copy \cite{Bern:2008qj,Bern:2019prr}, unitarity methods \cite{Bern:1994zx,Britto:2004nc}, and modern techniques for computing Feynman integrals \cite{Kotikov:1990kg,Bern:1993kr,Remiddi:1997ny,Gehrmann:1999as,Henn:2013pwa, Parra-Martinez:2020dzs}, originally developed in the context of collider physics to compute classical gravitational quantities. 

In the KMOC framework, one aims to compute various classical observables such as linear and angular impulses of a particle, or the radiative waveform itself, perturbatively in the gravitational coupling constant. Such classical observables are defined from classical limits of changes in the expectation value of an operator $\Op$ in the quantum theory
\begin{align}  \D   \Op &=\bra{\in}\Op^{\out} \ket{\in}-\bra{\in}\Op^{\in}\ket{\in}\label{kmoc0}\,,
\end{align}

where objects with \textit{in} and \textit{out} labels are defined at the asymptotic past and future infinity respectively, and $\ket{\in}$ are two-particle states with classically-peaked wavepackets. Such observables are \emph{in-in} quantities by definition, but in the KMOC framework these are expressed in terms of \emph{in-out} scattering amplitudes, using the relation  
\begin{equation}
    \Op^{\out} - \Op^{\in}=S^\dagger\Op^{\in} S - \Op^{\in} =  i [\Op^{\in},T] + T^\dagger [\Op^{\in} ,T]\,,
    \label{sos}
\end{equation}
where $S=1+i T$ is the familiar unitary S-matrix. However, from this perspective, the classical limit is not manifest, as the S matrix is semiclassically the exponential of the on-shell action
\begin{equation}
    S\sim e^{i \frac{I}{\hbar}}\,,
\end{equation}
and delicate manipulations are required to manifest the cancellation of singular contributions as $\hbar\to0$ between different terms in Eq.~\eqref{sos} \cite{Kosower2019, Herrmann2021}. Furthermore, despite classical observables being causal quantities by definition, when expressed in terms of the in-out scattering amplitudes, causality is hidden in the intermediate steps of the computation.

An alternative way of computing these observables is to directly use the \textit{in-in} or \textit{Schwinger-Keldysh} formalism \cite{Schwinger:1960qe,Keldysh:1964ud}, which is commonly utilized in worldline-based approaches to the two-body problem \cite{Kalin:2020mvi,Mogull:2020sak,Kalin:2020fhe,Kalin:2022hph,Dlapa:2021vgp,Jakobsen:2022psy,Dlapa:2022lmu, Damgaard:2023vnx, Driesse:2024xad}. In field theory, the relation between asymptotic observables and generalized scattering amplitudes has been discussed recently in \cite{CaronHuot2024}, where it was shown that a class of in-in \textit{response functions} \cite{Chou:1984es, CaronHuot2011}, computes KMOC-like expectation values such as the classical gravitational waveform. In this paper, we further explore this connection and show that other observables can also be computed from appropriate soft limits of response functions. We explain how the KMOC formula for other observables, such as the linear impulse and radiated momentum, $\Delta \mathbb{P}$, can be derived from a causal response function. 

A particularly interesting classical gravitational observable is the angular momentum loss for the binary system, $\Delta \mathbb{J}$. A position space calculation by Damour \cite{Damour:2020tta} showed that the angular momentum loss in General Relativity begins at  $\mathcal{O}\left(G^2\right)$, whereas the energy loss starts at $\mathcal{O}\left(G^3\right)$ \cite{Herrmann:2021lqe}. This result was reproduced in momentum space using Weinberg's soft theorem \cite{Manohar:2022dea, DiVecchia:2022owy}. However, the naive application of the KMOC framework produces an ambiguous result, as it requires the inclusion of a unitary cut involving an on-shell three-point amplitude, which is zero for real kinematics, times the singular soft-limit of a five-point amplitude. In this paper, we will show that the soft limit of the response function unveils new integration regions thereby 
enabling us to present an unambiguous calculation of the radiated angular momentum, within the KMOC framework.

Response functions are causal observables. Does manifest causality simplify the calculations of classical observables from amplitudes? In this work, we find an affirmative answer to this question by computing classical observables using causal Feynman rules, i.e., amplitudes with a retarded $i\e$ prescription, which appear naturally within the in-in formalism in the so-called Keldysh basis \cite{CaronHuot2011, Blanchet:2013haa}.
We show that the causal representation manifests various properties of certain classical observables, such as the linear impulse, the scattering waveform and its variance. In particular, we find that the retarded $i\e$ prescription ensures the manifest cancellation of ${\cal O}(1/\hbar^n)$ terms within classical observables, as well as the vanishing of the classical variance of the scattering waveform.

 The organization of this paper is as follows. In Section \ref{sec:review}, we review causal response functions and discuss their computation using the Schwinger-Keldysh closed time contour. We also review the relation between amputated response functions and the KMOC waveform from Ref.~\cite{CaronHuot2024}. In Section \ref{sec:sec2}, we present a derivation of the KMOC formulas for the linear impulse and radiated momentum from response functions, and explain the subtle calculation of the angular momentum loss. In Section \ref{sec:sec3}, we revisit the calculation of impulse, waveform, and variance up to one loop in the causal basis. Finally in Section \ref{sec:conclusion}, we summarize our results.
 
 For simplicity, we present example calculations in scalar QED, described by the Lagrangian
\begin{align}
   \mathcal{L}=-\frac{1}{4}F_{\mu\nu} F^{\mu\nu} +\sum_{i=1,2}\left[\left(D_\mu\F_i\right)\left(D^\mu\F_i\right)^\dagger-m_i^2\F_i\F_i^\dagger\right] \label{model}
\end{align}
where $D_\mu=\pd_\mu+ie\,Q_iA_\mu$ is the covariant derivative, $F_{\mu\nu} = \partial_\mu A_\nu - \partial_\nu A_\mu$ is the photon field strength, $Q_i$ and $m_i$ are the charges and masses of the fields $\F_i$, and $e$ is the gauge coupling. 
The generalization to the more interesting case of gravity is straightforward.


\section{Review: in-in formalism and asymptotic observables}
\label{sec:review}
In this section, we review causal response functions which are familiar objects in non-linear response theory. More details can be found in refs. \cite{Chou:1984es, Crossley:2015evo, CaronHuot2011, Blaizot:2001nr, Ghiglieri:2020dpq}. We present our discussion in terms of a generic field $\vf(x)$, as a stand-in for $\F_i$ and $A_\mu$. Then, we review the KMOC formula and its application for the scattering waveform from Refs.~\cite{Cristofoli:2021vyo,CaronHuot2024}, as well as the connection of the latter with a five-point causal response functions.

\subsection{Local response functions and in-in formalism}
\label{subsec:review1}
 Let us denote the expectation value of a local operator ${\cal O}(x_i)$ at some time $t=x_{i}^0$ by $\langle {\cal O} (x_i) \rangle = \bra{0} {{\cal O}}(x_i) \ket{0}$  where  $\ket{0}$ is the ground state of the theory.  A perturbation $\mathcal{H}_{j}=\int d^dx \vf(x) j(x)$ with source $j(x)$  is added to the system and the expectation value of the operator ${\cal O}(x)$ at $t=x^0$ as a result of the perturbation is given by \cite{CaronHuot2011} 
\begin{align}
   \langle  {\cal O}_j(x)\rangle&=\bra{0}\bar{\mathcal{T}}e^{i\int d^{D}x \vf(x)j(x)}{\cal O}(x)\mathcal{T}e^{-i\int d^{D}x \vf(x)j(x)}\ket{0} \label{exp_j}
\end{align}
where the time integral is from $x_i^0$ to $x^0$, $\mathcal{T}$ ($\bar{\mathcal{T}}$ ) is the time ordering (anti-time-ordering) symbol,  and the operator ${\cal O}(x)$ is in the interaction picture.  We shall describe the path-integral computation of the above quantity momentarily. The difference between $\langle {\cal O}_j(x) \rangle$ and $\langle {\cal O}(x) \rangle$ can be thought of as a measure of the response of the system to the external perturbation sourced by $j(x)$ \cite{Blaizot:2001nr}. The $n+1$-point response function is defined as 
\begin{align}
    R_{n+1} &[{\cal O}(x);\vf(x_1),\vf(x_2),\cdots,\vf(x_n)]=\frac{i^n\d^{n}}{\d j(x_1)\cdots\d j(x_n)}\langle {\cal O}_j(x)\rangle \nonumber\\
    &=\sum_{P}\th(x^0-x^0_{P(1)})\th(x_{P(2)}^0-x_{P(2)}^0)\cdots\th(x_{P(n-1)}^0-x_{P(n)}^0)\nonumber\\
    &\times\bra{0}\left[\left[\cdots\left[\left[{\cal O}(x),\vf(x_{P(1)})\right],\vf(x_{P(2)})\right]\cdots\right],\vf(x_{P(n)})\right]\ket{0}\,,
    \label{response}
\end{align}
where the sum is over all permutations of $(1,2,\cdots,n)$. The definition in Eq.~\eqref{response} is manifestly causal, and symmetric in $x_1,\cdots,x_n$. In particular, for the retarded 2-point function of the field one has
\begin{align}
    R_2[\vf(x_1);\vf(x_2)]=\th(x_1^0-x_2^0) \bra{0}[\vf(x_1),\vf(x_2)]\ket{0}
\end{align}
which is the usual retarded propagator. 
\begin{figure}[tbp]
    \centering    \begin{tikzpicture}[scale=0.8]
  \draw[darkbrown,-] (0,0)->(10,0) node[midway]{$\bm{>}$} ;
   \draw[darkbrown,-] (10,0)--(10,-0.4)  ;
    \draw[brown,-] (10,-0.4)--(0,-0.4) node[midway]{$<$}; 
%
 \node (l) at (-0.6,-0.2) {$-\infty$};
  \node (l2) at (10.4,-0.2) {$\infty$};
 \node (c1) at (0,0.4) {$\gamma_\I$};
 \node (c1) at (0,-0.8) {$\gamma_\II$};
 \node[inner sep=0pt] at(11.5,-0.2)(p){};
 \node[above =0.5 of p] (p1){$\Im t$};
 \node[right=0.5 of p] (p2){$\Re t$};
 \draw[->](p)--(p1);
 \draw[->](p)--(p2);
 \fill (p) circle(1 pt);
\end{tikzpicture}
    \caption{Closed-timed contour for computing in-in expectation values.}    \label{fig:sk_contour}
\end{figure}
Now, let us take $x_i^0\rightarrow -\infty$,  $x^0\rightarrow \infty$ and discuss the computation of the path integral in Eq.~\eqref{exp_j}. One can use a single closed time contour as shown in Fig.~\ref{fig:sk_contour}  with $\Gamma=\gamma^\I\cup \gamma^\II$ that goes from $\gamma^\I:\;-\infty\rightarrow \infty$ and back to $\gamma^\II:\;\infty\rightarrow -\infty$ along with two field variables $\varphi_\I,\, \varphi_\II$  to write the unitary time evolution operator and its conjugate as path-integrals.  These two fields are identified at $x^0=\infty$ i.e
\begin{align}
   \varphi^\I-\varphi^\II=0,\quad x^0=\infty. \label{bc}
\end{align}
The full partition function in terms of the doubled fields is given by  \cite{CaronHuot2011} 
\begin{align}
 Z= \int D\varphi^\I D\varphi^\II  \;e^{i \int\left(S[\varphi^\I]-S[\varphi^\II]\right)}\,.\label{path-int1}
\end{align}
Eq.~\eqref{path-int1} allows for the computation of contour-ordered correlation functions by inserting fields with contour labels $\I,\II$ on the path-integral in Eq.~\eqref{path-int1}. For example, an ordinary $n$-point time-ordered (or anti time-ordered) correlation function can be computed by inserting $n$ type $\I$ (or type $\II$) fields in Eq.~\eqref{path-int1}. It is useful to adopt notation of \cite{CaronHuot2024} to denote the contour ordering of various fields: for a string of operators ${\cal O}_1^{\I,\II},{\cal O}_2^{\I,\II},{\cal O}_3^{\I,\II},\cdots$, the contour-ordering symbol $\mathcal{C}$ sorts all type $\II$ fields to the left of type $\I$ fields and then time orders or anti-time orders the $\I$ and $\II$ fields respectively. For example,
$$\mathcal{C}\left\{\mathcal{O}_1^\II \mathcal{O}_2^\I \hat{\mathcal{O}}_3^\I \mathcal{O}_4^\I \mathcal{O}_5^\II\right\}=\bar{\mathcal{T}}\left\{\mathcal{O}_1\mathcal{O}_5\right\}\mathcal{T}\left\{\mathcal{O}_2\mathcal{O}_3\mathcal{O}_4\right\}.$$
The boundary condition in Eq.~\eqref{bc} implies that the correlation functions including all difference fields vanishes identically
\begin{align}
    \bra{0} \mathcal{C}\big\{(
    {\mathcal{O}}_1^\I-
    {\mathcal{O}}_1^\II)({\mathcal{O}}_2^\I-{\mathcal{O}}_2^\II)({\mathcal{O}}_3^\I-{\mathcal{O}}_3^\II)\cdots\big\} \ket{0} =0 \,,\label{largest}
\end{align}
which is known as the \textit{largest time equation}. Finally,  it can be shown that the response function in Eq.~\eqref{response} can be written as \cite{Chou:1984es}
\begin{align}
    R_{n+1} &[{\cal O}(x);\vf(x_1),\vf(x_2),\cdots,\vf(x_n)]\nonumber\\
   &=\bra{0} \mathcal{C}\left\{\frac{1}{2}\left({\cal O}^\I(x)+{\cal O}^\II(x)\right)(\vf^\I\left(x_1)-\vf^\II(x_1)\right)\cdots\left(\vf^\I(x_{n-1})-\vf^\II(x_n)\right)\right\}\ket{0}\nonumber\\
   &= \bra{0}\mathcal{C}\left\{{\cal O}^\I(x)(\vf^\I\left(x_1)-\vf^\II(x_1)\right)\cdots\left(\vf^\I(x_{n-1})-\vf^\II(x_n)\right)\right\}\ket{0}\;, \label{respone2} 
\end{align} 
using the largest time equation in Eq.~\eqref{largest}, and with the normalization condition of theta functions $\sum_P\th(x^0_{P(1)}-x^0_{P(2)})\cdots\th(x^0_{P(n-1)}-x^0_{P(n)})=1$. 

We conclude this section by describing the momentum space in-in Feynman rules to compute correlation functions in Eq.~\eqref{respone2} using the in-in path integral \cite{Caron-Huot:2022eqs, CaronHuot2024}:
\begin{itemize}
    \item Assign a label $\I$ to the response field $\mathcal{O}$. The source fields $\vf$ should be assigned all possible $\I/\II$ labels. External points must be connected with the same type of internal points.
     \item Include an additional minus sign for a type $\II$ vertex. 
    \item For an internal line connecting two type $\II$ fields, insert a factor of $i/(p^2-m^2+i\e)$, whereas for an internal line connecting two type $\I$ fields, insert a factor of $-i/(p^2-m^2-i\e)$. 
       \item Draw a positive energy cut for an internal line connecting a type $\I$ vertex to a type $\II$ vertex. Include a factor of $\hat{\d}_+(p^2-m^2)=\th(p^0)\hat{\d}(p^2-m^2)$ for the propagator.
\end{itemize}
\subsection{Waveforms from KMOC and response functions}
\label{sec:review_KMOC}
A systematic approach to computing classical observables from amplitudes was developed in  \cite{Kosower2019, Cristofoli:2021vyo} in the context of $2\rightarrow 2$ scattering. Suppose we are interested in computing the change in expectation value of some \textit{classical} observable $\Op$ due to the interaction between two massive objects mediated by some massless particle. The change in the expectation value of $\Op$ is defined as
\begin{align}    \D \Op&=\bra{\in}\Op^{\out} \ket{\in}-\bra{\in}\Op^{\in}\ket{\in} =\bra{\in}S^\dagger\Op^{\in} S\ket{\in}-\bra{\in}\Op^{\in}\ket{\in}\,,\label{kmoc1}
\end{align}
where, $\Op^{\out},\Op^{\in}$ are $\Op(t)$  evaluated at $t=\infty,t=\;-\infty$ respectively,   $S=1+iT$ is the unitary $S$ matrix.  The two-particle states, $\ket{\in}$, are carefully chosen so that they correspond to two classical particles separated by impact parameter $b^\mu=b_1^\mu-b_2^\mu$ in the following way
\begin{align}   
\ket{\in}&=\int \Big[\prod_{i,1,2} \frac{d^4p_i}{(2\pi)^4}\hat{\d}_+(p_i^2-m_i^2) e^{-ip_ib_i} \Ps_i(p_i) \Big]  \ket{1 2}\,, \\
\bra{\in}&=\int \Big[\prod_{i,1,2} \frac{d^4p_i'}{(2\pi)^4}\hat{\d}_+(p_i'^2-m_i^2) e^{ip_i'b_i} \Ps_i^*(p_i) \Big]  \bra{1'2'}\,,
 \end{align}
where $\ket{ij}=a^\dagger_{p_i}a^\dagger_{p_j}\ket{0}$ is a two-particle state,  $\hat{\d}_+(p_i^2-m_i^2) = (2\pi) \th(p_i^0) \d(p_i^2-m_i^2)$ the Lorentz invariant phase space measure, and $\Ps_i(p_i)$ are wavepackets sharply peaked around the classical momenta $p_i^\mu=m_iu_i^\mu$ so that for any function $f(p_i)$,
 \begin{align}
    f(p_i)\big|_{\rm cl.} = 
   \int   \Big[\prod_{i,1,2} \frac{d^4p_i'}{(2\pi)^4}\hat{\d}_+(p_i'^2-m_i^2)\Big] |\Ps_i(p_i)|^2 f(p_i)=f(m_iu_i) \,,\label{class-lim} \end{align}
where $m_i, u_i$ are the masses and four velocities of the particles, and $u_i^2=1$. 
Then, the change in an observable during the two-particle-scattering process is given by
\begin{align}
    \D \Op =  
    \bra{1'2'} \Op^{\out}-\Op^{\in}\ket{12}\big|_{\rm cl.} \,,  
\end{align}
where $\cdot\big|_{\rm cl.}$ denotes the integral against the classical wavefunctions as in Eq.~\eqref{class-lim}, with an appropriate impact parameter phase. KMOC then instruct us to write $\Op^{\out} = S^\dagger \Op^{\in} S$, and expand the S-matrix as $S=1+i T$, yielding
\begin{align}
    \D \Op &= i\bra{1'2'} [\Op^{\in},T]\ket{12} + \bra{1'2'}T^\dagger [\Op^{\in},T]\ket{12} \nonumber \\
    &= i\bra{1'2'} [\Op^{\in},T]\ket{12} + \sum_X \bra{1'2'}T^\dagger \ket{X}\bra{X} [\Op^{\in},T]\ket{12}
    \,,
\end{align}
upon the use of the unitarity of the S-matrix
\begin{equation}
    S S^\dagger =1 \quad \rightarrow \quad i(T-T^\dagger)   =  T^\dagger T\,.
\end{equation}
and inserting a complete set of states $\sum_X \ket{X}\bra{X}$.

Perhaps the simplest example of such an observable is the classical spectral waveform of the massless particle. This is defined as the change in expectation value of the annihilation operator $a^h(k)$  of the photon with helicity $h$ as
\begin{align}
 \mathcal{W}^h_k&= \bra{1'2'}a^{h\;\out}_k - a^{h\;\in}_k \ket{12}\big|_{\rm cl.} \,.
\end{align}
The absence of photons in the initial state, $a^{h\;\in}_k\ket{12}=0$, simplifies the calculation of the various terms
\begin{align}
    \bra{1'2'} [a_k^{h\,\in},T]\ket{12} =  \bra{1'2'} a_k^{h\,\in}T\ket{12}  = \bra{1'2' k} T\ket{12} = {\cal A}(12\to 1'2' k) \,.
\end{align}
The momentum conserving delta functions are implicit in the definition of the amplitude $\mathcal{A}$. Similarly,
\begin{equation}
    \sum_X \bra{1'2'}T^\dagger \ket{X}\bra{X} a_k^{h\,\in} T\ket{12} 
    = \sum_X  {\cal A}(1,2 \to X k) {\cal A}^*(X \to 1'2') \,.
\end{equation}
\\
Thus, by this procedure, we obtain a formula for the spectral waveform in terms of the scattering amplitudes which was  derived in Ref.~\cite{Cristofoli:2021jas}
\begin{align}
\label{waveform-master} 
    & \mathcal{W}^h_k=\int \prod_{i=1,2}\hat{d}^4q_i\;\hat{\d}_+\left(2p_1\cdot q_i+q_i^2\right)e^{-ib_i\cdot  q_i} 
     \hat{\delta}^4(k+q_1+q_2)
      \Bigg[\begin{tikzpicture}[baseline=.1ex,scale=0.3]
    \draw[fermion] (-2,-2) -- (0,0) node[pos=0.35] {} ;
      \node (a) at (-2.5,-2.5){$p_2$};
    \draw[fermion] (2,2) -- (0,0) node[pos=0.1] {} ;
      \node (a) at (3.3,2.7){$p_1+q_1$};
    \draw[fermion] (-2,2) -- (0,0) node[pos=0.35] {} ;
      \node (a) at (-2.5,2.5){$p_1$};
    \draw[fermion] (2,-2) -- (0,0) node[pos=0.35] {} ;
      \node (a) at (2.5,-2.5){$p_2+q_2$};   
    \node (k) at (0,4) {$k$};
    \draw[photon] (0,0)--(k);
    \node[minimum size=1.2cm, fill = gray!20, draw = black, ultra thick, circle] at (0,0){$i\mathcal{A}$};
\end{tikzpicture}  \\
   & +  \sum_X\int \prod_{i=1,2}\hat{d}^4\ell_i\;\hat{\d}_+\left(2p_1\cdot \ell_i+\ell_i^2\right)\;\hat{\d}(\ell_1+\ell_2+r_X)\begin{tikzpicture}[baseline=.1ex,scale=0.35]
\node[inner sep=0pt] (c) at (0,0){};
\node[above left=1 of c] (a) {$p_1$};
    \draw[fermion] (a) -- (c) ;
      \node (b)[below left=1 of c]{$p_2$};
    \draw[fermion] (b) -- (c) ;
   \node[above right=1 of c](d) {};
    \node[below right=1 of c](e){};
    
    \node[inner sep=0pt] (c) at (7,0){};
    \node[above left=1 of c] (a) {};
      \node (b)[below left=1 of c]{};
   \node[inner sep=0pt,above right=1 of c](d) {};
   \node[inner sep=0pt,above right=0.01 of d] (d2){$p_1+q_1$};
    \draw[fermion] (c) -- (d); 
    \node[inner sep=0pt, below right=1 of c](e){};
     \node[inner sep=0pt,below right=0.01 of e] (e2){$p_2+q_2$};
    \draw[fermion] (c) -- (e);
    \draw[photon] (0,0)--(7,0);
      \draw [fermion](6.5,0) arc (25:155:3.5);
     \draw [fermion](6.5,0) arc (335:205:3.5);
     \draw [photon](-0.1,0) arc (190:90:3.5); \node[minimum size=1.3cm, fill = gray!20, draw = black, ultra thick, circle] at (0,0){$i\mathcal{A}$};
   \node[minimum size=0.6cm, fill = gray!20, draw = black, ultra thick, circle] at (c){$-i\mathcal{A}^*$};
 \node (m1) at (5.1,3){$p_1+\ell_1$};
 \node (m2) at (5.1,-3){$p_2+\ell_2$};
 \node (m3) at (4.5,0.7){$r_X$};
  \node (m3) at (4,4.5){$k$};
 \draw[brown,fcut](3.3,4.8)--(3.3,-3.8);
\end{tikzpicture}\Bigg] \,. \nonumber    
\end{align}

Let us now review how this formula is derived from a local causal response function, as explained in Ref.~\cite{CaronHuot2024}.
Consider the LSZ amputations
\begin{align}
  \text{LSZ}_{x}^{\out} \F_i(x) &= + i \int d^4x e^{+i{p_i}x}\left(\pd^2+m^2\right)\F_i(x)= a_{p_i}^{\out}-a_{p_i}^\in \quad \;\; (\text{outgoing})\,,\nonumber\\
  \text{LSZ}_{x}^{\in\hphantom{t}} \F_i^\dagger(x) &=   -i \int d^4x e^{-i{p_i}x}\left(\pd^2+m^2\right)\F^\dagger_i(x)= a_{p_i}^{\dagger\out}-a_{p_i}^{\dagger\in} \quad \! (\text{incoming})\,.
 \end{align}
Similarly for the photon with helicity $h$
\begin{align}
  \text{LSZ}_{x}^{\out} A^h(x) &= + i \int d^4k e^{+i{k}x}\left(\pd^2+m^2\right)\ve_\mu^{h*}(k)A^\mu(x)= a_{k}^{h\;\out}-a_{k}^{h\;\in} \quad \!(\text{outgoing})\,,\nonumber\\
  \text{LSZ}_{x}^{\in\hphantom{t}} A^h(x) &=   -i \int d^4x e^{-i{k}x}\left(\pd^2+m^2\right)\ve_\mu^{h}(k)A^\mu(x)=  a_{k}^{h\dagger\;\out}-a_{k}^{h\dagger\;\in} \; (\text{incoming})\,.
 \end{align}
Then, by amputating the five-point response function we obtain the key relation
\begin{align}
&\text{LSZ}_{x}^{\out} \prod_{i=1,2} \text{LSZ}_{x_{i}'}^{\out} \; \text{LSZ}_{x_i}^{\in} R_{5} [A^h(x);\F_1^\dagger(x_1),\F_2^\dagger(x_2),\F_1(x_1'),\F_2(x_2')] \nonumber\\
&=\bra{0}\mathcal{C}\left\{(a_k^{\I \,h\,\out}-a_k^{\I\, h\,\in}) (a_{1'}^{\II\;\in}-a_{1'}^{\I\;\in}) (a_{2'}^{\II\;\in}-a_{2'}^{\I\;\in}) (a_{1}^{\II\;\dagger\in}-a_{1}^{\I\;\dagger\in}) (a_{2}^{\II\;\dagger\in}-a_{1}^{\I\;\dagger\in})  \right\}\ket{0}  \nonumber\\
    &=\bra{0}a_{1'}^{\in} a_{2'}^{\in} (a_{k}^{h\;\out}-a_{k}^{h\;\in}) a_{1}^{\in\dagger} a_{2}^{\in\dagger}\ket{0} = \bra{1'2'} a_{k}^{h\;\out}\ket{12} =  \mathcal{W}^h_k\,.    \label{5pt-waveform}
\end{align}
In the second line, we have adopted the contour ordering prescription in Eq.~\eqref{respone2}. As a result of the boundary condition in Eq.~\eqref{bc}, we obtain, for instance, from the amputated heavy fields \cite{CaronHuot2024}
\begin{align}
    \text{LSZ}_{x_i'}^{\out}\Big[\F^{\I}_i(x_i')-\F^{\II}_i(x_i')\Big]=a_{p_i'}^{\II\;\in}-a_{p_i'}^{\I\;\in}\,,\quad \text{LSZ}_{x_i}^{\in}\Big[\F^{\I\dagger}_i(x_i)-\F^{\II\dagger}_i(x_i)\Big]=a_{p_i}^{\II\;\in\;\dagger}-a_{p_i}^{\I\;\in\;\dagger}\,,\nonumber
\end{align}
and similarly for the photon.
Using the Feynman rules to compute the response functions outlined in Section \ref{subsec:review1}, it is easy to see that indeed, upon amputation, the five-point function reproduces precisely the KMOC formula for the spectral waveform in Eq.~\eqref{waveform-master}. The first term Eq.~\eqref{waveform-master} corresponds to diagrams with only $\I$ fields and the cuts in the second term correspond to diagrams which connect $\I$ and $\II$ fields. 

 
\section{KMOC formulas from causal response functions}
\label{sec:sec2}

In this section, we explain in detail the relationship between in-in expectation values of more general (composite) operators and causal response functions. This was briefly touched upon in \cite{CaronHuot2024}, but here we further clarify the precise relationship and explain various subtleties. In particular, we show that the change in the expectation value of an observable $\Op$ is given by a particular soft limit of a five-point response function.

We shall consider observables, $\Op$ which are integrals of a local density given by an operator ${\cal O}(x)$
\begin{align}
    \Op(t)=\int d^3\bm{x}\;\mathcal{O}(t,\bm{x})\,.
\end{align}
Indeed, such is the case for the linear impulse and radiated momentum, as well as the radiated angular momentum loss, which are suitable integrals of the stress-energy tensor, as we review below.

Just as for the waveform, following \cite{CaronHuot2024}, we can write the expectation value of the local density $\cal O$ in terms of an amputated response function
\begin{align}
    &\bra{1'2'} \mathcal{O} (x) \ket{12} = \bra{0}a_{1'}^{\in} a_{2'}^{\in} \mathcal{O}(x) a_{1}^{\in\dagger} a_{2}^{\in\dagger}\ket{0}\nonumber\\
%
    & =\prod_{i=1,2} \text{LSZ}_{x_{i}'}^{\out} \; \text{LSZ}_{x_i}^{\in}
 R_{5} [\mathcal{O}(x);\F_1^\dagger(x_1),\F_2^\dagger(x_2),\F_1(x_1'),\F_2(x_2')] \,.
    \label{5pt-responseO}
\end{align}
Let us now define the Fourier transform of this expectation value
\begin{align}
    \langle \mathcal{O}(\omega,\bm{k}) \rangle =\int d^4 x e^{i\omega t-i\bm{k}\bm{x}}\bra{1'2'} \mathcal{O} (t,\bm x) \ket{12}\,,
\end{align}
with $\omega>0$,
which at zero three-momentum, $\bm{k}$, yields
\begin{align}
    \lim_{\bm{k}\rightarrow 0}\langle \mathcal{O}(\omega,\bm{k}) \rangle= \int dt e^{i\omega t} \bra{1'2'} \mathbb{O} (t) \ket{12}\,.
\end{align}
Finally, by taking the zero-frequency limit we obtain the change of the observable
\begin{align}
    -\lim_{\o\rightarrow 0}\lim_{\bm{k}\rightarrow 0} i\o\; \langle \mathcal{O}(\omega,\bm{k}) \rangle &=\lim_{\o\rightarrow 0} \int dt\;e^{i\o t} \bra{1'2'} \pd_t\mathbb{O} (t) \ket{12}
    = \bra{1'2'} \Op^{\out}-\Op^{\in}\ket{12} .
\end{align}
where the first equality follows from integration by parts, which is carried out before taking the $\omega\to 0$ limit.  

Therefore, we find that the change of the observable between the \textit{in} and \textit{out} states is simply given by the soft limit of an amputated five-point local response function
\begin{align}
     &\bra{1'2'} \Op^{\out}-\Op^{\in}\ket{12} \label{kmoc-resp3}\\
     &=-\lim_{\o\rightarrow 0}\lim_{\bm{k}\rightarrow 0} i\o \int d^4 x\; e^{i\omega t-i\bm{k}\bm{x}}  \prod_{i=1,2} \text{LSZ}_{x_{i}'}^{\out} \; \text{LSZ}_{x_i}^{\in}
 R_{5} [\mathcal{O}(x);\F_1^\dagger(x_1),\F_2^\dagger(x_2),\F_1(x_1'),\F_2(x_2')] \nonumber . 
\end{align}
Eq.~\eqref{kmoc-resp3} is our key formula to compute the change in classical observables from causal response functions. Next, we will explain how this soft limit does indeed reproduce the KMOC formula for the linear impulse and reveals a subtlety in the computation of the angular impulse.


\subsection{Linear impulse}
\label{sec: impulse}
Let us first demonstrate how the KMOC  formula for the impulse of a particle undergoing scattering follows from Eq.~\eqref{kmoc-resp3} using the in-in Feynman rules to compute the response functions discussed in section \ref{subsec:review1}. 
In the KMOC framework the impulse is defined as the difference in expectation value of momentum of particle $1$ defined at $x^0\rightarrow -\infty$  from its value at $x^0\rightarrow \infty$, that is,
\begin{equation}
    \D \mathbb{P}_1^\mu = \bra{1'2'} \mathbb{P}_1^{\mu\, \out}-\mathbb{P}_1^{\mu\, \in}\ket{12}\big|_{\rm cl.}\,,
\end{equation}
with
\begin{align}
    \mathbb{P}_1^{\mu\, \in\hphantom{t}} &= \int \hat{d}^4p_1\; p_1^\mu\hat{\d}(p_1^2-m_1^2) a^{\in\dagger}_{p_1}a_{p_1}^{\in}\,, \\
    \mathbb{P}_1^{\mu\, \out}& = \int \hat{d}^4p_1\; p_1^\mu\hat{\d}(p_1^2-m_1^2) a^{\out\dagger}_{p_1}a_{p_1}^{\out}\,.
\end{align}
The KMOC impulse formula in terms of the in-out scattering amplitudes is \cite{Kosower2019}
\begin{align}
\label{impulse}  
   & \D \mathbb{P}_1^\mu=\int \hat{d}^4q\;\hat{\d}\left(2p_1\cdot  q+ q^2\right)\hat{\d}\left(2p_2\cdot q- q^2\right)\; e^{-ib\cdot  q}\; \Bigg[\;\;q^\mu\;\begin{tikzpicture}[baseline=.1ex,scale=0.35]

    \draw[fermion] (-2,-2) -- (0,0) node[pos=0.35] {} ;
      \node (a) at (-2.5,-2.5){$p_2$};
    \draw[fermion] (2,2) -- (0,0) node[pos=0.1] {} ;
      \node (a) at (3.3,2.7){$p_1+q$};
    \draw[fermion] (-2,2) -- (0,0) node[pos=0.35] {} ;
      \node (a) at (-2.5,2.5){$p_1$};
    \draw[fermion] (2,-2) -- (0,0) node[pos=0.35] {} ;
      \node (a) at (2.5,-2.5){$p_2-q$};

    \node[minimum size=1.2cm, fill = gray!20, draw = black, ultra thick, circle] at (0,0){$i\mathcal{A}$};

\end{tikzpicture} \\
   & +  \sum_X\int \prod_{i=1,2}\hat{d}^4\ell_i\;\hat{\d}_+\left(2p_1\cdot  \ell_i+\ell_i^2\right)\;\hat{\d}(\ell_1+\ell_2+r_X) \;\ell_1^\mu \begin{tikzpicture}[baseline=.1ex,scale=0.35]
\node[inner sep=0pt] (c) at (0,0){};
\node[above left=1 of c] (a) {$p_1$};
    \draw[fermion] (a) -- (c) ;
      \node (b)[below left=1 of c]{$p_2$};
    \draw[fermion] (b) -- (c) ;
   \node[above right=1 of c](d) {};
    \node[below right=1 of c](e){};
    
    \node[inner sep=0pt] (c) at (7,0){};
    \node[above left=1 of c] (a) {};
      \node (b)[below left=1 of c]{};
   \node[inner sep=0pt,above right=1 of c](d) {};
   \node[inner sep=0pt,above right=0.01 of d] (d2){$p_1+q$};
    \draw[fermion] (c) -- (d); 
    \node[inner sep=0pt, below right=1 of c](e){};
     \node[inner sep=0pt,below right=0.01 of e] (e2){$p_2-q$};
    \draw[fermion] (c) -- (e);
    \draw[photon] (0,0)--(7,0);
      \draw [fermion](6.5,0) arc (25:155:3.5);
     \draw [fermion](6.5,0) arc (335:205:3.5);
        \node[minimum size=1.3cm, fill = gray!20, draw = black, ultra thick, circle] at (0,0){$i\mathcal{A}$};
   \node[minimum size=0.6cm, fill = gray!20, draw = black, ultra thick, circle] at (c){$-i\mathcal{A}^*$};
 \node (m1) at (5.1,3){$p_1+\ell_1$};
 \node (m2) at (5.1,-3){$p_2+\ell_2$};
 \node (m3) at (4.5,0.7){$r_X$};
 \draw[brown,fcut](3.3,3.5)--(3.3,-3.5);
\end{tikzpicture}\Bigg]. \nonumber 
\end{align}
where the momentum  transfer $q$ is
\begin{align}
    q^\mu = p_1^{\prime\mu}-p_1^\mu =p_2^\mu - p_2^{\prime\mu}.
\end{align}
and is taken to be small in the classical limit, $q\ll p_i$.
We refer the reader to \cite{Kosower2019} for a detailed derivation.

Next, we will derive this formula from amputated response functions. The relevant local density for $\mathbb{P}_1$ is the momentum density
\begin{equation}
   \mathbb{P}_1^\mu(t) = \int d^3\bm{x} \mathcal{P}_1(t,\bm{x}) =  \int d^3\bm{x} T_1^{0\mu}(t,\bm{x})\,,
\end{equation}
given in terms of the stress-energy tensor of particle 1 
\begin{equation}
    T_1^{\mu\nu} = 2\pd^\mu\F_1\pd^\nu\F_1^\dagger-\eta^{\mu\nu}\left(\pd_\rho\F_1\pd^\rho\F_1^\dagger-m_1^2\F_1\F_i^\dagger\right)\,.
\end{equation}
Thus, to derive the linear impulse we consider the amputated response function 
\begin{equation}
    \langle \mathcal{P}_1^{\mu}(\omega,\bm{k}) \rangle  = \int d^4 x e^{i\omega t-i\bm{k}\bm{x}}\bra{1'2'} T_1^{0\mu} (t,\bm x) \ket{12}\,.
\end{equation}

The Feynman rule corresponding to the heavy-scalar stress-tensor insertion can be easily obtained
\begin{align}
 \raisebox{-20pt}{\begin{tikzpicture}[scale=.99]
    \node (a) at (0,0){};
    \node[right=2 of a] (b) {};
      \draw[fermion] (a) -- (b);  
      \node[inner sep=0pt,right=1 of a](x){};
      \node[above= 0.8 of x](y){};
     \draw [line width=1pt, double distance=1pt] (x)-- (y) ;  
       \draw[->] (0.2,-0.3)--(0.9,-0.3);
       \node(n1) at (0.6,-.6) {$\;\;p_1+k\quad$};
       \draw[->] (1.4,-0.3)--(2.1,-0.3);
       \node(n1) at (1.9,-.6) {$p_1$};
       \draw[->] (1.4,0.3)--(1.4,0.8) node[right]{$ k$};
       \end{tikzpicture}
       }= (p_1+k)^{(0} p_1^{\mu)} = 2p_1^0 p_1^\mu +\mathcal{O}(k)\,,
       \label{p-rule}
\end{align}
where we used an unnormalized symmetrization symbol $A^{(\mu}B^{\nu)} = A^{\mu}B^{\nu} + A^{\nu}B^{\mu}$ and on the second equality we show the terms which are relevant in the soft limit $k\to 0$. Importantly, the energy of the scalar legs is conserved only up to  $\omega=k^0$.
\begin{figure}
    \centering
\subfloat[Operator insertions on uncut diagrams \label{fig:i-ii basis (a)}]{
\begin{tikzpicture}[scale=0.4]
 \node[inner sep=0pt] (c) at (0,0){};
\node[above=0.2 of c] (a) {};
\node[left=1.1 of a](a2){$p_1$};
    \draw[fermion] (a) -- (a2) ;
 \node[inner sep=0pt,left=0.7 of a] (k1){};
       \node[above=0.5 of k1] (k2){$k$};
        \node[above right=0.05 of k1] (mk){$\textcolor{gray}{\I}$};
        \draw [line width=1.5pt, double distance=1pt] (k1)-- (k2) ; 
    \node[inner sep=0pt] (c) at (0,0){};
\node[below=0.2 of c] (b) {};
\node[left=1.1 of b](b2){$p_2$};
    \draw[fermion] (b) -- (b2) ;
  \node[above=0.2 of c](d1){};
   \node[ right=1.1 of d1](d) {{$p_1+q$}};
   \draw[fermion](d1)--(d);
   \node[below=0.2 of c](e1){};
   \node[ right=1.1 of e1](e) {{$p_2-q$}};
   \draw[fermion](e1)--(e);
             \node[minimum size=1.3cm, fill = gray!20, draw = black, ultra thick, circle] at (0,0){$i\mathcal{A}$};
   \node[inner sep=0pt, above right= 0.001 of a2] (na2) {$\textcolor{gray}{\I}$};
   \node[inner sep=0pt, above left= 0.001 of d] (nd1) {$\textcolor{gray}{\I}$};
   \node[inner sep=0pt, below right= 0.001 of b2] (nb2) {$\textcolor{gray}{\I}$};
   \node[inner sep=0pt, below left= 0.001 of e] (ne) {$\textcolor{gray}{\I}$};
 \node[inner sep=0pt] (c) at (11.5,0){};
\node[above=0.2 of c] (a) {};
\node[left=1.1 of a](a2){$p_1$};
    \draw[fermion] (a) -- (a2) ;
   \node[below=0.2 of c] (b) {};
\node[left=1.1 of b](b2){$p_2$};
    \draw[fermion] (b) -- (b2) ;
    \node[above=0.2 of c](d1){};
   \node[ right=1.1 of d1](d) {{$p_1+q$}};
    \draw[fermion] (d1) -- (d); 
     \node[inner sep=0pt,right=0.7 of d1] (k1){};
       \node[above=0.5 of k1] (k2){$k$};
        \node[above left=0.05 of k1] (mk){$\textcolor{gray}{\I}$};
        \draw [line width=1.5pt, double distance=1pt] (k1)-- (k2) ; 
   \node[below=0.2 of c](e1){};
   \node[ right=1.1 of e1](e) {{$p_2-q$}};
   \draw[fermion](e1)--(e);
             \node[minimum size=1.3cm, fill = gray!20, draw = black, ultra thick, circle] at (c){$i\mathcal{A}$};
     \node[inner sep=0pt, above right= 0.001 of a2] (na2) {$\textcolor{gray}{\I}$};
   \node[inner sep=0pt, above left= 0.001 of d] (nd1) {$\textcolor{gray}{\I}$};
   \node[inner sep=0pt, below right= 0.001 of b2] (nb2) {$\textcolor{gray}{\I}$};
   \node[inner sep=0pt, below left= 0.001 of e] (ne) {$\textcolor{gray}{\I}$};
  \node[inner sep=0pt] (c) at (21,0){};
  \node[above=0.2 of c] (a) {};
\node[left=0.8 of a](a2){};
    \draw[fermion] (a) -- (a2) ;
     \node[below=0.2 of c] (b) {};
\node[left=0.8 of b](b2){};
    \draw[fermion] (b) -- (b2) ;
   \node[above right=1 of c](d) {};
    \node[below right=1 of c](e){};
    
    \node[inner sep=0pt] (c) at (7+21,0){};
    \node[above left=1 of c] (a) {};
 
    \node[above=0.2 of c](d1){};
   \node[ right=0.8 of d1](d) {{}};
   \draw[fermion](d1)--(d);
   \node[below=0.2 of c](e1){};
   \node[ right=0.8 of e1](e) {{}};
   \draw[fermion](e1)--(e);
    \draw[photon] (0+21,0)--(7+21,0);
      \draw [fermion](6.5+21,0) arc (25:155:3.5);
     \draw [fermion](6.5+21,0) arc (335:205:3.5);
        \node[minimum size=1.3cm, fill = gray!20, draw = black, ultra thick, circle] at (0+21,0){$i\mathcal{A}$};
   \node[minimum size=1.3cm, fill = gray!20, draw = black, ultra thick, circle] at (c){$i\mathcal{A}$};
 \node (m1) at (4.3+21,2.6){$\ell$};
 \node (m2) at (5.1+21,-3){};
 \node (m3) at (4.5+21,0.7){};
 \node[inner sep=0pt] (k3) at (3.3+21,1.95){};
 \node[above=.5 of k3] (k4){$k$};
  \node[above left=0.05 of k3] (mk){$\textcolor{gray}{\I}$};
   \draw [line width=1.5pt, double distance=1pt] (k3)-- (k4) ;  
     \node[inner sep=0pt, above right= 0.001 of a2] (na2) {$\textcolor{gray}{\I}$};
   \node[inner sep=0pt, above left= 0.001 of d] (nd1) {$\textcolor{gray}{\I}$};
   \node[inner sep=0pt, below right= 0.001 of b2] (nb2) {$\textcolor{gray}{\I}$};
   \node[inner sep=0pt, below left= 0.001 of e] (ne) {$\textcolor{gray}{\I}$};
\end{tikzpicture}
}\\
    
    \subfloat[Cut diagrams with operator insertions\label{fig:i-ii basis (b)}]{
  \begin{tikzpicture}[scale=0.4]
    \node[inner sep=0pt] (c) at (-20,0){};
\node[above=0.28 of c] (a) {};
\node[left=1.1 of a](a2){$p_1$};
    \draw[fermion] (a) -- (a2) ;
       \node[inner sep=0pt,left=0.8 of a] (k1){};
       \node[above=0.5 of k1] (k2){$k$};
       \node[above right=0.05 of k1] (mk){$\textcolor{gray}{\I}$};
        \draw [line width=1.5pt, double distance=1pt] (k1)-- (k2) ; 
\node[below=0.2 of c] (b) {};
\node[left=1.1 of b](b2){$p_2$};
    \draw[fermion] (b) -- (b2) ;
   \node[above right=1 of c](d) {};
    \node[below right=1 of c](e){};
    
    \node[inner sep=0pt] (c) at (-13,0){};
    \node[above left=1 of c] (a) {};
      \node (b)[below left=1 of c]{};
    \node[above=0.2 of c](d1){};
   \node[ right=1.1 of d1](d) {{$p_1+q$}};
   \draw[fermion](d1)--(d);
   \node[below=0.2 of c](e1){};
   \node[ right=1.1 of e1](e) {{$p_2-q$}};
   \draw[fermion](e1)--(e);
    \draw[photon] (-20,0)--(-13,0);
      \draw [fermion](-13.5,0) arc (25:155:3.5);
     \draw [fermion](-13.5,0) arc (335:205:3.5);
        \node[minimum size=1.3cm, fill = gray!20, draw = black, ultra thick, circle] at (-20,0){$i\mathcal{A}$};
   \node[minimum size=0.6cm, fill = gray!20, draw = black, ultra thick, circle] at (c){$-i\mathcal{A}^*$};
 \node (m1) at (-14.9,3){$p_1+\ell_1$};
 \node (m2) at (-14.9,-3){$p_2+\ell_2$};
 \node (m3) at (-15.5,0.7){$r_X$};
 \draw[brown,fcut](-16.7,3.5)--(-16.7,-3.5);
   \node[inner sep=0pt, above right= 0.001 of a2] (na2) {$\textcolor{gray}{\I}$};
   \node[inner sep=0pt, above left= 0.001 of d] (nd1) {$\textcolor{gray}{\II}$};
   \node[inner sep=0pt, below right= 0.001 of b2] (nb2) {$\textcolor{gray}{\I}$};
   \node[inner sep=0pt, below left= 0.001 of e] (ne) {$\textcolor{gray}{\II}$};
   \node[inner sep=0pt,  left= 1.3 of ne] (ne2) {$\textcolor{gray}{\II}$};
   \node[inner sep=0pt,  left= 3.25 of ne] (ne3) {$\textcolor{gray}{\I}$};
   \node[inner sep=0pt,  left= 1.3 of nd1] (nd2) {$\textcolor{gray}{\II}$};
  \node[inner sep=0pt,  left= 3.25 of nd1] (nd3) {$\textcolor{gray}{\I}$}; 
   \node[inner sep=0pt] (c) at (0,0){};
  \node[above=0.2 of c] (a) {};
\node[left=1.1 of a](a2){$p_1$};
    \draw[fermion] (a) -- (a2) ;
     \node[below=0.2 of c] (b) {};
\node[left=1.1 of b](b2){$p_2$};
    \draw[fermion] (b) -- (b2) ;
   \node[above right=1 of c](d) {};
    \node[below right=1 of c](e){};
    
    \node[inner sep=0pt] (c) at (7,0){};
    \node[above left=1 of c] (a) {};
 
    \node[above=0.2 of c](d1){};
   \node[ right=1.1 of d1](d) {{$p_1+q$}};
   \draw[fermion](d1)--(d);
   \node[below=0.2 of c](e1){};
   \node[ right=1.1 of e1](e) {{$p_2-q$}};
   \draw[fermion](e1)--(e);
    \draw[photon] (0,0)--(7,0);
      \draw [fermion](6.5,0) arc (25:155:3.5);
     \draw [fermion](6.5,0) arc (335:205:3.5);
        \node[minimum size=1.3cm, fill = gray!20, draw = black, ultra thick, circle] at (0,0){$i\mathcal{A}$};
   \node[minimum size=0.6cm, fill = gray!20, draw = black, ultra thick, circle] at (c){$-i\mathcal{A}^*$};
 \node (m1) at (5.1,3){$p_1+\ell_1$};
 \node (m2) at (5.1,-3){$p_2+\ell_2$};
 \node (m3) at (4.5,0.7){$r_X$};
 \draw[brown,fcut](3.3,3.5)--(3.3,-3.5);
 \node[inner sep=0pt] (k3) at (2.5,1.9){};
 \node[above=.5 of k3] (k4){$k$};
  \node[above left=0.05 of k3] (mk){$\textcolor{gray}{\I}$};
   \draw [line width=1.5pt, double distance=1pt] (k3)-- (k4) ;  
     \node[inner sep=0pt, above right= 0.001 of a2] (na2) {$\textcolor{gray}{\I}$};
   \node[inner sep=0pt, above left= 0.001 of d] (nd1) {$\textcolor{gray}{\II}$};
   \node[inner sep=0pt, below right= 0.001 of b2] (nb2) {$\textcolor{gray}{\I}$};
   \node[inner sep=0pt, below left= 0.001 of e] (ne) {$\textcolor{gray}{\II}$};
   \node[inner sep=0pt,  left= 1.3 of ne] (ne2) {$\textcolor{gray}{\II}$};
   \node[inner sep=0pt,  left= 3.25 of ne] (ne3) {$\textcolor{gray}{\I}$};
   \node[inner sep=0pt,  left= 1.3 of nd1] (nd2) {$\textcolor{gray}{\II}$};
  \node[inner sep=0pt,  left= 3.25 of nd1] (nd3) {$\textcolor{gray}{\I}$}; 
     
\end{tikzpicture}
}
    \caption{Two different types of diagrams with the momentum operator insertions are shown. We have taken $p_1,\;p_2$ to be incoming and $p_1+q,\;p_2-q,\; k$ to be outgoing.}
    \label{fig:I_II-basis}
\end{figure}

In writing the diagrams for this response function we assign a type $\I$ vertex for the operator insertion with energy $\omega$ flowing \textit{out} of the diagrams and all possible assignments of $\I/\II$ vertices for the remaining vertices. The in-in Feynman rules lead to various classes diagrams with all type $\I$ fields as well as cut diagrams involving type $\I$ and type $\II$ fields as shown in Fig.~\ref{fig:I_II-basis}. Consider the types of diagrams in Fig.~\ref{fig:i-ii basis (a)} where $\mathcal{O}$ is inserted on the external legs. The first diagram corresponds to multiplying four-point the amplitude $i\mathcal{A}$ without the soft operator insertion by an additional propagator. Taking the soft limit we find
\begin{align}
 \lim_{\omega\rightarrow 0}(-2i\omega p_1^0p_1^\mu)\frac{i}{(\omega-p_1^0)^2-\bm{p}_1^2-m_1^2+i\e}=\lim_{\omega\rightarrow 0}\frac{2\omega p_1^0p_1^\mu}{-2p^0\omega+\omega^2+i\e}=  -p_1^\mu\,,
\end{align}
where in the second line we have used the on-shell condition $p_1^2=m_1^2$.
Similarly, the second diagram with the momentum operator inserted on the outgoing external line produces a factor of $p_1^\mu+q^\mu$. These two classes of diagrams combined yield
\begin{equation}
    -\lim_{\o\rightarrow 0}\lim_{\bm{k}\rightarrow 0} i\o \Bigg[\!\!
    \raisebox{-16pt}{
    \begin{tikzpicture}[scale=0.4]
 \node[inner sep=0pt] (c) at (0,0){};
\node[above=0.2 of c] (a) {};
\node[left=1.1 of a](a2){};
    \draw[fermion] (a) -- (a2) ;
 \node[inner sep=0pt,left=0.7 of a] (k1){};
       \node[above=0.5 of k1] (k2){$k$};
        \draw [line width=1.5pt, double distance=1pt] (k1)-- (k2) ; 
    \node[inner sep=0pt] (c) at (0,0){};
\node[below=0.2 of c] (b) {};
\node[left=1.1 of b](b2){};
    \draw[fermion] (b) -- (b2) ;
  \node[above=0.2 of c](d1){};
   \node[ right=1.1 of d1](d) {};
   \draw[fermion](d1)--(d);
   \node[below=0.2 of c](e1){};
   \node[ right=1.1 of e1](e) {};
   \draw[fermion](e1)--(e);
             \node[minimum size=1.3cm, fill = gray!20, draw = black, ultra thick, circle] at (0,0){$i\mathcal{A}$};
   \end{tikzpicture}}
   \;\;+
\raisebox{-16pt}{
\begin{tikzpicture}
 \node[inner sep=0pt] (c) at (11.5,0){};
\node[above=0.2 of c] (a) {};
\node[left=1.1 of a](a2){};
    \draw[fermion] (a) -- (a2) ;
   \node[below=0.2 of c] (b) {};
\node[left=1.1 of b](b2){};
    \draw[fermion] (b) -- (b2) ;
    \node[above=0.2 of c](d1){};
   \node[ right=1.1 of d1](d) {};
    \draw[fermion] (d1) -- (d); 
     \node[inner sep=0pt,right=0.7 of d1] (k1){};
       \node[above=0.5 of k1] (k2){$k$};
        \draw [line width=1.5pt, double distance=1pt] (k1)-- (k2) ; 
   \node[below=0.2 of c](e1){};
   \node[ right=1.1 of e1](e) {};
   \draw[fermion](e1)--(e);
             \node[minimum size=1.3cm, fill = gray!20, draw = black, ultra thick, circle] at (c){$i\mathcal{A}$};
\end{tikzpicture}
    } \!\! \Bigg]
    = \;\; q^\mu \hspace{-4pt}
    \raisebox{2pt}{
    \begin{tikzpicture}[baseline=.1ex,scale=0.4]

    \draw[fermion] (-2,-2) -- (0,0) node[pos=0.35] {} ;
      \node (a) at (-2.5,-2.5){$p_2$};
    \draw[fermion] (2,2) -- (0,0) node[pos=0.1] {} ;
      \node (a) at (3.3,2.7){$p_1+q$};
    \draw[fermion] (-2,2) -- (0,0) node[pos=0.35] {} ;
      \node (a) at (-2.5,2.5){$p_1$};
    \draw[fermion] (2,-2) -- (0,0) node[pos=0.35] {} ;
      \node (a) at (2.5,-2.5){$p_2-q$};

    \node[minimum size=1.2cm, fill = gray!20, draw = black, ultra thick, circle] at (0,0){$i\mathcal{A}$};
    \end{tikzpicture}\,.
    }
    \label{softamp}
\end{equation}
which reproduces the first term in the KMOC formula in Eq.~\eqref{impulse}.
The uncut diagram on the right of Fig.~\ref{fig:i-ii basis (a)}, which features the operator inserted on an internal line with momentum $\ell$ does not contribute in the soft limit 
\begin{align}
 &\lim_{\o\rightarrow0}\frac{2i\omega \ell^0\ell^\mu}{\left((\omega+\ell^0)^2-\bm{\ell}^2-m_1^2+i\e\right)\left(\ell^2-m_1^2+i\e\right)}=0
 \label{partial-fraction}
\end{align}
Finally, let us consider the cut diagrams obtained by connecting a type $\I$ vertex with a type $\II$ vertex in Fig.~\ref{fig:i-ii basis (b)}. The stability condition, which requires the vanishing of the three-particle cuts, and the positivity of energy flow along the cuts guarantee that only the unitarity cuts survive as depicted in Fig.~\ref{fig:I_II-basis}. Using the identity in Eq.~\ref{softamp} on the left-hand side of the cut one easily finds that these two classes of diagrams combined give the cut term in the second line of Eq.~\eqref{impulse}.  All together, we have reproduced the KMOC formula for the linear impulse from the soft limit of a causal response function of the stress tensor.
\subsection{Radiated linear momentum}
\begin{figure}[t!]
\subfloat[\label{fig:i-ii-photon(a)}]{

\raisebox{15pt}{\begin{tikzpicture}[scale=0.4]
     \node[inner sep=0pt] (c) at (0,0){};
  \node[above=0.2 of c] (a) {};
\node[left=0.8 of a](a2){};
    \draw[fermion] (a) -- (a2) ;
     \node[below=0.2 of c] (b) {};
\node[left=0.8 of b](b2){};
    \draw[fermion] (b) -- (b2) ;
   \node[above right=1 of c](d) {};
     \node[below right=1 of c](e){};
    \node[inner sep=0pt] (c2) at (7,0){};
    \node[above left=1 of c2] (a) {};
    \node[above=0.2 of c2](d1){};
   \node[ right=0.8 of d1](d) {{}};
   \draw[fermion](d1)--(d);
   \node[below=0.2 of c2](e1){};
   \node[ right=0.8 of e1](e) {{}};
   \draw[fermion](e1)--(e);
    \draw[photon] (c)--(c2);
      \draw [fermion](6.5,0) arc (25:155:3.5);
     \draw [fermion](6.5,0) arc (335:205:3.5);
        \node[minimum size=1.3cm, fill = gray!20, draw = black, ultra thick, circle] at (0,0){$i\mathcal{A}$};
   \node[minimum size=01.3cm, fill = gray!20, draw = black, ultra thick, circle] at (c2){$i\mathcal{A}$};
 \node[inner sep=0pt] (k3) at (3.42,-.13){};
  \node[above=.6 of k3] (k4){};
    \draw [line width=1.5pt, double distance=1pt] (k3)-- (k4); 
  \node[above left=0.05 of k3] (mk){$\textcolor{gray}{\I}$};
   \draw [line width=1.5pt, double distance=1pt] (k3)-- (k4) ;  
     \node[inner sep=0pt, above right= 0.001 of a2] (na2) {$\textcolor{gray}{\I}$};
   \node[inner sep=0pt, above left= 0.001 of d] (nd1) {$\textcolor{gray}{\I}$};
   \node[inner sep=0pt, below right= 0.001 of b2] (nb2) {$\textcolor{gray}{\I}$};
   \node[inner sep=0pt, below left= 0.001 of e] (ne) {$\textcolor{gray}{\I}$};
     \node[inner sep=0pt, below left= 0.001 of e] (ne) {$\textcolor{gray}{\I}$};
    \node[inner sep=0pt, right=3.05 of nb2] (nb3) {};
     \node[inner sep=0pt, below=0.001 of nb3] (nb4) {$\textcolor{gray}{\I}$};
      \node[inner sep=0pt, above=1.21 of nb3] (nb5) {$\textcolor{gray}{\I}$};
    \node[inner sep=0pt, left=1.75 of nb4] (nb4) {$\textcolor{gray}{\I}$};
    \node[inner sep=0pt, left=1.75 of nb5] (nb4) {$\textcolor{gray}{\I}$};
    \end{tikzpicture}}}
\subfloat[\label{fig:i-ii-photon(b)}]{
\begin{tikzpicture}[scale=0.4]
   \node[inner sep=0pt] (c) at (15,0){};
  \node[above=0.2 of c] (a) {};
\node[left=0.8 of a](a2){};
    \draw[fermion] (a) -- (a2) ;
     \node[below=0.2 of c] (b) {};
\node[left=0.8 of b](b2){};
    \draw[fermion] (b) -- (b2) ;
   \node[above right=1 of c](d) {};
     \node[below right=1 of c](e){};
    \node[inner sep=0pt] (c2) at (7+15,0){};
    \node[above left=1 of c2] (a) {};
    \node[above=0.2 of c2](d1){};
   \node[ right=0.8 of d1](d) {{}};
   \draw[fermion](d1)--(d);
   \node[below=0.2 of c2](e1){};
   \node[ right=0.8 of e1](e) {{}};
   \draw[fermion](e1)--(e);
    \draw[photon] (c)--(c2);
      \draw [fermion](6.5+15,0) arc (25:155:3.5);
     \draw [fermion](6.5+15,0) arc (335:205:3.5);
        \node[minimum size=1.3cm, fill = gray!20, draw = black, ultra thick, circle] at (c){$i\mathcal{A}$};
   \node[minimum size=0.6cm, fill = gray!20, draw = black, ultra thick, circle] at (c2){$-i\mathcal{A}^*$};
 \node[inner sep=0pt] (k3) at (3.42+15,-.13){};
  \node[above=.6 of k3] (k4){};
    \draw [line width=1.5pt, double distance=1pt] (k3)-- (k4); 
\node (x1) at (4.2+15,3.3){};
\node[below=2.5 of x1](x2){};
\draw[fcut,brown] (x1)--(x2);
  \node[above left=0.05 of k3] (mk){$\textcolor{gray}{\I}$};
   \draw [line width=1.5pt, double distance=1pt] (k3)-- (k4) ;  
     \node[inner sep=0pt, above right= 0.001 of a2] (na2) {$\textcolor{gray}{\I}$};
   \node[inner sep=0pt, above left= 0.001 of d] (nd1) {$\textcolor{gray}{\II}$};
   \node[inner sep=0pt, below right= 0.001 of b2] (nb2) {$\textcolor{gray}{\I}$};
   \node[inner sep=0pt, below left= 0.001 of e] (ne) {$\textcolor{gray}{\II}$};
    \node[inner sep=0pt, right=3.05 of nb2] (nb3) {};
     \node[inner sep=0pt, below=0.001 of nb3] (nb4) {$\textcolor{gray}{\II}$};
      \node[inner sep=0pt, above=1.21 of nb3] (nb5) {$\textcolor{gray}{\II}$};
    \node[inner sep=0pt, left=1.6 of nb4] (nb4) {$\textcolor{gray}{\I}$};
    \node[inner sep=0pt, left=1.6 of nb5] (nb4) {$\textcolor{gray}{\I}$};
\end{tikzpicture}}
    \caption{Two types of diagrams contributing to the causal response function for the photon stress tensor.}
    \label{fig:I_II-basis-photon}
\end{figure}

It is easy to generalize the above discussion to give a formula for the radiated momentum, which is carried away by the photons:
\begin{equation}
    \D \mathbb{P}_\gamma^\mu = \bra{1'2'} \mathbb{P}_\gamma^{\mu\, \out}-\mathbb{P}_\gamma^{\mu\, \in}\ket{12}\big|_{\rm cl.} =  \bra{1'2'} \mathbb{P}_\gamma^{\mu\, \out}\ket{12}\big|_{\rm cl.}\,,
\end{equation}
where we used the absence of photons in the initial state to set $\mathbb{P}_\gamma^{\mu\, \in}\ket{12}=0$ and 
\begin{align}
    \mathbb{P}_\gamma^{\mu\, \out}& = \sum_h\int \hat{d}^4p\; p^\mu\hat{\d}(p^2) a^{h\,\out\dagger}_{p}a_{p}^{h\,\out}\,.
\end{align}
This can be written in terms of the local density
\begin{equation}
   \mathbb{P}_\gamma^\mu(t) = \int d^3\bm{x} \mathcal{P}^\mu_\gamma(t,\bm{x}) =  \int d^3\bm{x} T_\gamma^{0\mu}(t,\bm{x})\,,
\end{equation}
where the stress energy tensor of the photon is
\begin{equation}
    T_{\gamma}^{\mu\nu}=F^{\m\r}F_\r^\n-\frac14 \eta^{\nu\nu}F^{\r\s}F_{\r\s}\,.
    \label{ph-stress-tensor}
\end{equation}
The Feynman rule corresponding to the insertion of this operator is
 \begin{align}
 \raisebox{-15pt}{\begin{tikzpicture}[scale=.99]
    \node (a) at (0,0){$^\alpha$};
    \node[right=1.9 of a] (b) {$^\beta$};
      \draw[photon] (a) -- (b);  
      \node[inner sep=0pt,right=.9 of a](x){};
      \node[above= 0.8 of x](y){};
       \draw [line width=1pt, double distance=1pt] (x)-- (y) ;  
       \draw[->] (0.2,-0.3)--(0.9,-0.3);
       \node(n1) at (0.6,-.6) {$\;\;\ell+k\quad$};
       \draw[->] (1.4,-0.3)--(2.1,-0.3);
       \node(n1) at (1.9,-.6) {$\ell$};
       \draw[->] (1.4,0.3)--(1.4,0.8) node[right]{$ k$};
              \end{tikzpicture}
       }&= (\ell+k)^\a\ell^{(\m}\eta^{0)\b}+\ell^\b\eta^{\a(\m}(\ell+k)^{0)}\nonumber\\
       &-\eta^{\a (\m}\eta^{0)\b}\ell\cdot(\ell+k)-\eta^{\a\b}\ell^{(\m}(\ell+k)^{0)}\,.
          \label{p-rule-em}
\end{align}
This simplifies considerably when one of the legs is put on-shell and contracted with a transverse polarization
 \begin{align}
 \raisebox{-25pt}{\begin{tikzpicture}[scale=.99]
    \node (a) at (0,0){$^\alpha$};
    \node[right=2.4 of a] (b) {};
      \draw[photon] (a) -- (b);  
      \node[inner sep=0pt,right=.9 of a](x){};
      \node[above= 0.8 of x](y){};
       \draw [line width=1pt, double distance=1pt] (x)-- (y) ;  
       \draw[->] (0.2,-0.3)--(0.9,-0.3);
       \node(n1) at (0.6,-.6) {$\;\;\ell+k\quad$};
       \draw[->] (1.4,-0.3)--(2.1,-0.3);
       \node(n1) at (1.9,-.6) {$\ell$};
       \draw[->] (1.4,0.3)--(1.4,0.8) node[right]{$k$};
       \node[right=1.9 of a] (x0) {};
       \node[above=0.8 of x0] (x1) {};
\node[below=1.6 of x1](x2){};
\draw[fcut,brown] (x1)--(x2);
              \end{tikzpicture}
       }&= - 2 \ell^0 \ell^\mu \varepsilon^\alpha + \ell^\alpha \varepsilon^{(\mu} \ell^{0)}  +{\cal O}(k) \,,
       \label{p-rule-em-soft}
\end{align}
where we only display terms that remain in the soft limit $k\to 0$.

The absence of photons in the initial state implies that the only classes of diagrams that can contribute to the response function of the photon stress tensor are those in Fig.~\ref{fig:I_II-basis-photon}. The diagrams with only $\I$ or $\II$ fields in Fig.~\ref{fig:i-ii-photon(a)} vanish in the soft limit by a manipulation analogous to Eq.~\eqref{partial-fraction}. On the other hand, the diagram with both $\I$ or $\II$ fields yields the cut diagrams of the form of Fig.~\ref{fig:i-ii-photon(a)} where on the support of the on-shell delta function for the photon, $\delta(\ell^2)$, we can use the completeness relation $\eta^{\beta \delta} = \sum_h \varepsilon_h^\beta \varepsilon^{*\,\delta}_h$, which holds up to terms that vanish due to the Ward-Takahashi identity. In the soft limit these diagrams produce terms on the form
\begin{align}
 \lim_{\omega\rightarrow 0} &(-i\omega)  {\cal A}_\alpha \frac{-i (- 2 \ell^0 \ell^\mu \varepsilon^\alpha + \ell^\alpha \varepsilon^{(\mu} \ell^{\,0)})}{(\omega+\ell^0)^2-\bm{\ell}^2+i\e} = \lim_{\omega\rightarrow 0} {\cal A}_\alpha \varepsilon^\alpha \frac{2\omega \ell^0 \ell^\mu}{2\omega\ell^0+\omega^2+i\e} =  \ell^\mu {\cal A}\,,
\end{align}
where ${\cal A} = \varepsilon^\alpha{\cal A}_\alpha $ denotes the amplitude on the left of the cut, and we used its on-shell Ward identity, $\ell^\alpha {\cal A}_\alpha=0$. Thus, from the soft limit of the causal response function we obtain the KMOC formula for the radiated momentum 
\begin{align}
   & \D \mathbb{P}_\gamma^\mu= \int \hat{d}^4q\;\hat{\d}\left(2p_1\cdot  q+ q^2\right)\hat{\d}\left(2p_2\cdot q- q^2\right)\; e^{-ib\cdot  q}\; \Bigg[\\
   &\sum_X\int \prod_{i=1,2}\hat{d}^4\ell_i\;\hat{\d}_+\left(2p_1\cdot  \ell_i+\ell_i^2\right)\;\hat{\d}(\ell_1+\ell_2+r_X) \;r_X^\mu \begin{tikzpicture}[baseline=.1ex,scale=0.35]
\node[inner sep=0pt] (c) at (0,0){};
\node[above left=1 of c] (a) {$p_1$};
    \draw[fermion] (a) -- (c) ;
      \node (b)[below left=1 of c]{$p_2$};
    \draw[fermion] (b) -- (c) ;
   \node[above right=1 of c](d) {};
    \node[below right=1 of c](e){};
    
    \node[inner sep=0pt] (c) at (7,0){};
    \node[above left=1 of c] (a) {};
      \node (b)[below left=1 of c]{};
   \node[inner sep=0pt,above right=1 of c](d) {};
   \node[inner sep=0pt,above right=0.01 of d] (d2){$p_1+q$};
    \draw[fermion] (c) -- (d); 
    \node[inner sep=0pt, below right=1 of c](e){};
     \node[inner sep=0pt,below right=0.01 of e] (e2){$p_2-q$};
    \draw[fermion] (c) -- (e);
    \draw[photon] (0,0)--(7,0);
      \draw [fermion](6.5,0) arc (25:155:3.5);
     \draw [fermion](6.5,0) arc (335:205:3.5);
        \node[minimum size=1.3cm, fill = gray!20, draw = black, ultra thick, circle] at (0,0){$i\mathcal{A}$};
   \node[minimum size=0.6cm, fill = gray!20, draw = black, ultra thick, circle] at (c){$-i\mathcal{A}^*$};
 \node (m1) at (5.1,3){$p_1+\ell_1$};
 \node (m2) at (5.1,-3){$p_2+\ell_2$};
 \node (m3) at (4.5,0.7){$r_X$};
 \draw[brown,fcut](3.3,3.5)--(3.3,-3.5);
\end{tikzpicture}\Bigg]. \nonumber \label{momrad}  
\end{align}
where we recall that $r_X = \sum_{a} \ell_a$ is the sum over the momenta of all photons in the cut.

Finally, we note that in deriving the above results, we assumed $\omega\ll \ell$, and expanded the Feynman rules and propagators using the method of regions \cite{Smirnov:2002pj}. For the linear impulse and momentum loss there are no other regions in the expansion that contribute, but we will see next that this is not true when considering the angular momentum loss.
\subsection{Radiated angular momentum}
\label{sec: angular-impulse}

In this section, we compute angular momentum loss in the form of photons during scattering, from a response function. A key difference between this calculation and that of the linear impulse or energy loss is that the soft limit $\omega \to 0$ of the response function is subtle. 

We take the canonical definition of the angular momentum operator for a massive or massless particle as an integral of the corresponding density
\begin{align}
   \mathbb{J}_\gamma^{ij}(t)= \int d^3 \bm{x}\, \mathcal{J}^{ij}(t,\bm x) = \int d^3 \bm{x}\, \bm{x}^{[i}T_\gamma^{0j]}(x)= \int d^3 \bm{x}\, (\bm{x}^iT_\gamma^{0j}(x)-\bm{x}^jT_\gamma^{0i}(x)) \,,
\end{align}
where $T^{\mu\nu}$ is the appropriate energy-momentum tensor, and $i,j$ are spatial indices. Thus we conclude that the angular impulse is given as the soft limit of a derivative of the same amputated causal response function
\begin{equation}
 \D \mathbb{J}_\gamma^{ij} =  -\lim_{\o\rightarrow 0}\lim_{\bm{k}\rightarrow 0} i \o\;  \langle \mathcal{J}_\gamma^{ij}(\omega,\bm{k}) \rangle\big|_{\rm cl.} = -\lim_{\o\rightarrow 0}\lim_{\bm{k}\rightarrow 0} \o\; \partial_{\bm k}^{[i} \langle \mathcal{P}_\gamma^{j]}(\omega,\bm{k}) \rangle\big|_{\rm cl.} \,.\label{angular-momentum}
\end{equation}

Naively, for this computation, we can simply consider the diagrams in Fig.~\ref{fig:I_II-basis-photon} and compute the appropriate derivatives and soft limit as prescribed in Eq.~\eqref{angular-momentum}. But the set of diagrams that contribute will be slightly different. In the calculation of the radiated momentum $\D \mathbb{P}_\gamma$ we tacitly discarded the diagrams in Fig.~\ref{cut3}, because in the soft limit, they yield an on-shell three-point amplitude, which vanishes for real kinematics.
\begin{figure}[t]
    \subfloat[\label{fig:A3A5(a)}]{
\begin{tikzpicture}[scale=0.4]
     \node[inner sep=0pt] (c) at (0,0){};
  \node[left=1 of c] (a) {};
  \draw[fermion] (a) -- (c) ;
     \node[below=0.2 of c] (b) {};
   \node[above right=1 of c](d) {};
     \node[below right=1 of c](e){};
    \node[inner sep=0pt] (c2) at (7,0){};
    \draw[photon] (c)--(c2);
   \node[above right=1 of c2](a2){};
   \node[below right=1 of c2](b2){};
   \node[below left=1 of c2](d2){};
   \draw[fermion] (c2)--(a2);
   \draw[fermion] (c2)--(b2);
   \draw[fermion] (c2)--(d2);   
     \draw [fermion](6.5,0) arc (25:155:3.5);
        \node[minimum size=.4cm, fill = gray!20, draw = black, ultra thick, circle] at (c){$i\mathcal{A}_3$};
   \node[minimum size=1.3cm, fill = gray!20, draw = black, ultra thick, circle] at (c2){$i\mathcal{A}$};
 \node[inner sep=0pt] (k3) at (3.42,0){};
  \node[below=.6 of k3] (k4){$k$};
    \draw [line width=1.5pt, double distance=1pt] (k3)-- (k4); 
    \node[inner sep=0pt,above= 0.1 of k3] (nk3) {$\textcolor{gray}{\I}$};
    \node[inner sep=0pt,left= 1.9 of nk3] (nk4) {$\textcolor{gray}{\I}$};
    \node[inner sep=0 pt, above= .7 of c](piv){};
    \node[inner sep=0pt,right= 2.15 of piv](p1) {$\textcolor{gray}{\;\I}$};
    \node[inner sep=0pt,right= 3 of piv](p2) {$\textcolor{gray}{\;\I}$};
    \node[inner sep=0pt,below= 1.3 of p1](p11) {$\textcolor{gray}{\;\I}$};
    \node[inner sep=0pt,below= 1.3 of p2](p22) {$\textcolor{gray}{\;\I}$};
    \node[inner sep=0 pt, above right=0.3 of c] (na)  {};
    \node[inner sep=0 pt, above=0.3 of na] (na2)  {$\textcolor{gray}{\;\I}$};
    \end{tikzpicture}
    }
\subfloat[\label{fig:A3A5(b)}]{    
\begin{tikzpicture}[scale=0.4]
   \node[inner sep=0pt] (c) at (0,0){};
  \node[above=0.2 of c] (a) {};
\node[left=1 of c](a2){};
    \draw[fermion] (c) -- (a2) ;
   \node[above right=1 of c](d) {};
     \node[below right=1 of c](e){};
     \node[inner sep=0pt] (c2) at (7,0){};
    \draw[photon] (c)--(c2);
      \node[above right=1 of c2](a2){};
   \node[below right=1 of c2](b2){};
   \node[below left=1 of c2](d2){};
   \draw[fermion] (c2)--(a2);
   \draw[fermion] (c2)--(b2);
   \draw[fermion] (c2)--(d2); 
      \draw [fermion](6.5,0) arc (25:155:3.5);
        \node[minimum size=.4cm, fill = gray!20, draw = black, ultra thick, circle] at (c){$i\mathcal{A}_3$};
   \node[minimum size=0.6cm, fill = gray!20, draw = black, ultra thick, circle] at (c2){$-i\mathcal{A}^*$};
 \node[inner sep=0pt] (k3) at (3.42,0){};
  \node[below=.6 of k3] (k4){$k$};
    \draw [line width=1.5pt, double distance=1pt] (k3)-- (k4); 
\node (x1) at (4.2,3.1){};
\node[below=2 of x1](x2){};
\draw[fcut,brown] (x1)--(x2);
    \node[inner sep=0pt,above= 0.1 of k3] (nk3) {$\textcolor{gray}{\I}$};
    \node[inner sep=0pt,left= 1.9 of nk3] (nk4) {$\textcolor{gray}{\I}$};
    \node[inner sep=0 pt, above= .25 of a](piv){};
    \node[inner sep=0pt,right= 2.15 of piv](p1) {$\textcolor{gray}{\II}$};
    \node[inner sep=0pt,right= 3 of piv](p2) {$\textcolor{gray}{\II}$};
    \node[inner sep=0pt,below= 1.3 of p1](p11) {$\textcolor{gray}{\II}$};
    \node[inner sep=0pt,below= 1.3 of p2](p22) {$\textcolor{gray}{\II}$};
    \node[inner sep=0 pt, above right=0.1 of a] (na)  {$\textcolor{gray}{\I}$};
\end{tikzpicture}
}
    \caption{Diagrams relevant for the computation of the angular momentum loss.}
    \label{cut3}
\end{figure}
However, to reach this conclusion we expanded the integrand in the limit $\omega \ll \ell$, which implicitly assumes that there are no relevant regions in which the loop momentum is or order $\ell \sim \omega$. This assumption is valid for the class of diagrams without cut propagators in Fig.~\ref{fig:A3A5(a)}, which indeed vanish; but it fails when considering the diagrams in Fig.~\ref{fig:A3A5(b)}. The region $\ell \sim \omega$ of the loop integration in these diagrams in fact completely captures the leading contribution to the angular momentum loss.
The mechanism for this is that the integrals develop a pole in $\omega$ in the soft limit, thanks to the cut photon propagator
\begin{align*}
 \lim_{\k\to 0}  \frac{\hat{\d}_+(\ell^2)\hat{\d}\left((p_1-k-\ell)^2-m_1^2\right)}{(\ell+k)^2+i\e}=\frac{\hat{\d}_+(\ell^2)\hat{\d}\left(2p_1\cdot\ell-2p_1^0\omega\right)}{2\ell^0\o+\o^2+i\e}\,.
\end{align*}
In the region $\o\sim \ell$, one must not drop the $p_1^0\omega$ term in the delta function, and power-counting the measure as $d^4\ell \sim \omega^4$ we find that the result is indeed of order $1/\omega$. The shift in the on-shell delta function for the massive particle can be interpreted as it being off-shell by an infinitesimal amount of ${\cal O}(\omega)$. We will see momentaritly that this off-shell-ness makes the evaluation of related integrals completely unambiguous. 
Furthermore, since we are interested in small $\o\sim\ell$ we can write $(q_2-\ell)^2\sim q_2^2$, so the amplitude on the other side of the cut effectively becomes a contact term, as it cannot be resolved by waves with small frequency of ${\cal O}(\omega)$.
In hindsight, this justifies the calculations of the angular momentum loss in refs. \cite{Manohar:2022dea, DiVecchia:2022owy} using Weinberg's soft theorem.

\begin{figure}[t]
    \centering
    
\begin{tikzpicture}[scale=1]
\node (a) at (-.8,0){$p_1$};
\node[right=3 of a](b){$p_1+q_1$};
\node[below=1.22 of a](c){$p_2$};
\node[right=3 of c](d){$p_2+q_2$};
\draw[fermion] (a)--(b);
\draw[fermion] (c)--(d);
\node[inner sep=0pt,right=0.75 of a](x1){};
\node[inner sep=0pt, right=0.75 of c](x2){};
\draw[photon](x1)--(x2);
\node[inner sep=0pt,left=0.25 of b](x3){};
\node[inner sep=0pt, left=0.25 of d](x4){};
\draw[photon](x3)--(x4);
\node[inner sep=0pt, below=0.47 of x1](w){};
\node[left=0.45 of w](w1){$k$};
  \draw [line width=1.5pt, double distance=1pt] (w)-- (w1) ; 
  \node[inner sep=0 pt] (c1) at (0.8-0.8,-1.4){};
  \node[inner sep=0 pt, above right=2.4 of c1] (c2){};
  \draw[fcut, brown]  (c1)--(c2);
  \node (l1) at (0.6-0.8,-1.6){};
  \node[above=0.7 of l1](l2){};
  \draw[<-](l1)--(l2);
  \node (n) at (0.3-0.8,-1.2){$\ell$};

\node (a) at (4.3,0){};
\node[right=3 of a](b){};
\node[below=1.5 of a](c){};
\node[right=3 of c](d){};
\draw[fermion] (a)--(b);
\draw[fermion] (c)--(d);
\node[inner sep=0pt,right=0.75 of a](x1){};
\node[inner sep=0pt, right=0.83 of c](x2){};
\node[inner sep=0pt,left=0.27 of b](x3){};
\node[inner sep=0pt, left=0.25 of d](x4){};
\draw[photon](x3)--(x2) node[midway](m){};
\draw[white,fill=white] (m) circle (0.15);
\draw[photon](x1)--(x4);
\node[inner sep=0 pt] at (4.3+.75+.72,-.52)(w){};
\node[left=0.45 of w](w1){};
  \draw [line width=1.5pt, double distance=1pt] (w)-- (w1) ; 
\draw[white,fill=white] (a) circle (.5);
\draw[white,fill=white] (c) circle (.5);
  \node[inner sep=0 pt] (c1) at (1.4+4.3,-1){};
  \node[inner sep=0 pt, above right=1.8 of c1] (c2){};
  \draw[fcut, brown]  (c1)--(c2);
  \node (l1) at (3+4.3,-1.6){};
  \node[above left=0.8 of l1](l2){};
  \draw[<-](l1)--(l2);
  \node (n) at (2.7+4.3,-.9){$\ell$};
\node (a) at(8.3,0){};
\node[right=3 of a](b){ };
\node[below=1.5 of a](c){ };
\node[right=3 of c](d){ };
\draw[fermion] (c)--(d);
\node[inner sep=0pt,right=.6 of a](x1){};
\node[inner sep=0pt, right=.6 of c](x2){};
\node[inner sep=0pt, right=1.5 of a](m1){};
\node[inner sep=0pt, right=1.5 of c](m2){};
\node[inner sep=0pt,left=0.25 of b](x3){};
\node[inner sep=0pt, left=0.25 of d](x4){};
\draw [decorate, decoration={snake, segment length=2.8 mm, amplitude=.5 mm, pre length=0.7 mm, post length=0.01 mm}](x3) arc (345:185:1.15);
\node[above=0.7 of m2](v){}; 
\draw[white,fill=white] (v) circle (0.15);
\node (v2) at (8.3+.62,0.13){};
\draw[white,fill=white] (v2) circle (0.1);
\draw[photon](m1)--(m2);
\draw[fermion] (a)--(b);
\node[inner sep=0 pt] at (8.3+.93,-.53)(w){};
\node[left=0.45 of w](w1){};
  \draw [line width=1.5pt, double distance=1pt] (w)-- (w1) ; 
  \node[inner sep=0 pt] (c1) at (8.3+1.2,.3){};
  \node[inner sep=0 pt, below=1.35 of c1] (c2){};
  \draw[fcut, brown]  (c1)--(c2);
  \node (l1) at (8.8+1.4,-1.7){};
  \node[above=0.45 of l1](l2){};
  \draw[<-](l1)--(l2);
  \node (n) at (9.3+1.2,-1.35){$q_2$};
  \node (v1) at (1.2,-2.2){$(a)$};
  \node (v2) at (2+4.3,-2.2){$(b)$};
  \node (v3) at (2+8.3,-2.2){$(c)$};
\end{tikzpicture}
    \caption{Three diagrams contributing to the angular impulse are shown. We take the  $p_1,p_2$ incoming and $p_1+q_1,p_2+q_2,\;k$ outgoing.}
    \label{fig:ang-impulse-1}
\end{figure}

Let us now compute this quantity at leading order in electromagnetism. 
The leading diagrams  we can draw are the cut diagrams shown in Fig.~\ref{fig:ang-impulse-1}.
These can be straightforwardly computed using the Feynman rule in Eq.~\eqref{p-rule-em}. Although the latter looks rather complicated, the calculation simplifies greatly by using power counting to read off the contributing terms after applying the derivatives. For more details, we refer the reader to Appendix~\ref{sec: angular-impulse-scalar} where we present the calculation in detail for a scalar toy model. In electromagnetism, summing up the diagrams in Fig.~\ref{fig:ang-impulse-1} we get
\begin{align}
    \mathcal{J}_{ph,1}^{ij}&=\frac{ie^4}{4}\lim_{\o\rightarrow 0}e^4Q_1Q_2\;p_1\cdot p_2\int \frac{\hat{d}^4q}{q^2}e^{ib\cdot q}\;\hat{\d}\left(p_1\cdot q\right)\hat{\d}\left(p_2\cdot q\right)\times\nonumber\\
    &\times p_1^{[i}\bigg[Q_1Q_2\left(4p_1\cdot p_2 \mathcal{K}_{12}^{j]}+2\mathcal{I}_{12}q^{j]}\right)
    -Q_1^2\left(4m_1^2\mathcal{K}_{11}^{j]}+2\mathcal{I}_{11}q^{j]}\right)\bigg]\,,
    \label{em-imp1}
 \end{align}
 where we have defined the following integrals
 \begin{align}     
    &\mathcal{K}_{ab}^{\mu}=\int \hat{d}^4\ell\frac{\hat{\d}_+(\ell^2)\hat{\d}'(p_a\ell-p_1^0\o)\;q\cdot\ell}{\left(p_b\ell+ i\e\right)^2}\ell^\mu 
    \,,
    \\
    &\mathcal{I}_{ab}=\int \hat{d}^4\ell\frac{\hat{\d}_+(\ell^2)\hat{\d}(p_a\ell-p_a^0\o)}{\left(p_b\ell+ i\e\right)}+\int \hat{d}^4\ell\frac{\hat{\d}_+(\ell^2)\hat{\d}'(p_a\ell-p_a^0\o)(p_a\cdot \ell)}{\left(p_b\ell+ i\e\right)} 
    \,.  
 \end{align}
These can be computed by using integration by parts identities (IBP) \cite{Chetyrkin:1981qh,Tkachov:1981wb} to reduce them to a basis of master integrals
\begin{align}
    I_a  &=\int \hat{d}^4\ell\hat{\d}_+(\ell^2)\hat{\d}(p_a\cdot\ell-p_a^0\o)\,, \\
     I_{ab}&=\int \hat{d}^4\ell\frac{\hat{\d}_+(\ell^2)\hat{\d}(p_a\cdot\ell-p_a^0\o)}{p_b\cdot\ell + i\e}\,,
\end{align}
which can be unambiguously evaluated. For instance, the first master integral yields
\begin{align}
    I_1  &=\int \hat{d}^4\ell\hat{\d}_+(\ell^2)\hat{\d}(p_1\cdot\ell-p_1^0\o)=\frac{1}{4\pi^2 m_1}\int_0^\infty \frac{d^3\bm\ell}{2|\bm{\ell}|}{\d}(|\bm{\ell}|-\o)=\frac{\omega}{2\pi m_1}\,.
\end{align}
More details about the integration are given in Appendix~\ref{sec:integrals}. Here we quote the results:
\begin{align}     
\mathcal{I}_{12}&=\frac{1}{\pi m_1m_2}\frac{ \cosh^{-1}\g}{\sqrt{\g^2-1}} \,,
&\mathcal{I}_{11} &=\frac{1}{\pi m_1^2}\,,   
\\
    \mathcal{K}_{12}^{\mu}&= \frac{q^\mu}{2\pi  m_1^2 m_2^2}\left[\frac{1}{\g^2-1}-\frac{\g \cosh ^{-1}\g}{\left(\g^2-1\right)^{3/2}}\right] \,, 
&\mathcal{K}^{\mu}_{11} &=- \frac{q^\mu}{6\pi m_1^4}  
    \,.
 \end{align}
 where $\gamma = p_1\cdot p_2/m_1m_2$ is the relative Lorentz factor of the particles.

There are three additional diagrams shown below that can be evaluated by simply changing labels $1\leftrightarrow 2$:
\begin{align}
\mathcal{J}_{\gamma,2}^{ij}&=\raisebox{-35pt}{ \begin{tikzpicture}[scale=1]
\node (a) at (-.8,0){};
\node[right=3 of a](b){};
\node[below=1.5 of a](c){};
\node[right=3 of c](d){};
\draw[fermion] (a)--(b);
\draw[fermion] (c)--(d);
\node[inner sep=0pt,right=0.75 of a](x1){};
\node[inner sep=0pt, right=0.75 of c](x2){};
\draw[photon](x1)--(x2);
\node[inner sep=0pt,left=0.25 of b](x3){};
\node[inner sep=0pt, left=0.25 of d](x4){};
\draw[photon](x3)--(x4);
\node[inner sep=0pt, below=1.05 of x1](w){};
\node[left=0.45 of w](w1){};
  \draw [line width=1.5pt, double distance=1pt] (w)-- (w1) ; 
  \node[inner sep=0 pt] (c1) at (0.7-0.8,-.4){};
  \node[inner sep=0 pt, below right=2.2 of c1] (c2){};
  \draw[fcut, brown]  (c1)--(c2);

\node (a) at (3,0){};
\node[right=3 of a](b){};
\node[below=1.5 of a](c){};
\node[right=3 of c](d){};
\draw[fermion] (a)--(b);
\draw[fermion] (c)--(d);
\node[inner sep=0pt,right=0.75 of a](x1){};
\node[inner sep=0pt, right=0.83 of c](x2){};
\node[inner sep=0pt,left=0.27 of b](x3){};
\node[inner sep=0pt, left=0.25 of d](x4){};
\draw[photon](x3)--(x2) node[midway](m){};
\draw[white,fill=white] (m) circle (0.15);
\draw[photon](x1)--(x4);
\node[inner sep=0 pt] at (3+2+.75,-1.2)(w){};
\node[left=0.45 of w](w1){};
  \draw [line width=1.5pt, double distance=1pt] (w)-- (w1) ; 
\draw[white,fill=white] (a) circle (.6);
\draw[white,fill=white] (c) circle (.6);
  \node[inner sep=0 pt] (c1) at (1.25+3,-2){};
  \node[inner sep=0 pt, above right=1.8 of c1] (c2){};
  \draw[fcut, brown]  (c1)--(c2);
\node (a) at(7,0){};
\node[right=3 of a](b){ };
\node[below=1.5 of a](c){ };
\node[right=3 of c](d){ };
\draw[fermion] (c)--(d);
\node[inner sep=0pt,right=.6 of a](x1){};
\node[inner sep=0pt, right=.6 of c](x2){};
\node[inner sep=0pt, right=1.5 of a](m1){};
\node[inner sep=0pt, right=1.5 of c](m2){};
\node[inner sep=0pt,left=0.25 of b](x3){};
\node[inner sep=0pt, left=0.25 of d](x4){};
\draw [decorate, decoration={snake, segment length=2.8 mm, amplitude=.5 mm, pre length=0.7 mm, post length=0.01 mm}](x4) arc (15:175:1.15);
\node[above=0.7 of m2](v){}; 
\draw[white,fill=white] (v) circle (0.15);
\node (v2) at (7+.6,-0.13-1.8){};
\draw[white,fill=white] (v2) circle (0.15);
\draw[photon](m1)--(m2);
\draw[fermion] (a)--(b);
\draw[fermion] (c)--(d);
\node[inner sep=0 pt] at (7+.93,-.7-.53)(w){};
\node[left=0.45 of w](w1){};
  \draw [line width=1.5pt, double distance=1pt] (w)-- (w1) ; 
  \node[inner sep=0 pt] (c1) at (7+1.2,-.5-1.5){};
  \node[inner sep=0 pt, above=1.35 of c1] (c2){};
  \draw[fcut, brown]  (c1)--(c2);
 \node (v1) at (3,-.9){$+$};
  \node (v1) at (6.5,-.9){$+$};
\end{tikzpicture}}\nonumber\\
&=\mathcal{J}_{\gamma,1}^{ij}\left(1\leftrightarrow 2\right)\,.
\end{align}
Furthermore, simple power counting in $\omega$ shows that the triangle diagrams do not contribute to angular impulse. Finally, using the expression for the leading order electromagnetic linear impulse of the heavy particle 1 \cite{Kosower2019}
\begin{equation}
    \D \mathbb{P}_1^{i}=-\D \mathbb{P}_2^i={ie^2Q_1Q_2}p_1\cdot p_2\int \frac{\hat{d}^4q}{q^2}\hat{\d}\left(p_1\cdot q\right)\hat{\d}\left(p_2\cdot q\right)q^{i}e^{-iq\cdot b} = -\frac{e^2Q_1Q_2}{2\pi}\frac{\g}{\sqrt{\g^2-1}}\frac{b^i}{b^2}\,,
\end{equation} 
we find the leading order electromagnetic angular momentum loss
\begin{align}
    \D\mathbb{J}_{\gamma}^{ij}&=\mathcal{J}_{\gamma,1}^{ij}+\mathcal{J}_{\gamma,2}^{ij}\nonumber\\
    &=\frac{e^2}{2\pi}\left[-\frac{2Q_1^2}{ 3m_1^2}+\frac{Q_1Q_2}{ m_1m_2}\left(\frac{\gamma}{\gamma ^2-1}-\frac{ \cosh ^{-1}\gamma }{\left(\gamma ^2-1\right)^{3/2}}\right)\right]  p_1^{[i}\;\D \mathbb{P}_1^{j]}+(1\leftrightarrow 2).
\end{align}
which agrees with \cite{Saketh:2021sri}.



\section{Classical observables in the causal basis}
\label{sec:sec3}
In this section, we will present examples of classical observables computed using manifestly causal diagrams, and point out the various simplifications that this method provides.

\subsection{Amputated response functions in causal basis}
Since we are interested in causal observables,  we shall now choose a basis in the in-in path integral in which the causal structure of these quantities becomes manifest. This basis, called the Keldysh basis or $r/a$ basis, is defined via a rotation of the $\I/\II$ basis in the path integral in Eq.~\eqref{path-int1} as
\begin{align}
    \varphi^r=\frac{1}{2}(\varphi^\I+\varphi^\II)\,,\qquad  \varphi^a=\varphi^\I-\varphi^\II\,,
\end{align}
and for a generic operator, ${\cal O}$, we also define
\begin{align}
    {\cal O}^r=\frac{1}{2}({\cal O}^\I+{\cal O}^\II),\qquad {\cal O}^a= {\cal O}^\I-{\cal O}^\II . 
\end{align}
The response function in Eq.~\eqref{respone2} takes a simple form in this basis
\begin{align}
     R_{n+1}[{\cal O}(x);\vf(x_1),\cdots,\vf(x_n)]=  \langle \mathcal{C}\left\{ {\cal O}^r(x)\vf^a(x_1)\cdots\vf^a (x_n)\right\}\rangle. \label{response3}
\end{align}
More generally, one can insert $n_r$ number of $r$-type fields and $n-n_r$ number of $a$-type fields to construct general $n$-point causal response function 
\begin{align}
   & R_n[{\cal O}^r(x_1), \cdots,{\cal O}^r(x_{n_r});\vf^a(x_{n_r+1}),\cdots,\vf^a(x_n)] \\
    &=  \langle \mathcal{C}\left\{ {\cal O}^r(x_1) \cdots {\cal O}^r(x_{n_r})\vf^a(x_{n_r+1})\cdots\vf^a (x_n)\right\}\rangle\,.\nonumber
\end{align}
The boundary condition in Eq.~\eqref{bc} becomes $\varphi^a(t=+\infty)=0$ in the causal basis.  
Thanks to this, the largest time equation in Eq.~\eqref{largest} simply becomes the vanishing of correlation functions of all $a$ fields. 

The $r/a$ basis serves as a natural basis for taking the semi-classical limit. One can think of the difference in the fields along two contours to be much smaller compared to their sum i.e. $\varphi^a= (\varphi^\I-\varphi^\II)\ll \frac{1}{2}\left(\varphi^\I+\varphi^\II\right)=\vf^r $ in this limit. Thus, the action can be expanded in small $\varphi^a$

\begin{equation}
  S[\varphi^\I]-S[\varphi^\II] =  \frac{\d S[\varphi]}{\d\varphi}\Big|_{\varphi=\varphi_r}\varphi_a + \frac{\d^3 S[\varphi]}{\d\varphi^3}\Big|_{\varphi=\varphi_r}\varphi_a^3 + \cdots\,.
\end{equation}
At the leading order in $\vf^a$, we have
\begin{align}
  \int D\varphi_r D\varphi_a \;e^{i \frac{\d S[\varphi]}{\d\varphi}\big|_{\varphi=\varphi_r}\varphi_a} \,.
  \label{path-int2}
\end{align}
The path integral over $\varphi^a$ can be performed exactly and it imposes the equation of motion for $\varphi^r$. For this reason, we will refer to all vertices linear in $a$ fields as \emph{classical vertices}. Since the exponent in Eq.~\eqref{path-int1} has to be odd in $\varphi^a$, the next-order term is cubic in $\varphi^a$, etc, which we will call \emph{quantum vertices}.

The free-field propagators in the causal basis are
\cite{Mueller:2002gd}
\begin{align}
    \begin{pmatrix}
        G_{rr}(p) & G_{ra}(p)\\
        G_{ar}(p) & G_{aa}(p)
    \end{pmatrix}=
    \begin{pmatrix}
        \frac{1}{2}\hat{\d}(p^2-m_i^2) & \frac{i}{p^2-m_i^2+i\e p^0}\\
       \frac{i}{p^2-m_i^2-i\e p^0}  & 0
    \end{pmatrix}\,.
    \label{prop}
\end{align}
The vanishing of the $G_{aa}$ propagator follows non-perturbatively from the largest time equation. 
Note the appearance of causal, that is, advanced and retarded, propagators
\begin{align}
    G_R(p)=G_A(-p)=
 \raisebox{-5pt}{
\begin{tikzpicture}
\node (a) {};
\node[right=2 of a] (b){};
\draw[fermion] (a)--(b) node[midway]{$\blacktriangleright$};
\draw[->] (0.5,0.3)--(1.5,0.3) node[right]{$p$};
 \end{tikzpicture}
 }      
    =\frac{i}{p^2-m_i^2+i\e p^0} \,,
\end{align}
which  make the causal properties of the amplitude manifest, as well as the ``cut'' (or Hadamard) propagator
\begin{align}
    G_{rr}(p)=
    \raisebox{-15pt}{
    \begin{tikzpicture}
        \node(c2){};
\node[right=.9 of c2](d2){};
\draw[fermion] (c2)--(d2) node[pos=0.01]
{$\blacktriangleleft$};
\node[above=.25 of d2](x2){};
\node[below=.25 of d2](y2){};
\draw[fcut](x2)-- (y2);
\node[right=0.9 of d2](e2){};
\draw[fermion] (d2)--(e2) node[pos=0.9]{$\blacktriangleright$};
    \end{tikzpicture}
    }
    =
    \frac{1}{2} \hat \delta(p^2-m_i^2)\,,
\end{align}
which we note does not include a positive-energy condition. 

\begin{figure}
    \centering
    \subfloat[Bare propagators]{
    \begin{tikzpicture}[scale=.8]
\node (a) at (0,0){$G_R$};
\node[right=2 of a](b){};
\draw[fermion] (a)--(b) node[midway]{$\blacktriangleright$} node[right]
{$\;=\frac{i}{p^2-m_i^2+i\epsilon p^0}$};
\draw[->] (1,0.3)--(2,0.3) node[right]{$p$};

\node[below=.65 of a](c){$G_{rr}$};
\node[right=0.9 of c](d){};
\draw[fermion] (c)--(d) node[pos=0.02]
{$\blacktriangleleft$};
\node[above=.25 of d](x){};
\node[below=.25 of d](y){};
\draw[fcut](x)-- (y);
\node[right=0.9 of d](e){};
\draw[fermion] (d)--(e) node[pos=0.9]{$\blacktriangleright$} node[right]{$\;=\frac{1}{2}\hat{\delta}(p^2-m_i^2)$};

\node[inner sep=3pt] (a2) at (8.5,0){$G_R\;\;\mu$};
\node[right=2 of a2](b2){$\nu$};
\draw[photon] (a2)--(b2) node[midway]{$\blacktriangleright$} node[right]
{$\quad=\frac{-i\eta^{\mu\nu}}{p^2+i\epsilon p^0}$};
\draw[->] (8.8+1,0.3)--(8.8+2,0.3) node[right]{$p$};

\node[inner sep=3pt,below=.65 of a2](c2){$G_{rr}\;\;\mu\;$};
\node[right=.9 of c2](d2){};
\draw[photon] (c2)--(d2) node[pos=0.01]
{$\blacktriangleleft$};
\node[above=.25 of d2](x2){};
\node[below=.25 of d2](y2){};
\draw[fcut](x2)-- (y2);
\node[right=0.9 of d2](e2){$\;\nu$};
\draw[photon] (d2)--(e2) node[pos=0.9]{$\blacktriangleright$} node[right]{$\quad=-\frac{\eta^{\mu\nu}}{2}\hat{\delta}(p^2)$};
\end{tikzpicture}
    }\\
\subfloat[Classical vertices]{
\begin{tikzpicture}[scale=.95]
\node[] (a) {$\mu$};
\node[inner sep=0pt,right=.8 of a] (x){};
\fill (x) circle (1pt);
\draw[photon] (a)--(x) node[midway]{$\blacktriangleright$};   
\node[ above right=.8 of x] (b){$p'$};
\draw[fermion] (x)--(b) node[pos=.95]{$\rotatebox[origin=c]{45}{$\blacktriangleright$}$};
\node[below right=.8 of x] (c){$p$};
\draw[fermion] (x)--(c) node[midway]{$\rotatebox[origin=c]{45}{$\blacktriangle$}$} ;
%

\node (a2) at (3,0){$\mu$};
\node[inner sep=0pt,right=.8 of a2] (x2){};
\fill (x2) circle (1pt);
\draw[photon] (a2)--(x2) node[pos=0.05]{$\blacktriangleleft$};   
\node[ above right=.8 of x2] (b2){$p'$};
\draw[fermion] (x2)--(b2) node[midway]{$\rotatebox[origin=c]{45}{$\blacktriangleleft$}$};
\node[below right=.8 of x2] (c2){$p$};
\draw[fermion] (x2)--(c2) node[midway]{$\rotatebox[origin=c]{45}{$\blacktriangle$}$} ;
\node[ below=1 of a2] (v2) {$=-ieQ_i(p+p')^{\mu}$};
%
\node[inner sep=0pt] (x4) at (8,0){};
\fill (x4) circle(1 pt);
\node[above left= .8 of x4] (a4){$\mu$};
\draw[photon] (x4)--(a4) node[midway]{$\rotatebox[origin=c]{45}{$\blacktriangledown$}$};
\node[above right= .8 of x4] (b4){$\nu$};
\draw[photon] (x4)--(b4) node[pos= 0.9]{$\rotatebox[origin=c]{45}{$\blacktriangleright$}$};
\node[below left= .8 of x4] (c4){};
\draw[fermion] (x4)--(c4) node[midway]{$\rotatebox[origin=c]{45}{$\blacktriangleright$}$};
\node[below right= .8 of x4] (d4){};
\draw[fermion] (x4)--(d4) node[midway]{$\rotatebox[origin=c]{45}{$\blacktriangle$}$};

\node[inner sep=0pt] (x5) at (11,0){};
\fill (x5) circle(1 pt);
\node[above left= .8 of x5] (a5){$\mu$};
\draw[photon] (x5)--(a5) node[midway]{$\rotatebox[origin=c]{45}{$\blacktriangledown$}$};
\node[above right= .8 of x5] (b5){$\nu$};
\draw[photon] (x5)--(b5) node[midway]{$\rotatebox[origin=c]{45}{$\blacktriangleleft$}$};
\node[below left= .8 of x5] (c5){};
\draw[fermion] (x5)--(c5) node[midway]{$\rotatebox[origin=c]{45}{$\blacktriangleright$}$};
\node[below right= .8 of x5] (d5){};
\draw[fermion] (x5)--(d5) node[pos=0.9]{$\rotatebox[origin=c]{45}{$\blacktriangledown$}$};
\node[below=1.8 of a5]{$=2ie^2Q_i^2\eta^{\mu\nu}\quad\quad$};
\draw[->] (2.15-.15,-0.6)--(1.7-.15,-0.1);
\draw[->] (1.7-.15,0.1)--(2.15-.15,0.6);
\draw[->] (2.15-.15+3,-0.6)--(1.7-.15+3,-0.1);
\draw[->] (1.7-.15+3,0.1)--(2.15-.15+3,0.6);
\draw[->] (0.55+7.0,-4.25+3.55)--(0.99+7.0,-3.79+3.55);
\draw[->] (1.1+7.0,-3.85+3.55)--(1.45+7.0,-4.25+3.55);
\draw[->] (0.55+10.0,-4.25+3.55)--(0.99+10.0,-3.79+3.55);
\draw[->] (1.1+10.0,-3.85+3.55)--(1.45+10.0,-4.25+3.55);
\end{tikzpicture} 
} \\   
\subfloat[Quantum vertices]{
\begin{tikzpicture}
\node (a3) at (0,0){$\mu$};
\node[inner sep=0pt,right=.8 of a3] (x3){};
\fill (x3) circle (1pt);
\draw[photon] (a3)--(x3) node[pos=0.05]{$\blacktriangleleft$};   
\node[ above right=.8 of x3] (b3){$p'$};
\draw[fermion] (x3)--(b3) node[pos=0.95]{$\rotatebox[origin=c]{45}{$\blacktriangleright$}$};
\node[below right=.8 of x3] (c3){$p$};
\draw[fermion] (x3)--(c3) node[pos=.95]{$\rotatebox[origin=c]{45}{$\blacktriangledown$}$} ;
\node[ below=1 of x3] (v3) {$=-\frac{ieQ_i}{4}(p+p')^{\mu}$};
\node[inner sep=0pt] (x6) at (5,0){};
\fill (x6) circle(1 pt);
\node[above left= .8 of x6] (a6){$\mu$};
\draw[photon] (x6)--(a6) node[midway]{$\rotatebox[origin=c]{45}{$\blacktriangledown$}$};
\node[above right= .8 of x6] (b6){$\nu$};
\draw[photon] (x6)--(b6) node[pos= 0.9]{$\rotatebox[origin=c]{45}{$\blacktriangleright$}$};
\node[below left= .8 of x6] (c6){};
\draw[fermion] (x6)--(c6) node[pos=.9]{$\rotatebox[origin=c]{45}{$\blacktriangleleft$}$};
\node[below right= .8 of x6] (d6){};
\draw[fermion] (x6)--(d6) node[pos=0.9]{$\rotatebox[origin=c]{45}{$\blacktriangledown$}$};
   \node[inner sep=0pt] (x6) at (8,0){};
\fill (x6) circle(1 pt);
\node[above left= .8 of x6] (a6){$\mu$};
\draw[photon] (x6)--(a6) node[pos=0.9]{$\rotatebox[origin=c]{45}{$\blacktriangle$}$};
\node[above right= .8 of x6] (b6){$\nu$};
\draw[photon] (x6)--(b6) node[pos= 0.9]{$\rotatebox[origin=c]{45}{$\blacktriangleright$}$};
\node[below left= .8 of x6] (c6){};
\draw[fermion] (x6)--(c6) node[midway]{$\rotatebox[origin=c]{45}{$\blacktriangleright$}$};
\node[below right= .8 of x6] (d6){};
\draw[fermion] (x6)--(d6) node[pos=0.9]{$\rotatebox[origin=c]{45}{$\blacktriangledown$}$};
\node[below=1.8 of a6]{$=\frac{ie^2}{2}Q_i^2\eta^{\mu\nu}$}; 
\draw[->] (2.15-.15,-0.6)--(1.7-.15,-0.1);
\draw[->] (1.7-.15,0.1)--(2.15-.15,0.6);
\draw[->] (0.55+4.0,-4.25+3.55)--(0.99+4.0,-3.79+3.55);
\draw[->] (1.1+4.0,-3.85+3.55)--(1.45+4.0,-4.25+3.55);
\draw[->] (0.55+7.0,-4.25+3.55)--(0.99+7.0,-3.79+3.55);
\draw[->] (1.1+7.0,-3.85+3.55)--(1.45+7.0,-4.25+3.55);
\end{tikzpicture}
}
    \caption{Feynman rules for scalar QED are shown above. The arrow on the retarded propagators $G_R$ corresponds to the direction of the causal flow. The $G_{rr}$ type propagators are represented by cuts. The $a$ and $r$-type fields in the vertices are represented by outgoing and incoming arrows respectively. The directions of momenta going in or out of the vertices are aligned with the directions of particle flows shown by the thin arrows.  The quantum vertices get an additional factor of $\frac{1}{4}$ owing to two additional $a$ fields.}
    \label{fig:feynrules}
\end{figure}

We are interested in momentum-space $n$-point amputated causal amplitudes with $n_r$ number of $r$-type external fields, which can be computed using the following  diagrammatic rules: 
On each diagram,
\begin{itemize}
    \item Put $r, a$ labels to the $r$ or $a$-type external fields. The internal lines should be dressed with all possible assignments of $r/a$ labels. 
    \item  Draw arrows such that all the causal flows are from $a$ to $r$. In doing so, $G_{rr}$ propagators can be treated as sources for the arrows.  
    \item Align the directions of momenta of the heavy lines with the directions of the particle flow unless otherwise specified. Amputate the external lines. 
\end{itemize}

For the case of electrodynamics, the corresponding Feynman rules are given in Fig.~\ref{fig:feynrules}. The calculation of the observables in the $r/a$ basis is simplified by a very useful counting rule derived in \cite{CaronHuot2011} which we quote without proof: 
  for a $n$-point $L$-loop diagram with $n_r$ number of $r$-type external legs, let us denote the number of $rr$  propagators by $P_{rr}$, and the number of \textit{additional} $a$-type fields at the internal vertices  by $n_a$ (which is zero for the classical vertex, and always even for the quantum ones). 
  These are related by \cite{CaronHuot2011}
\begin{equation}
    L+n_r-1=P_{rr}+n_a\,. \label{counting}
\end{equation}
Crucially, the above counting only depends on the number of $a$ and $r$-type fields, but not on their species (i.e whether they come from massive or massless fields in our context).
It was argued in \cite{CaronHuot2011} using Eq.~\eqref{counting} that a $n$-point response function does not receive any contribution from the quantum vertices up to one loop. Indeed, inserting a cubic all-$a$ vertex in a 1-loop diagram results in a closed loop of retarded propagators which vanishes due to causality.  Thus, for the causal amplitudes with $n_r=1$, Eq.~\eqref{counting} implies that the tree amplitudes are connected and the one-loop amplitudes have a single cut. More generally, it implies that diagrams with $n_a=0$, that is, those which only feature the classical vertex (i.e., the vertex linear in $\varphi^a$) in Eq.~\eqref{path-int2}, have at least one cut propagator per loop.

Finally, let us comment on the $i\e$ prescription of the causal massive propagators. A crucial feature of the computations in the classical limit is the soft expansion of the matter propagators \cite{Herrmann2021,Bern2022} in the soft limit $\ell \sim q\sim \hbar \ll p_i$, with $\ell$ being a loop momentum. For a massive retarded propagator we get
\begin{align}
    \frac{1}{(p_i+\ell-q)^2-m_i^2+i\e(p_i^0+\ell^0-q^0)}&=\frac{1}{2 p_i\cdot\ell+i\e\, p_i^0} + \cdots\,.
    \label{soft-exp}
\end{align}

The key observation is that in the soft limit, the sign of $(p_i^0+\ell^0+q^0)$ is determined by the sign of the massive particle energy, $p_i^0$, since $\ell^0,\; q^0\ll p_i^0$. Thus, the $i\e$ prescription for the massive propagator is fixed by the causal flow of the heavy particle in the classical limit.

\subsection{Linear impulse}
In this section, we compute the impulse of particle 1 in the $r/a$ basis. The corresponding response function was described in detail in Section~\ref{sec: impulse}. Let us begin by stating the Feynman rules for inserting the momentum operator into the internal and external lines in Fig.~\ref{fig:loop-insertion-ra}.
\begin{figure}[htbp]
    \centering
    \subfloat[\label{fig:5a}Operator insertion on the external lines]{
\begin{tikzpicture}[scale=.99]
    \node (a) at (0,0){$p_1$};
    \node[right=2 of a] (b) {$p_1-k$};
      \draw[fermion] (a) -- (b) node[pos=0.25] {$\blacktriangleright$} node[pos=0.75] {$\blacktriangleleft$} ;  
      \node[inner sep=0pt,right=1 of a](x){};
      \node[above= 0.5 of x](y){$k$};
       \draw [line width=1pt, double distance=1pt] (x)-- (y) ;  
       \node[right=0.2 of b](n){$=-p_1^\mu$};
       \node[inner sep=0pt,above left=0.28 of x](p){};
       \node[left=0.5 of p] (q){};
       \draw[<-](p)--(q);
       \node[inner sep=0pt,above right=0.28 of x](p2){};
       \node[right=0.5 of p2] (q2){};
       \draw[->](p2)--(q2);
       \node (a) at (6,0){$p_1+k$};
    \node[right=2 of a] (b) {$p_1$};
      \draw[fermion] (a) -- (b) node[pos=0.25] {$\blacktriangleright$} node[pos=0.75] {$\blacktriangleleft$} ;  
      \node[inner sep=0pt,right=1 of a](x){};
      \node[above= 0.5 of x](y){$k$};
       \draw [line width=1pt, double distance=1pt] (x)-- (y) ;  
         \node[right=0.01 of b](n){$=p_1^\mu$};
       \node[inner sep=0pt,above left=0.28 of x](p){};
       \node[left=0.5 of p] (q){};
       \draw[<-](p)--(q);
       \node[inner sep=0pt,above right=0.28 of x](p2){};
       \node[right=0.5 of p2] (q2){};
       \draw[->](p2)--(q2);
\end{tikzpicture}
}\\
 \subfloat[\label{fig:5b}Operator insertion on an internal line]{
\begin{tikzpicture}[scale=.99]
         \node (a) at (0,0){$\ell+k$};
    \node[right=2 of a] (b) {$\ell$};
      \draw[fermion] (a) -- (b) node[pos=0.25] {$\blacktriangleright$} node[pos=0.75] {$\blacktriangleleft$} ;  
      \node[inner sep=0pt,right=1 of a](x){};
      \node[above= 0.5 of x](y){$k$};
       \draw [line width=1pt, double distance=1pt] (x)-- (y) ;  
    \node[below=0.1 of x](n){$=\ell^\mu\hat{\d}\left(\ell^2-m_1^2\right)$};
       \node[inner sep=0pt,above left=0.28 of x](p){};
       \node[left=0.5 of p] (q){};
       \draw[<-](p)--(q);
       \node[inner sep=0pt,above right=0.28 of x](p2){};
       \node[right=0.5 of p2] (q2){};
       \draw[->](p2)--(q2);
       
       \node (a) at (5,0){$\ell+k$};
              \node[inner sep=1pt,right=1.35 of a](c1){$\;$};
         \draw[fermion](a)--(c1) node[pos=0.25] {$\blacktriangleright$} node[pos=0.75] {$\blacktriangleleft$} ;
      \node[above=0.25 of c1] (c2){};
      \node[below=0.25 of c1] (c3){};
      \draw[fcut] (c2)--(c3);
    \node[right=.45 of c1] (b) {$\ell$};
      \draw[fermion] (c1) -- (b) node[pos=0.9] {$\blacktriangleright$} ;          
      \node[inner sep=0pt,right=.65 of a](x){};
      \node[above= 0.5 of x](y){$k$};
       \draw [line width=1pt, double distance=1pt] (x)-- (y) ; 
       \node[below=0.1 of x](n){$=\frac{\ell^\mu}{2}\hat{\d}\left(\ell^2-m_1^2\right)$};
       \node[inner sep=0pt,above left=0.28 of x](p){};
       \node[left=0.5 of p] (q){};
       \draw[<-](p)--(q);
       \node[inner sep=0pt,above right=0.28 of x](p2){};
       \node[right=0.4 of p2] (q2){};
       \draw[->](p2)--(q2);
        \node (a) at (10,0){$\ell+k$};
           \node[inner sep=1pt,right=0.4 of a](c1){$\;$};
         \draw[fermion](a)--(c1) node[pos=0.1]{$\blacktriangleleft$};
      \node[above=0.25 of c1] (c2){};
      \node[below=0.25 of c1] (c3){};
      \draw[fcut] (c2)--(c3);
    \node[right=1.6 of c1] (b) {$\ell$};
      \draw[fermion] (c1) -- (b) node[pos=0.25] {$\blacktriangleright$} node[pos=0.75] {$\blacktriangleleft$} ;  
         
      \node[inner sep=0pt,right=1.3 of a](x){};
      \node[above= 0.5 of x](y){$k$};
       \draw [line width=1pt, double distance=1pt] (x)-- (y) ; 
       \node[below=0.1 of x](n){$=-\frac{\ell^\mu}{2}\hat{\d}\left(\ell^2-m_1^2\right)$};  
       \node[inner sep=0pt,above left=0.28 of x](p){};
       \node[left=0.5 of p] (q){};
       \draw[<-](p)--(q);
       \node[inner sep=0pt,above right=0.28 of x](p2){};
       \node[right=0.5 of p2] (q2){};
       \draw[->](p2)--(q2);
\end{tikzpicture}
}
    \caption{Five possible cases are shown when the momentum vertex is inserted on the external and internal lines. The directions of momenta of the heavy lines are denoted by the thin arrows.  We take  $k^\mu=(\omega,\bm{0})$ outgoing.}
    \label{fig:loop-insertion-ra}
\end{figure}

The derivation of these rules is similar to the ones described in sec \ref{sec:sec2}. Notice, that the operator, represented by a double line, is $r$-type since causal arrows flow towards it. According to Eq.~\eqref{counting}, we have single-cut diagrams at one-loop one external $r$ leg. The simplest are the triangle diagrams 
\begin{align}
    \begin{tikzpicture}[baseline=-0.8 cm,scale=0.45]
 \node (a) at (0,0){};
 \node[inner sep=0pt,right=3.7 of a] (b){};
 \draw[fermion] (a)--(b) node[pos=0.15] {$\blacktriangleright$} node[pos=0.65] {$\blacktriangleright$}
 node[pos=0.9] {$\blacktriangleleft$}; 
 \node[inner sep=0pt, right= 1.75 of a] (x){};
 \fill (x) circle (1pt);
 \node[inner sep=0pt,below left=1.8 of x] (c){};
 \fill (c) circle (1pt);
 \node[inner sep=0pt,below right=1.8 of x] (d){};
 \fill (d) circle (1pt);
 \draw[photon] (x)--(c) node[pos=0.4]{$\rotatebox[origin=c]{45}{$\blacktriangleright$}$}; 
 \draw[photon] (x)--(d) node[pos=0.4]{$\rotatebox[origin=c]{45}{$\blacktriangle$}$}; 
 \node[right= 1.2 of c](cx){};
 \node[above=.25 of cx] (x1){};
 \node[below=.25 of cx] (x2){};
 \draw[fcut] (x1)--(x2);
 \draw[fermion] (c)--(cx) node [midway]{$\blacktriangleleft$};
 \draw[fermion] (d)--(cx)  node [midway]{$\blacktriangleright$};
 \node[right=1 of c] (cx){};
 \node[inner sep=0pt,right=0.01 of c](c2) {};
 \node[inner sep=0pt,left=0.5 of c2](c3){};
 \draw[fermion] (c2)--(c3)  node[midway]{$\blacktriangleright$};
  \node[inner sep=0pt,left=0.01 of d](d2) {};
 \node[inner sep=0pt,right=0.6 of d2](d3){};
 \draw[fermion] (d2)--(d3) node[midway]{$\blacktriangleleft$};
  \node[inner sep=0 pt, left=0.8 of b] (b2){};
  \node[above= 0.5 of b2](y){};
       \draw [line width=1pt, double distance=1pt] (b2)-- (y) ;
\node (a) at (10,0){};
 \node[inner sep=0pt,right=3.7 of a] (b){};
 \draw[fermion] (a)--(b) node[pos=0.15] {$\blacktriangleright$} node[pos=0.3] {$\blacktriangleleft$}
 node[pos=0.85] {$\blacktriangleleft$}; 
 \node[inner sep=0pt, right= 1.75 of a] (x){};
 \fill (x) circle (1pt);
 \node[inner sep=0pt,below left=1.8 of x] (c){};
 \fill (c) circle (1pt);
 \node[inner sep=0pt,below right=1.8 of x] (d){};
 \fill (d) circle (1pt);
 \draw[photon] (x)--(c) node[pos=0.4]{$\rotatebox[origin=c]{45}{$\blacktriangleright$}$}; 
 \draw[photon] (x)--(d) node[pos=0.4]{$\rotatebox[origin=c]{45}{$\blacktriangle$}$}; 
 \node[right= 1.2 of c](cx){};
 \node[above=.25 of cx] (x1){};
 \node[below=.25 of cx] (x2){};
 \draw[fcut] (x1)--(x2);
 \draw[fermion] (c)--(cx) node [midway]{$\blacktriangleleft$};
 \draw[fermion] (d)--(cx)  node [midway]{$\blacktriangleright$};
 \node[right=1 of c] (cx){};
 \node[inner sep=0pt,right=0.01 of c](c2) {};
 \node[inner sep=0pt,left=0.5 of c2](c3){};
 \draw[fermion] (c2)--(c3)  node[midway]{$\blacktriangleright$};
  \node[inner sep=0pt,left=0.01 of d](d2) {};
 \node[inner sep=0pt,right=0.6 of d2](d3){};
 \draw[fermion] (d2)--(d3) node[midway]{$\blacktriangleleft$};
  \node[inner sep=0 pt, right=0.8 of a] (b2){};
  \node[above= 0.5 of b2](y){};
       \draw [line width=1pt, double distance=1pt] (b2)-- (y) ;
\node(n1) at (0.7,-2.5){};
\node[above right=1 of n1] (n2){};
\draw[->] (n1)--(n2);
\node (n3) at (1.1,-1.2){$\ell$};
\node(n1) at (10+0.7,-2.5){};
\node[above right=1 of n1] (n2){};
\draw[->] (n1)--(n2);
\node (n3) at (10+1.1,-1.2){$\ell$};
\node (n3) at (9,-1.5){$+$};
\end{tikzpicture}=4i e^4Q_1^2Q_2^2p_1^2 \int \hat{d}^4\ell\;\frac{\hat{\d}(2p_2\cdot\ell)q^\mu}{\ell^2(\ell-q)^2}\,.
\end{align}
We have dropped the $i\e$ for the photon propagators since they can not go on-shell at one-loop. For the same reason, we do not need to consider the diagrams where the photon propagators are cut.  The inverted triangle can be found by replacing $1\leftrightarrow 2$. However, for the inverted diagram, we have two additional diagrams that sum to zero:
\begin{align}
    \begin{tikzpicture}[baseline=-0.8 cm,scale=0.45]
 \node (a) at (10,0){};
 \node[inner sep=0pt,right=3.6 of a] (b){};
 \draw[fermion] (a)--(b) node[pos=0.1] {$\blacktriangleright$}
 node[pos=0.25] {$\blacktriangleleft$}
 node[pos=0.5] {$\blacktriangleright$}
  node[pos=0.73] {$\blacktriangleleft$}
 node[pos=0.95] {$\blacktriangleleft$}; 
 \node[inner sep=0pt, right= 0.5 of a] (x){};
 \fill (x) circle (1pt);
 \node[inner sep=0pt,below right=1.8 of x] (c){};
 \fill (c) circle (1pt);
 \node[inner sep=0pt,above right=1.8 of c] (e){};
 \fill (d) circle (1pt);
 \draw[photon] (x)--(c) node[pos=0.4]{$\rotatebox[origin=c]{45}{$\blacktriangledown$}$}; 
 \draw[photon] (c)--(e) node[pos=0.6]{$\rotatebox[origin=c]{45}{$\blacktriangleright$}$}; 
 \node[left=1.8 of c] (c1){};
  \node[right=1.8 of c] (c2){};
 \draw[fermion] (c1)--(c2)  node[pos=0.2]{$\blacktriangleright$} node[pos=0.8]{$\blacktriangleleft$};
  \node[inner sep=0 pt, right=2.2 of a ] (b2){};
  \node[above= 0.5 of b2](y){};
       \draw [line width=1pt, double distance=1pt] (b2)-- (y) ;
 \node[right= 1.2 of a](cx){};
  \fill[white] (cx) circle (7pt);
 \node[above=.25 of cx] (x1){};
 \node[below=.25 of cx] (x2){};
 \draw[fcut] (x1)--(x2);
 
 \node (a) at (0,0){};
 \node[inner sep=0pt,right=3.6 of a] (b){};
 \draw[fermion] (a)--(b) node[pos=0.1] {$\blacktriangleright$}
 node[pos=0.27] {$\blacktriangleright$}
 node[pos=0.47] {$\blacktriangleleft$}
  node[pos=0.73] {$\blacktriangleright$}
 node[pos=0.95] {$\blacktriangleleft$}; 
 \node[inner sep=0pt, right= 0.5 of a] (x){};
 \fill (x) circle (1pt);
 \node[inner sep=0pt,below right=1.8 of x] (c){};
 \fill (c) circle (1pt);
 \node[inner sep=0pt,above right=1.8 of c] (e){};
 \fill (d) circle (1pt);
 \draw[photon] (x)--(c) node[pos=0.4]{$\rotatebox[origin=c]{45}{$\blacktriangle$}$}; 
 \draw[photon] (c)--(e) node[pos=0.6]{$\rotatebox[origin=c]{45}{$\blacktriangleleft$}$}; 
 \node[left=1.8 of c] (c1){};
  \node[right=1.8 of c] (c2){};
 \draw[fermion] (c1)--(c2)  node[pos=0.2]{$\blacktriangleright$} node[pos=0.8]{$\blacktriangleleft$};
  \node[inner sep=0 pt, right=1.35 of a] (b2){};
  \node[above= 0.5 of b2](y){};
       \draw [line width=1pt, double distance=1pt] (b2)-- (y) ;
 \node[left= 1.3 of b](cx){};
  \fill[white] (cx) circle (7pt);
 \node[above=.25 of cx] (x1){};
 \node[below=.25 of cx] (x2){};
 \draw[fcut] (x1)--(x2);
\node(n1) at (3,-2.5){};
\node[above left=1 of n1] (n2){};
\draw[->] (n1)--(n2);
\node (n3) at (1.4,-1.6){$\ell$};
\node(n1) at (10+3,-2.5){};
\node[above left=1 of n1] (n2){};
\draw[->] (n1)--(n2);
\node (n3) at (10+1.4,-1.6){$\ell$};
\node (n3) at (9,-1.5){$+$};
\end{tikzpicture}=0, \label{zero-traingle}
\end{align}
thanks to the last two Feynman rules in Fig.~\ref{fig:5b}. Next, we consider the box and crossed-box diagrams.  When the operator is inserted on the external lines, we have
\begin{align}
\quad\quad& \begin{tikzpicture}
  \node[inner sep=0 pt] (a) at (0,0){};
\node[inner sep=0 pt, left=.1 of a] (a1){};
 \node[right=2.7 of a] (b){};
 \node[inner sep=0pt, right= 1.75 of a] (x){};
 \fill (x) circle (1pt);
 \node[inner sep=0pt, right= 0.3 of a] (x2){};
 \fill (x2) circle (1pt);
 \node[inner sep=0pt,below left=2 of x] (c){};
 \fill (c) circle (1pt);
 \node[inner sep=0pt,below right=2 of x2] (d){};
 \fill (d) circle (1pt);
 \draw[photon] (x)--(d) node[pos=0.45]{$\blacktriangle$}; 
 \draw[photon] (x2)--(c) node[pos=0.45]{$\blacktriangledown$}; 
 \node[right= 0.9 of a](cx){};
 \node[above=.25 of cx] (f1){};
 \node[below=.25 of cx] (f2){};
 \draw[fcut] (f1)--(f2);
 \draw[fermion] (cx)--(a1) node [pos=0.25]{$\blacktriangleleft$} node[pos=0.75]{$\blacktriangleright$};
 \draw[fermion] (cx)--(b)  node [pos=.2]{$\blacktriangleright$} node[pos=0.58]{$\blacktriangleright$} node[pos=0.8]{$\blacktriangleleft$};
 \node[right=1 of c] (cx){};
 \node[inner sep=0pt,right=0.01 of c](c2) {};
 \node[left=0.5 of c2](c3){};
 \draw[fermion] (c2)--(c3)  node[midway]{$\blacktriangleright$};
  \node[inner sep=0pt,left=0.01 of d](d2) {};
 \node[inner sep=0pt,right=0.6 of d2](d3){};
 \draw[fermion] (d2)--(d3) node[midway]{$\blacktriangleleft$};
 \draw[fermion] (c3)--(d3) node[midway]{$\blacktriangleright$};
  \node[inner sep=0 pt, left=0.45 of b] (b2){};
  \node[above= 0.5 of b2](y){};
       \draw [line width=1pt, double distance=1pt] (b2)-- (y) ; 
       
\node[inner sep=0 pt] (a) at (3.5,0){};
 \node[inner sep=0 pt, left=0.04 of a] (a1){};
 \node[right=2.7 of a] (b){};
 
 \node[inner sep=0pt, right= 1.75 of a] (x){};
 \fill (x) circle (1pt);
 \node[inner sep=0pt, right= 0.3 of a] (x2){};
 \fill (x2) circle (1pt);
 \node[inner sep=0pt,below left=2 of x] (c){};
 \fill (c) circle (1pt);
 \node[inner sep=0pt,below right=2 of x2] (d){};
 \fill (d) circle (1pt);
  \draw[photon] (x)--(c) node[pos=0.2]{$\rotatebox[origin=c]{45}{$\blacktriangleright$}$}; 
  \node[inner sep=2pt, below right=0.9 of x2](td){};
 \draw[photon] (x2)--(td) node[pos=0.6]{$\rotatebox[origin=c]{45}{$\blacktriangledown$}$};
 \draw[photon] (d)--(td);
 \node[right= 0.9 of a](cx){};
 \node[above=.25 of cx] (f1){};
 \node[below=.25 of cx] (f2){};
 \draw[fcut] (f1)--(f2);
 \draw[fermion] (cx)--(a1) node [pos=0.25]{$\blacktriangleleft$} node[pos=0.75]{$\blacktriangleright$};
 \draw[fermion] (cx)--(b)  node [pos=.2]{$\blacktriangleright$} node[pos=0.58]{$\blacktriangleright$} node[pos=0.8]{$\blacktriangleleft$};
 \node[right=1 of c] (cx){};
 \node[inner sep=0pt,right=0.01 of c](c2) {};
 \node[left=0.5 of c2](c3){};
 \draw[fermion] (c2)--(c3)  node[midway]{$\blacktriangleright$};
  \node[inner sep=0pt,left=0.01 of d](d2) {};
 \node[inner sep=0pt,right=0.6 of d2](d3){};
 \draw[fermion] (d2)--(d3) node[midway]{$\blacktriangleleft$};
  \draw[fermion] (c3)--(d3) node[midway]{$\blacktriangleleft$};
  \node[inner sep=0 pt, left=0.45 of b] (b2){};
  \node[above= 0.5 of b2](y){};
       \draw [line width=1pt, double distance=1pt] (b2)-- (y) ; 

  \node[inner sep=0 pt] (a) at (7.5,0){};
\node[inner sep=0 pt, left=.5 of a] (a1){};
 \node[right=2.2 of a] (b){};
 \node[inner sep=0pt, right= 1.75 of a] (x){};
 \fill (x) circle (1pt);
 \node[inner sep=0pt, right= 0.3 of a] (x2){};
 \fill (x2) circle (1pt);
 \node[inner sep=0pt,below left=2 of x] (c){};
 \fill (c) circle (1pt);
 \node[inner sep=0pt,below right=2 of x2] (d){};
 \fill (d) circle (1pt);
 \draw[photon] (x)--(d) node[pos=0.45]{$\blacktriangledown$}; 
 \draw[photon] (x2)--(c) node[pos=0.45]{$\blacktriangle$}; 
 \node[right= 0.9 of a](cx){};
 \node[above=.25 of cx] (f1){};
 \node[below=.25 of cx] (f2){};
 \draw[fcut] (f1)--(f2);
 \draw[fermion] (a1)--(cx) node [pos=0.15]{$\blacktriangleright$} node[pos=0.4]{$\blacktriangleleft$} node[pos=0.8]{$\blacktriangleleft$};
 \draw[fermion] (cx)--(b)  node [pos=.3]{$\blacktriangleright$} node[pos=0.8]{$\blacktriangleleft$};
 \node[right=1 of c] (cx){};
 \node[inner sep=0pt,right=0.01 of c](c2) {};
 \node[left=0.5 of c2](c3){};
 \draw[fermion] (c2)--(c3)  node[midway]{$\blacktriangleright$};
  \node[inner sep=0pt,left=0.01 of d](d2) {};
 \node[inner sep=0pt,right=0.6 of d2](d3){};
 \draw[fermion] (d2)--(d3) node[midway]{$\blacktriangleleft$};
 \draw[fermion] (c3)--(d3) node[midway]{$\blacktriangleleft$};
  \node[inner sep=0 pt, right=0.35 of a1] (b2){};
  \node[above= 0.5 of b2](y){};
       \draw [line width=1pt, double distance=1pt] (b2)-- (y) ; 
       
\node[inner sep=0 pt] (a) at (11,0){};
 \node[inner sep=0 pt, left=0.5 of a] (a1){};
 \node[right=2.2 of a] (b){};
 
 \node[inner sep=0pt, right= 1.75 of a] (x){};
 \fill (x) circle (1pt);
 \node[inner sep=0pt, right= 0.3 of a] (x2){};
 \fill (x2) circle (1pt);
 \node[inner sep=0pt,below left=2 of x] (c){};
 \fill (c) circle (1pt);
 \node[inner sep=0pt,below right=2 of x2] (d){};
 \fill (d) circle (1pt);
  \draw[photon] (x)--(c) node[pos=0.3]{$\rotatebox[origin=c]{45}{$\blacktriangleleft$}$}; 
  \node[inner sep=2pt, below right=0.9 of x2](td){};
 \draw[photon] (x2)--(td) node[pos=0.6]{$\rotatebox[origin=c]{45}{$\blacktriangle$}$};
 \draw[photon] (d)--(td);
 \node[right= 0.9 of a](cx){};
 \node[above=.25 of cx] (f1){};
 \node[below=.25 of cx] (f2){};
 \draw[fcut] (f1)--(f2);
\draw[fermion] (a1)--(cx) node [pos=0.15]{$\blacktriangleright$} node[pos=0.4]{$\blacktriangleleft$} node[pos=0.8]{$\blacktriangleleft$};
 \draw[fermion] (cx)--(b)  node [pos=.3]{$\blacktriangleright$} node[pos=0.8]{$\blacktriangleleft$};
 \node[right=1 of c] (cx){};
 \node[inner sep=0pt,right=0.01 of c](c2) {};
 \node[left=0.5 of c2](c3){};
 \draw[fermion] (c2)--(c3)  node[midway]{$\blacktriangleright$};
  \node[inner sep=0pt,left=0.01 of d](d2) {};
 \node[inner sep=0pt,right=0.6 of d2](d3){};
 \draw[fermion] (d2)--(d3) node[midway]{$\blacktriangleleft$};
  \draw[fermion] (c3)--(d3) node[midway]{$\blacktriangleright$};
  \node[inner sep=0 pt, right=0.35 of a1] (b2){};
  \node[above= 0.5 of b2](y){};
       \draw [line width=1pt, double distance=1pt] (b2)-- (y) ; 
\node(n1) at (0.05,-1.3){};
\node[above=0.8 of n1] (n2){};
\draw[->] (n1)--(n2);
\node (n3) at (-.2,-0.75){$\ell$};
\node(p1) at(3.1,-.8){$+$};
\node(n1) at (3.6,-0.1){};
\node[below right=0.7 of n1] (n2){};
\draw[<-] (n1)--(n2);
\node (n3) at (3.8,-0.6){$\ell$};
\node(p1) at(6.5,-.8){$+$};
\node(n1) at (7.5+0.05,-1.3){};
\node[above=0.8 of n1] (n2){};
\draw[->] (n1)--(n2);
\node (n3) at (7.5-.2,-0.75){$\ell$};
\node(p1) at(7.2+3.1,-.8){$+$};
\node(n1) at (7.5+3.6,-0.1){};
\node[below right=0.7 of n1] (n2){};
\draw[<-] (n1)--(n2);
\node (n3) at (7.5+3.8,-0.6){$\ell$};
 \end{tikzpicture} \nonumber\\
&=-i {\cal N} \int \hat{d}^4\ell\;\frac{\hat{\d}(2p_1\cdot \ell)}{\ell^2(\ell-q)^2}\left[\frac{(p_1+q)^\mu 2q\cdot\ell}{(2p_2\cdot \ell-i\e)^2}-\frac{p_1^\mu 2q\cdot\ell}{(2p_2\cdot \ell+i\e)^2}\right]\,, \label{causal_imp1}
\end{align}
with ${\cal N} = 8e^4Q_1^2Q_2^2 (p_1\cdot p_2)^2$.
Notice that leading-order terms from the expansion of the propagators cancel automatically due to the causal $i\e$ prescription. We have also dropped all the scaleless integrals from the expansion for simplicity. Similarly,
\begin{align}
&\quad\quad\begin{tikzpicture}
 \node[inner sep=0 pt] (a) at (0,0){};
\node[inner sep=0 pt, left=0.04 of a] (a1){};
 \node[right=2.7 of a] (b){};
 \draw[fermion] (a1)--(b) node[pos=0.1] {$\blacktriangleright$} node[pos=0.45] {$\blacktriangleright$} node[pos=0.75] {$\blacktriangleright$}  node[pos=0.9] {$\blacktriangleleft$}; 
 \node[inner sep=0pt, right= 1.75 of a] (x){};
 \fill (x) circle (1pt);
 \node[inner sep=0pt, right= 0.3 of a] (x2){};
 \fill (x2) circle (1pt);
 \node[inner sep=0pt,below left=2 of x] (c){};
 \fill (c) circle (1pt);
 \node[inner sep=0pt,below right=2 of x2] (d){};
 \fill (d) circle (1pt);
 \draw[photon] (x)--(d) node[pos=0.45]{$\blacktriangle$}; 
 \draw[photon] (x2)--(c) node[pos=0.45]{$\blacktriangle$}; 
 \node[right= 0.55 of c](cx){};
 \node[above=.25 of cx] (x1){};
 \node[below=.25 of cx] (x2){};
 \draw[fcut] (x1)--(x2);
 \draw[fermion] (c)--(cx) node [midway]{$\blacktriangleleft$};
 \draw[fermion] (d)--(cx)  node [midway]{$\blacktriangleright$};
 \node[right=1 of c] (cx){};
 \node[inner sep=0pt,right=0.01 of c](c2) {};
 \node[left=0.5 of c2](c3){};
 \draw[fermion] (c2)--(c3)  node[midway]{$\blacktriangleright$};
  \node[inner sep=0pt,left=0.01 of d](d2) {};
 \node[inner sep=0pt,right=0.6 of d2](d3){};
 \draw[fermion] (d2)--(d3) node[midway]{$\blacktriangleleft$};
  \node[inner sep=0 pt, left=0.45 of b] (b2){};
  \node[above= 0.5 of b2](y){};
       \draw [line width=1pt, double distance=1pt] (b2)-- (y) ;  
 
 \node[inner sep=0 pt] (a) at (3.5,0){};
\node[inner sep=0 pt, left=0.04 of a] (a1){};
 \node[right=2.7 of a] (b){};
 \draw[fermion] (a1)--(b) node[pos=0.1] {$\blacktriangleright$} node[pos=0.45] {$\blacktriangleright$} node[pos=0.75] {$\blacktriangleright$}  node[pos=0.9] {$\blacktriangleleft$}; 
 \node[inner sep=0pt, right= 1.75 of a] (x){};
 \fill (x) circle (1pt);
 \node[inner sep=0pt, right= 0.3 of a] (x2){};
 \fill (x2) circle (1pt);
 \node[inner sep=0pt,below left=2 of x] (c){};
 \fill (c) circle (1pt);
 \node[inner sep=0pt,below right=2 of x2] (d){};
 \fill (d) circle (1pt);
  \draw[photon] (x)--(c) node[pos=0.2]{$\rotatebox[origin=c]{45}{$\blacktriangleright$}$}; 
  \node[inner sep=2pt, below right=0.9 of x2](td){};
 \draw[photon] (x2)--(td) node[pos=0.5]{$\rotatebox[origin=c]{45}{$\blacktriangle$}$};
 \draw[photon] (d)--(td);
 \node[right= 0.55 of c](cx){};
 \node[above=.25 of cx] (x1){};
 \node[below=.25 of cx] (x2){};
 \draw[fcut] (x1)--(x2);
 \draw[fermion] (c)--(cx) node [midway]{$\blacktriangleleft$};
 \draw[fermion] (d)--(cx)  node [midway]{$\blacktriangleright$};
 \node[right=1 of c] (cx){};
 \node[inner sep=0pt,right=0.01 of c](c2) {};
 \node[left=0.5 of c2](c3){};
 \draw[fermion] (c2)--(c3)  node[midway]{$\blacktriangleright$};
  \node[inner sep=0pt,left=0.01 of d](d2) {};
 \node[inner sep=0pt,right=0.6 of d2](d3){};
 \draw[fermion] (d2)--(d3) node[midway]{$\blacktriangleleft$};
  \node[inner sep=0 pt, left=0.45 of b] (b2){};
  \node[above= 0.5 of b2](y){};
       \draw [line width=1pt, double distance=1pt] (b2)-- (y) ;  

 \node[inner sep=0 pt] (a) at (7.5,0){};
\node[inner sep=0 pt, left=0.5 of a] (a1){};
 \node[right=2.2 of a] (b){};
 \draw[fermion] (a1)--(b) node[pos=0.1] {$\blacktriangleright$} node[pos=0.25] {$\blacktriangleleft$} node[pos=0.55] {$\blacktriangleleft$}  node[pos=0.9] {$\blacktriangleleft$}; 
 \node[inner sep=0pt, right= 1.75 of a] (x){};
 \fill (x) circle (1pt);
 \node[inner sep=0pt, right= 0.3 of a] (x2){};
 \fill (x2) circle (1pt);
 \node[inner sep=0pt,below left=2 of x] (c){};
 \fill (c) circle (1pt);
 \node[inner sep=0pt,below right=2 of x2] (d){};
 \fill (d) circle (1pt);
 \draw[photon] (x)--(d) node[pos=0.45]{$\blacktriangle$}; 
 \draw[photon] (x2)--(c) node[pos=0.45]{$\blacktriangle$}; 
 \node[right= 0.55 of c](cx){};
 \node[above=.25 of cx] (x1){};
 \node[below=.25 of cx] (x2){};
 \draw[fcut] (x1)--(x2);
 \draw[fermion] (c)--(cx) node [midway]{$\blacktriangleleft$};
 \draw[fermion] (d)--(cx)  node [midway]{$\blacktriangleright$};
 \node[right=1 of c] (cx){};
 \node[inner sep=0pt,right=0.01 of c](c2) {};
 \node[left=0.5 of c2](c3){};
 \draw[fermion] (c2)--(c3)  node[midway]{$\blacktriangleright$};
  \node[inner sep=0pt,left=0.01 of d](d2) {};
 \node[inner sep=0pt,right=0.6 of d2](d3){};
 \draw[fermion] (d2)--(d3) node[midway]{$\blacktriangleleft$};
  \node[inner sep=0 pt, right=0.45 of a1] (b2){};
  \node[above= 0.5 of b2](y){};
       \draw [line width=1pt, double distance=1pt] (b2)-- (y) ;  
 \node[inner sep=0 pt] (a) at (11,0){};
\node[inner sep=0 pt, left=0.5 of a] (a1){};
 \node[right=2.2 of a] (b){};
 \draw[fermion] (a1)--(b) node[pos=0.1] {$\blacktriangleright$} node[pos=0.25] {$\blacktriangleleft$} node[pos=0.55] {$\blacktriangleleft$}  node[pos=0.9] {$\blacktriangleleft$};  
 \node[inner sep=0pt, right= 1.75 of a] (x){};
 \fill (x) circle (1pt);
 \node[inner sep=0pt, right= 0.3 of a] (x2){};
 \fill (x2) circle (1pt);
 \node[inner sep=0pt,below left=2 of x] (c){};
 \fill (c) circle (1pt);
 \node[inner sep=0pt,below right=2 of x2] (d){};
 \fill (d) circle (1pt);
  \draw[photon] (x)--(c) node[pos=0.2]{$\rotatebox[origin=c]{45}{$\blacktriangleright$}$}; 
  \node[inner sep=2pt, below right=0.9 of x2](td){};
 \draw[photon] (x2)--(td) node[pos=0.5]{$\rotatebox[origin=c]{45}{$\blacktriangle$}$};
 \draw[photon] (d)--(td);
 \node[right= 0.55 of c](cx){};
 \node[above=.25 of cx] (x1){};
 \node[below=.25 of cx] (x2){};
 \draw[fcut] (x1)--(x2);
 \draw[fermion] (c)--(cx) node [midway]{$\blacktriangleleft$};
 \draw[fermion] (d)--(cx)  node [midway]{$\blacktriangleright$};
 \node[right=1 of c] (cx){};
 \node[inner sep=0pt,right=0.01 of c](c2) {};
 \node[left=0.5 of c2](c3){};
 \draw[fermion] (c2)--(c3)  node[midway]{$\blacktriangleright$};
  \node[inner sep=0pt,left=0.01 of d](d2) {};
 \node[inner sep=0pt,right=0.6 of d2](d3){};
 \draw[fermion] (d2)--(d3) node[midway]{$\blacktriangleleft$};
   \node[inner sep=0 pt, right=0.45 of a1] (b2){};
  \node[above= 0.5 of b2](y){};
       \draw [line width=1pt, double distance=1pt] (b2)-- (y) ;  
\node(n1) at (0.05,-1.3){};
\node[above=0.8 of n1] (n2){};
\draw[->] (n1)--(n2);
\node (n3) at (-.2,-0.75){$\ell$};
\node(p1) at(3.1,-.8){$+$};
\node(n1) at (1.6+4,-0.1){};
\node[below left=0.7 of n1] (n2){};
\draw[<-] (n1)--(n2);
\node (n3) at (1.4+4,-0.6){$\ell$};
\node(p1) at(6.1,-.8){$+$};
\node(n1) at (7.5+0.05,-1.3){};
\node[above=0.8 of n1] (n2){};
\draw[->] (n1)--(n2);
\node (n3) at (7.5-.2,-0.75){$\ell$};
\node(p1) at(7.5+3.1,-.8){$+$};
\node(n1) at (7.5+1.6+4,-0.1){};
\node[below left=0.7 of n1] (n2){};
\draw[<-] (n1)--(n2);
\node (n3) at (7.5+1.4+4,-0.6){$\ell$};
\end{tikzpicture}  \nonumber\\
&=-{\cal N} \int \hat{d}^4\ell\;\frac{\hat{\d}(2p_2\cdot\ell)}{\ell^2(\ell-q)^2}\left[\hat{\d}(2p_1\cdot\ell)(2p_1^\mu+q^\mu)+i\left(\frac{(p_1+q)^\mu 2q\cdot\ell}{(2p_1\cdot\ell-i\e)^2}-\frac{p_1^\mu 2q\cdot\ell}{(2p_1\cdot\ell+i\e)^2}\right)\right]\,, \label{causal_imp2}
\end{align}
where we have used the distributional identity
\begin{equation}
    \frac{1}{x-i\e}-\frac{1}{x+i\e}=i\hat{\d}(x)\,.\label{delta}
\end{equation}
Next, we have diagrams from inserting the momentum operator on the internal lines:
\begin{align}    \quad\quad& \begin{tikzpicture}
  \node[inner sep=0 pt] (a) at (0,0){};
\node[inner sep=0 pt, left=.1 of a] (a1){};
 \node[right=2.2 of a] (b){};
 \node[inner sep=0pt, right= 1.75 of a] (x){};
 \fill (x) circle (1pt);
 \node[inner sep=0pt, right= 0.3 of a] (x2){};
 \fill (x2) circle (1pt);
 \node[inner sep=0pt,below left=2 of x] (c){};
 \fill (c) circle (1pt);
 \node[inner sep=0pt,below right=2 of x2] (d){};
 \fill (d) circle (1pt);
 \draw[photon] (x)--(d) node[pos=0.45]{$\blacktriangle$}; 
 \draw[photon] (x2)--(c) node[pos=0.45]{$\blacktriangledown$}; 
 \node[right= 0.72 of a](cx){};
 \node[above=.25 of cx] (f1){};
 \node[below=.25 of cx] (f2){};
 \draw[fcut] (f1)--(f2);
 \draw[fermion] (cx)--(a1) node [pos=0.25]{$\blacktriangleleft$} node[pos=0.75]{$\blacktriangleright$};
 \draw[fermion] (cx)--(b)  node [pos=.2]{$\blacktriangleright$}
 node[pos=0.5]{$\blacktriangleleft$} node[pos=0.8]{$\blacktriangleleft$};
 \node[right=1 of c] (cx){};
 \node[inner sep=0pt,right=0.01 of c](c2) {};
 \node[left=0.5 of c2](c3){};
 \draw[fermion] (c2)--(c3)  node[midway]{$\blacktriangleright$};
  \node[inner sep=0pt,left=0.01 of d](d2) {};
 \node[inner sep=0pt,right=0.6 of d2](d3){};
 \draw[fermion] (d2)--(d3) node[midway]{$\blacktriangleleft$};
 \draw[fermion] (c3)--(d3) node[midway]{$\blacktriangleright$};
  \node[inner sep=0 pt, left=0.78 of b] (b2){};
  \node[above= 0.5 of b2](y){};
       \draw [line width=1pt, double distance=1pt] (b2)-- (y) ; 
       
\node[inner sep=0 pt] (a) at (3.5,0){};
 \node[inner sep=0 pt, left=0.04 of a] (a1){};
 \node[right=2.2 of a] (b){};
 
 \node[inner sep=0pt, right= 1.78 of a] (x){};
 \fill (x) circle (1pt);
 \node[inner sep=0pt, right= 0.2 of a] (x2){};
 \fill (x2) circle (1pt);
 \node[inner sep=0pt,below left=2 of x] (c){};
 \fill (c) circle (1pt);
 \node[inner sep=0pt,below right=2 of x2] (d){};
 \fill (d) circle (1pt);
  \draw[photon] (x)--(c) node[pos=0.2]{$\rotatebox[origin=c]{45}{$\blacktriangleright$}$}; 
  \node[inner sep=2pt, below right=0.9 of x2](td){};
 \draw[photon] (x2)--(td) node[pos=0.6]{$\rotatebox[origin=c]{45}{$\blacktriangledown$}$};
 \draw[photon] (d)--(td);
\node[right= 0.72 of a](cx){};
 \node[above=.25 of cx] (f1){};
 \node[below=.25 of cx] (f2){};
 \draw[fcut] (f1)--(f2);
 \draw[fermion] (cx)--(a1) node [pos=0.25]{$\blacktriangleleft$} node[pos=0.75]{$\blacktriangleright$};
 \draw[fermion] (cx)--(b)  node [pos=.2]{$\blacktriangleright$}
 node[pos=0.5]{$\blacktriangleleft$} node[pos=0.8]{$\blacktriangleleft$};
 \node[right=1 of c] (cx){};
 \node[inner sep=0pt,right=0.01 of c](c2) {};
 \node[left=0.5 of c2](c3){};
 \draw[fermion] (c2)--(c3)  node[midway]{$\blacktriangleright$};
  \node[inner sep=0pt,left=0.01 of d](d2) {};
 \node[inner sep=0pt,right=0.6 of d2](d3){};
 \draw[fermion] (d2)--(d3) node[midway]{$\blacktriangleleft$};
  \draw[fermion] (c3)--(d3) node[midway]{$\blacktriangleleft$};
  \node[inner sep=0 pt, left=0.78 of b] (b2){};
  \node[above= 0.5 of b2](y){};
       \draw [line width=1pt, double distance=1pt] (b2)-- (y) ; 

  \node[inner sep=0 pt] (a) at (7.5,0){};
\node[inner sep=0 pt, left=.01 of a] (a1){};
 \node[right=2.2 of a] (b){};
 \node[inner sep=0pt, right= 1.75 of a] (x){};
 \fill (x) circle (1pt);
 \node[inner sep=0pt, right= 0.3 of a] (x2){};
 \fill (x2) circle (1pt);
 \node[inner sep=0pt,below left=2 of x] (c){};
 \fill (c) circle (1pt);
 \node[inner sep=0pt,below right=2 of x2] (d){};
 \fill (d) circle (1pt);
 \draw[photon] (x)--(d) node[pos=0.45]{$\blacktriangledown$}; 
 \draw[photon] (x2)--(c) node[pos=0.45]{$\blacktriangle$}; 
 \node[right= 1.15 of a](cx){};
 \node[above=.25 of cx] (f1){};
 \node[below=.25 of cx] (f2){};
 \draw[fcut] (f1)--(f2);
 \draw[fermion] (a1)--(cx) node [pos=0.15]{$\blacktriangleright$} node[pos=0.5]{$\blacktriangleright$} node[pos=0.8]{$\blacktriangleleft$};
 \draw[fermion] (cx)--(b)  node [pos=.3]{$\blacktriangleright$} node[pos=0.75]{$\blacktriangleleft$};
 \node[right=1 of c] (cx){};
 \node[inner sep=0pt,right=0.01 of c](c2) {};
 \node[left=0.5 of c2](c3){};
 \draw[fermion] (c2)--(c3)  node[midway]{$\blacktriangleright$};
  \node[inner sep=0pt,left=0.01 of d](d2) {};
 \node[inner sep=0pt,right=0.6 of d2](d3){};
 \draw[fermion] (d2)--(d3) node[midway]{$\blacktriangleleft$};
 \draw[fermion] (c3)--(d3) node[midway]{$\blacktriangleleft$};
  \node[inner sep=0 pt, right=0.75 of a1] (b2){};
  \node[above= 0.5 of b2](y){};
       \draw [line width=1pt, double distance=1pt] (b2)-- (y) ; 
       
\node[inner sep=0 pt] (a) at (11,0){};
 \node[inner sep=0 pt, left=0.01 of a] (a1){};
 \node[right=2.2 of a] (b){};
 
 \node[inner sep=0pt, right= 1.78 of a] (x){};
 \fill (x) circle (1pt);
 \node[inner sep=0pt, right= 0.2 of a] (x2){};
 \fill (x2) circle (1pt);
 \node[inner sep=0pt,below left=2 of x] (c){};
 \fill (c) circle (1pt);
 \node[inner sep=0pt,below right=2 of x2] (d){};
 \fill (d) circle (1pt);
  \draw[photon] (x)--(c) node[pos=0.3]{$\rotatebox[origin=c]{45}{$\blacktriangleleft$}$}; 
  \node[inner sep=2pt, below right=0.9 of x2](td){};
 \draw[photon] (x2)--(td) node[pos=0.6]{$\rotatebox[origin=c]{45}{$\blacktriangle$}$};
 \draw[photon] (d)--(td);
 \node[right= 1.15 of a](cx){};
 \node[above=.25 of cx] (f1){};
 \node[below=.25 of cx] (f2){};
 \draw[fcut] (f1)--(f2);
 \draw[fermion] (a1)--(cx) node [pos=0.15]{$\blacktriangleright$} node[pos=0.5]{$\blacktriangleright$} node[pos=0.8]{$\blacktriangleleft$};
 \draw[fermion] (cx)--(b)  node [pos=.3]{$\blacktriangleright$} node[pos=0.75]{$\blacktriangleleft$};
 \node[right=1 of c] (cx){};
 \node[inner sep=0pt,right=0.01 of c](c2) {};
 \node[left=0.5 of c2](c3){};
 \draw[fermion] (c2)--(c3)  node[midway]{$\blacktriangleright$};
  \node[inner sep=0pt,left=0.01 of d](d2) {};
 \node[inner sep=0pt,right=0.6 of d2](d3){};
 \draw[fermion] (d2)--(d3) node[midway]{$\blacktriangleleft$};
  \draw[fermion] (c3)--(d3) node[midway]{$\blacktriangleright$};
  \node[inner sep=0 pt, right=0.75 of a1] (b2){};
  \node[above= 0.5 of b2](y){};
       \draw [line width=1pt, double distance=1pt] (b2)-- (y) ; 
\node(n1) at (0.05,-1.3){};
\node[above=0.8 of n1] (n2){};
\draw[->] (n1)--(n2);
\node (n3) at (-.2,-0.75){$\ell$};
\node(p1) at(3.1,-.8){$+$};
\node(n1) at (3.5,-0.1){};
\node[below right=0.7 of n1] (n2){};
\draw[<-] (n1)--(n2);
\node (n3) at (3.7,-0.7){$\ell$};
\node(p1) at(6.5,-.8){$+$};
\node(n1) at (7.5+0.05,-1.3){};
\node[above=0.8 of n1] (n2){};
\draw[->] (n1)--(n2);
\node (n3) at (7.5-.2,-0.75){$\ell$};
\node(p1) at(7.2+3.1,-.8){$+$};
\node(n1) at (7.5+3.5,-0.1){};
\node[below right=0.7 of n1] (n2){};
\draw[<-] (n1)--(n2);
\node (n3) at (7.5+3.7,-0.7){$\ell$};
 \end{tikzpicture}\nonumber\\
&= {\cal N}  \int \hat{d}^4\ell\;\frac{\hat{\d}(2p_1\cdot\ell)}{\ell^2(\ell-q)^2}\hat{\d}'(2p_2\cdot\ell)(2q\cdot\ell)(p_1^\mu+\ell^\mu)\,,
\end{align}
where we have used
\begin{equation}
    \frac{i}{(x-i\e)^2}-\frac{i}{(x+i\e)^2}=i\hat{\d}'(x)\,.\label{delta2}
\end{equation}
Notice that the leading-order terms cancel similar to Eq.~\eqref{causal_imp1}. Finally, we have
\begin{align}
&\raisebox{-35pt}{\begin{tikzpicture}[scale=1]
 \node[inner sep=0 pt] (a) at (0,0){};
\node[inner sep=0 pt, left=0.04 of a] (a1){};
 \node[right=2.2 of a] (b){};
 \draw[fermion] (a1)--(b) node[pos=0.1] {$\blacktriangleright$} node[pos=0.3] {$\blacktriangleleft$} node[pos=0.68] {$\blacktriangleright$}  node[pos=0.9] {$\blacktriangleleft$}; 
 \node[inner sep=0pt, right= 1.75 of a] (x){};
 \fill (x) circle (1pt);
 \node[inner sep=0pt, right= 0.3 of a] (x2){};
 \fill (x2) circle (1pt);
 \node[inner sep=0pt,below left=2 of x] (c){};
 \fill (c) circle (1pt);
 \node[inner sep=0pt,below right=2 of x2] (d){};
 \fill (d) circle (1pt);
 \draw[photon] (x)--(d) node[pos=0.45]{$\blacktriangle$}; 
 \draw[photon] (x2)--(c) node[pos=0.45]{$\blacktriangle$}; 
 \node[right= 0.55 of c](cx){};
 \node[above=.25 of cx] (x1){};
 \node[below=.25 of cx] (x2){};
 \draw[fcut] (x1)--(x2);
 \draw[fermion] (c)--(cx) node [midway]{$\blacktriangleleft$};
 \draw[fermion] (d)--(cx)  node [midway]{$\blacktriangleright$};
 \node[right=1 of c] (cx){};
 \node[inner sep=0pt,right=0.01 of c](c2) {};
 \node[left=0.5 of c2](c3){};
 \draw[fermion] (c2)--(c3)  node[midway]{$\blacktriangleright$};
  \node[inner sep=0pt,left=0.01 of d](d2) {};
 \node[inner sep=0pt,right=0.6 of d2](d3){};
 \draw[fermion] (d2)--(d3) node[midway]{$\blacktriangleleft$};
  \node[inner sep=0 pt, right=1 of a] (b2){};
  \node[above= 0.5 of b2](y){};
       \draw [line width=1pt, double distance=1pt] (b2)-- (y) ;  
 
 \node[inner sep=0 pt] (a) at (3.5,0){};
\node[inner sep=0 pt, left=0.04 of a] (a1){};
 \node[right=2.2 of a] (b){};
 \draw[fermion] (a1)--(b) node[pos=0.1] {$\blacktriangleright$} node[pos=0.3] {$\blacktriangleleft$} node[pos=0.68] {$\blacktriangleright$}  node[pos=0.9] {$\blacktriangleleft$};
 \node[inner sep=0pt, right= 1.75 of a] (x){};
 \fill (x) circle (1pt);
 \node[inner sep=0pt, right= 0.3 of a] (x2){};
 \fill (x2) circle (1pt);
 \node[inner sep=0pt,below left=2 of x] (c){};
 \fill (c) circle (1pt);
 \node[inner sep=0pt,below right=2 of x2] (d){};
 \fill (d) circle (1pt);
  \draw[photon] (x)--(c) node[pos=0.2]{$\rotatebox[origin=c]{45}{$\blacktriangleright$}$}; 
  \node[inner sep=2pt, below right=0.9 of x2](td){};
 \draw[photon] (x2)--(td) node[pos=0.5]{$\rotatebox[origin=c]{45}{$\blacktriangle$}$};
 \draw[photon] (d)--(td);
 \node[right= 0.55 of c](cx){};
 \node[above=.25 of cx] (x1){};
 \node[below=.25 of cx] (x2){};
 \draw[fcut] (x1)--(x2);
 \draw[fermion] (c)--(cx) node [midway]{$\blacktriangleleft$};
 \draw[fermion] (d)--(cx)  node [midway]{$\blacktriangleright$};
 \node[right=1 of c] (cx){};
 \node[inner sep=0pt,right=0.01 of c](c2) {};
 \node[left=0.5 of c2](c3){};
 \draw[fermion] (c2)--(c3)  node[midway]{$\blacktriangleright$};
  \node[inner sep=0pt,left=0.01 of d](d2) {};
 \node[inner sep=0pt,right=0.6 of d2](d3){};
 \draw[fermion] (d2)--(d3) node[midway]{$\blacktriangleleft$};
  \node[inner sep=0 pt, right=1 of a] (b2){};
  \node[above= 0.5 of b2](y){};
       \draw [line width=1pt, double distance=1pt] (b2)-- (y) ;  
\node(n1) at (0.05,-1.3){};
\node[above=0.8 of n1] (n2){};
\draw[->] (n1)--(n2);
\node (n3) at (-.2,-0.75){$\ell$};
\node(p1) at(2.8,-.8){$+$};
\node(n1) at (1.6+4,-0.1){};
\node[below left=0.7 of n1] (n2){};
\draw[<-] (n1)--(n2);
\node (n3) at (1.4+4,-0.6){$\ell$};
\end{tikzpicture}} 
 \\
& = {\cal N}
\int \hat{d}^4\ell\;\frac{\hat{\d}(2p_2\cdot\ell)}{\ell^2(\ell-q)^2}
\left[\hat{\d}(2p_1\cdot\ell)(2p_1+q)^\mu + \hat{\d}'(2p_1\cdot\ell)(p_1+q-\ell)^\mu (2q\cdot\ell) \right]
\,. \nonumber 
\end{align}
Adding all the contributions together, we see that the super-classical terms cancel manifestly. The sub-leading terms give the next to leading-order impulse
\begin{align}
  \Delta\mathbb{P}_{1,\text{NLO}}^\mu&=\frac{ie^4}{2}Q_1^2Q_2^2\int \frac{\hat{d}^4\ell}{\ell^2(\ell-q)^2}\hat{d}^4q\;\hat{\d}\left(u_1\cdot q\right) \hat{\d}\left(u_2\cdot q\right)e^{-ib\cdot q}\times\nonumber\\
  &\times\Bigg[q^\mu\left(\frac{\hat{\d}\left(u_1\cdot \ell\right)}{m_1}+\frac{\hat{\d}\left(u_2\cdot \ell\right)}{m_2}-\g^2\left(\frac{\hat{\d}\left(u_2\cdot\ell \right)}{m_1}\frac{q\cdot \ell}{\left(u_1\cdot \ell+i\e\right)^2}+\frac{\hat{\d}\left(u_1\cdot\ell \right)}{m_2}\frac{q\cdot \ell}{\left(u_2\cdot \ell-i\e\right)^2}\right)\right)\nonumber\\
  &+i\g^2\ell^\mu\;q\cdot\ell\left(\frac{\hat{\d}'\left(u_1\cdot\ell\right)\left(u_2\cdot\ell\right)}{m_1}-\frac{\hat{\d}'\left(u_2\cdot\ell\right)\left(u_1\cdot\ell\right)}{m_2}\right)\Bigg],
\end{align}
which agrees with \cite{Kosower2019} modulo scaleless integrals.
\subsection{Waveform and its variance}
In this section,  we will discuss the leading order waveform and its variance in the causal basis. As discussed in \ref{sec:review_KMOC}, the waveform is given by the amputated five-point response function in Eq.~\eqref{5pt-waveform}. The leading-order waveform in the  $\I\II$ basis involves $\mathcal{O}\left(q^{-3}\right)$ super-classical terms upon the expansion of the massive propagators, the cancellation of which  formally requires including a cut with a three-particle amplitude:
\begin{align}    &\begin{tikzpicture}[baseline=-0.8 cm,scale=.7]    
\node (a) at (0,0){$p_1$};
  \node[right=2.7 of a ] (b){$p_1+q-k$};
  \draw[fermion] (a) -- (b) ;
 \node[inner sep=0pt,right=.9 of a] (x){};
\fill (x) circle (1pt);
 \node[inner sep=0pt,right=1.7 of a] (y){};
\fill (y) circle (1pt);
\node[above right= 1 of y] (c){$k$};
\draw[photon] (y)--(c);
\node[below=1 of a] (d){$p_2$};
\node[right=2.7 of d](e){$p_2-q$};
\draw[fermion] (d)--(e);
\node[inner sep=0pt, right= .9 of d](z){};
\fill (z) circle(1pt);
\draw[photon](x)--(z);
\node[inner sep=0pt, above=0.1 of x](nx){$\textcolor{gray}{\I}$};
\node[inner sep=0pt, right=0.55 of nx](ny){$\textcolor{gray}{\I}$};
\node[inner sep=0pt, below=0.1 of z](nz){$\textcolor{gray}{\I}$};
 \node (a2) at (7.5,0){};
  \node[right=2.7 of a2 ] (b2){};
  \draw[fermion] (a2) -- (b2);
 \node[inner sep=0pt,right=.9 of a2] (x2){};
\fill (x2) circle (1pt);
 \node[inner sep=0pt,right=1.7 of a2] (y2){};
\fill (y2) circle (1pt);
\node[above right= 1 of x2] (c2){};
\draw[photon] (x2)--(c2);
\node[below=1.2 of a2] (d2){};
\node[right=2.7 of d2](e2){};
\draw[fermion] (d2)--(e2);
\node[inner sep=0pt, right= 1.7 of d2](z2){};
\fill (z2) circle(1pt);
\draw[photon](y2)--(z2);
\node[inner sep=0pt, above=0.1 of y2](nx){$\textcolor{gray}{\I}$};
\node[inner sep=0pt, left=0.7 of nx](ny){$\textcolor{gray}{\I}$};
\node[inner sep=0pt, below=0.1 of z2](nz){$\textcolor{gray}{\I}$};

\node (n1) at (4.2+2.5,-1.1) {$+$};

 \node (a2) at (12.5,0){};
  \node[right=2.7 of a2 ] (b2){};
  \draw[fermion] (a2) -- (b2);
 \node[inner sep=0pt,right=.5 of a2] (x2){};
\fill (x2) circle (1pt);
 \node[inner sep=0pt,right=1.7 of a2] (y2){};
\fill (y2) circle (1pt);
\node[above right= 1 of x2] (c2){};
\draw[photon] (x2)--(c2);
\node[below=1.2 of a2] (d2){};
\node[right=2.7 of d2](e2){};
\draw[fermion] (d2)--(e2);
\node[inner sep=0pt, right= 1.7 of d2](z2){};
\fill (z2) circle(1pt);
\draw[photon](y2)--(z2);
\draw[fcut,brown](1.99+9+3.5,1.2)--(1.99+9+3.5,-1.2);
\node[inner sep=0pt, above=0.1 of y2](nx){$\textcolor{gray}{\II}$};
\node[inner sep=0pt, left=1.05 of nx](ny){$\textcolor{gray}{\I}$};
\node[inner sep=0pt, below=0.1 of z2](nz){$\textcolor{gray}{\II}$};
\node (n1) at (4.2+4.5+3.5,-1.1) {$+$};
\end{tikzpicture} \\
&= -8ie^3Q_1^2Q_2\frac{(p_1\cdot p_2)(p_1\cdot\ve^h)}{q^2}\left[\frac{1}{2p_1\cdot k+i\e}+\frac{1}{-2p_1\cdot k+i\e} - \hat{\d}\left(2p_1\cdot k\right)\right] + {\cal O}(q^{-2})\nonumber\\
&= {\cal O}(q^{-2})\,, \nonumber
\end{align}
where we have used the on-shell condition $2p_1\cdot q=2p_1\cdot k+\mathcal{O}\left(q^2\right)$, and Eq.~\eqref{delta}.
The analogous calculation in the $r/a$  basis  is given by the five-point amputated response function of one $r$-type photon field and four $a-$ type heavy fields 
\begin{align}
 \mathcal{W}^h_k=\text{LSZ}_{x}^{\out} \prod_{i=1,2} \text{LSZ}_{x_{i}'}^{\out} \; \text{LSZ}_{x_i}^{\in} R_{5} [A^{r\;h}(x);\F_1^{a\dagger}(x_1),\F_2^{a\dagger(x_2)},\F_1^a(x_1'),\F_2^a(x_2')]. 
\end{align}
At the leading order, we have the following diagrams
\begin{align}\mathcal{W}_k^h(q,k)= \begin{tikzpicture}[baseline=0.1 ex,scale=0.35]
\draw[fermion] (-2,-2) -- (0,0) node[pos=0.35] {$\rotatebox[origin=c]{45}{$\blacktriangleright$}$} ;
      \node (a) at (-2.5,-2.5){$p_2$};
    \draw[fermion] (2,2) -- (0,0) node[pos=0.35] {$\rotatebox[origin=c]{45}{$\blacktriangleleft$}$} ;
      \node (a) at (3.5,2.7){$p_1+q-k$};
    \draw[fermion] (-2,2) -- (0,0) node[pos=0.35] {$\rotatebox[origin=c]{45}{$\blacktriangledown$}$} ;
      \node (a) at (-2.5,2.5){$p_1$};
    \draw[fermion] (2,-2) -- (0,0) node[pos=0.35] {$\rotatebox[origin=c]{45}{$\blacktriangle$}$} ;
      \node (a) at (2.5,-2.5){$p_2-q$};
    
    \draw[photon] (0,0) -- (0,3) node[pos=0.9] {$\blacktriangle$} ;
      \node (a) at (0,3.6){$k$};
    
    \node[minimum size=0.8cm, fill = gray!20, draw = black, ultra thick, circle] at (0,0){};
\end{tikzpicture}&=  \begin{tikzpicture}[baseline=-0.8 cm ,scale=.7]    
\node (a) at (0,0){};
  \node[right=2.7 of a ] (b){};
  \draw[fermion] (a) -- (b) node[pos=0.2]{$\blacktriangleright$} node[pos=0.8] {$\blacktriangleleft$} node[pos=.52]{$\blacktriangleright$};
 \node[inner sep=0pt,right=.9 of a] (x){};
\fill (x) circle (1pt);
 \node[inner sep=0pt,right=1.7 of a] (y){};
\fill (y) circle (1pt);
\node[above right= 1 of y] (c){};
\draw[photon] (y)--(c) node[pos=0.95]{$\rotatebox[origin=c]{45}{$\blacktriangleright$}$};
\node[below=1.2 of a] (d){};
\node[right=2.7 of d](e){};
\draw[fermion] (d)--(e) node[pos=0.2]{$\blacktriangleright$} node[pos=0.8]{$\blacktriangleleft$};
\node[inner sep=0pt, right= .9 of d](z){};
\fill (z) circle(1pt);
\draw[photon](x)--(z) node[midway]{$\blacktriangle$};
 \node (a2) at (4.5,0){};
  \node[right=2.7 of a2 ] (b2){};
  \draw[fermion] (a2) -- (b2) node[pos=0.2]{$\blacktriangleright$} node[pos=0.8] {$\blacktriangleleft$} node[pos=.5]{$\blacktriangleleft$};
 \node[inner sep=0pt,right=.9 of a2] (x2){};
\fill (x2) circle (1pt);
 \node[inner sep=0pt,right=1.7 of a2] (y2){};
\fill (y2) circle (1pt);
\node[above right= 1 of x2] (c2){};
\draw[photon] (x2)--(c2) node[pos=0.95]{$\rotatebox[origin=c2]{45}{$\blacktriangleright$}$};
\node[below=1.2 of a2] (d2){};
\node[right=2.7 of d2](e2){};
\draw[fermion] (d2)--(e2) node[pos=0.2]{$\blacktriangleright$} node[pos=0.8]{$\blacktriangleleft$};
\node[inner sep=0pt, right= 1.7 of d2](z2){};
\fill (z2) circle(1pt);
\draw[photon](y2)--(z2) node[midway]{$\blacktriangle$};

 \node (a3) at (9,0){};
  \node[right=2.4 of a3 ] (b3){};
  \draw[fermion] (a3) -- (b3) node[pos=0.2]{$\blacktriangleright$} node[pos=0.8] {$\blacktriangleleft$};
 \node[inner sep=0pt,right=1.2 of a3] (x3){};
\fill (x3) circle (1pt);
\node[above right= 1 of x3] (c3){};
\draw[photon] (x3)--(c3) node[pos=0.95]{$\rotatebox[origin=c3]{45}{$\blacktriangleright$}$};
\node[below=1.2 of a3] (d3){};
\node[right=2.4 of d3](e3){};
\draw[fermion] (d3)--(e3) node[pos=0.2]{$\blacktriangleright$} node[pos=0.8]{$\blacktriangleleft$};
\node[inner sep=0pt, right= 1.2 of d3](z3){};
\fill (z3) circle(1pt);
\draw[photon](x3)--(z3) node[midway]{$\blacktriangle$};
\node (n1) at (4.2,-0.9) {$+$};
\node (n2) at (4.2+4.5,-0.9) {$+$};
\end{tikzpicture}  
\nonumber\\
&\quad \quad \quad \quad \quad \quad \quad \quad \quad \quad \quad \quad\quad \quad \quad \;+\;(1\leftrightarrow 2)\,.\label{waveform_tree}
\end{align} 
The ${\cal O}\left(q^{-3}\right)$ terms in Eq.~\eqref{waveform_tree} are
\begin{align}    -8ie^3Q_1^2Q_2\frac{(p_1\cdot p_2)(p_1\cdot\ve^h)}{q^2}\left[\frac{1}{2p_1\cdot q+i\e}+\frac{1}{-2p_1\cdot k-i\e}\right]=0.
\end{align}
where we have used the on-shell condition $2p_1\cdot q=2p_1\cdot k+{\cal O}\left(q^2\right)$.
Similar to the computation of the linear impulse in the causal basis, the ${\cal O}\left(q^{-3}\right)$ terms, which are singular as $\hbar \to 0$, manifestly cancel, thanks to the causal $i\e$ prescription. We have also verified that the subleading terms of these diagrams in the soft expansion of the waveform yield precisely the classical answer.

So far we have considered causal response functions with one $r$ type field. However, correlation functions with multiple $r$-type insertions which measure correlations between different measurements after a system is perturbed, are also familiar objects in non-linear response theory. The simplest of them in our context is the six-point amputated response function with two on-shell $r$-type photon fields and four $a$-type massive fields. Explicitly, we want to compute 
\begin{align}
     &\bra{1'2'}a_{k_1}^{h_1\,\out}a_{k_2}^{h_2\,\out}\ket{12}\nonumber \\    &=\prod_{i=1}^2\text{LSZ}^{\in}_{x'_i}\prod_{j=1}^2\text{LSZ}^{\out}_{x_j} \prod_{k=1}^2\text{LSZ}^{\out}_{z_k} R_{6}\left[A^{rh_1}(z_1),A^{rh_2}(z_2); \F_1^{a\dagger}(x_1),\F_2^{a\dagger}(x_2),\F_1^a(x_1'),\F_2^a(x_2')\right].
      \label{variance} 
\end{align}
The expectation value in Eq.~\eqref{variance} is used to compute the variance of the waveform in \cite{Cristofoli:2021jas} which can be defined as
\begin{align}
    \mathcal{V}(k_1,k_2)= \bra{1'2'}a_{k_1}^{h_1\,\out}a_{k_2}^{h_2\,\out}\ket{12}\big|_{\rm cl.}-\bra{1'2'}a_{k_1}^{h_1\,\out}\ket{12}\big|_{\rm cl.}\bra{1'2'}a_{k_2}^{h_2\,\out}\ket{12}\big|_{\rm cl.}.\label{var}
\end{align}
appropriately integrated against classical wavepackets for the heavy particles. We refer the reader to \cite{Cristofoli:2021jas} for details. Here, we concentrate on the properties of the six-point amputated response function in Eq.~\eqref{variance} in the classical limit. It was shown in \cite{Cristofoli:2021jas} that at the tree-level the expectation value in Eq.~\eqref{var} is quantum. More specifically, working in a gauge 
\begin{equation}
    p_1\cdot \e^{h_i}(k_i)=0,\quad i=1,2.
\end{equation}
it was shown that the ${\cal O}\left(q^{-4}\right)$, ${\cal O}\left(q^{-3}\right)$ diagrams in the six-point tree amplitudes cancel, and the leading order six-point connected tree amplitude is, in fact,
${\cal O}\left(q^{-2}\right)$. The particular gauge choice was crucial to simplify the analysis in \cite{Cristofoli:2021jas}. 

In this section, we demonstrate the advantage of the causal $r/a$ basis for the calculation of the variance. We will present our result in full generality without adopting any particular gauge. The counting rule in Eq.~\eqref{counting} for $n_r=2$ and $n_a=0$ gives $P_{rr}=1$ at the tree level. Therefore, at leading order one can only draw various single-cut diagrams that isolate a three-point on-shell amplitude or diagrams with horizontal cuts as shown below.
\begin{align}
    \label{var_main}\begin{tikzpicture}[baseline=0.1 ex,scale=0.25]
  \node[inner sep=0pt] (c) at (0,0){};
\node[above=0.2 of c] (a) {};
\node[left=1.1 of a](a2){$p_1$};
    \draw[fermion] (a) -- (a2) node[pos=0.7]{$\blacktriangleright$} ;
     \node[inner sep=0pt] (c) at (0,0){};
\node[below=0.2 of c] (b) {};
\node[left=1.1 of b](b2){$p_2$};
    \draw[fermion] (b) -- (b2) node[pos=0.7]{$\blacktriangleright$} ; ;
  \node[above=0.2 of c](d1){};
   \node[ right=1.1 of d1](d) {$p'_1$};
   \draw[fermion](d1)--(d) node[pos=0.7]{$\blacktriangleleft$} ;
   \node[below=0.2 of c](e1){};
   \node[ right=1.1 of e1](e) {$p'_2$};
   \draw[fermion](e1)--(e) node[pos=0.7]{$\blacktriangleleft$} ;
            
    \node(p1) at (-1.8,0){};
    \node[above right=1.7 of p1] (p2){$k_1$};
    \draw[photon] (p1) -- (p2) node[pos=0.99] {$\rotatebox[origin=c]{45}{$\blacktriangleright$}$};
    \node(q1) at (0,0){};
    \node[above right=1.6 of q1] (q2){$k_2$};
    \draw[photon] (q1) -- (q2) node[pos=0.99] {$\rotatebox[origin=c]{45}{$\blacktriangleright$}$};
      \node[minimum size=1.5cm, fill = gray!20, draw = black, ultra thick, circle] at (0,0){};
    
\end{tikzpicture}&=\begin{tikzpicture}[baseline=-0.8 cm ,scale=.8]
\node (a) at (0,0){};
\node[right=1 of a](a2){};
\draw[fermion] (a)--(a2) node[pos=.2]{$\blacktriangleright$} node[pos=.7]{$\blacktriangleleft$};
\node[above=0.25 of a2](c1){};
\node[below=0.25 of a2](c2){};
\draw[fcut] (c1)--(c2);
  \node[right=1.7 of a2 ] (b2){};
  \draw[fermion] (a2) -- (b2) node[pos=0.15]{$\blacktriangleright$} node[pos=0.8] {$\blacktriangleleft$} node[pos=.4]{$\blacktriangleleft$};
 \node[inner sep=0pt,right=.4 of a2] (x2){};
\fill (x2) circle (1pt);
 \node[inner sep=0pt,right=1 of a2] (y2){};
\fill (y2) circle (1pt);
\node[above right= 1 of x2] (c2){};
\draw[photon] (x2)--(c2) node[pos=0.95]{$\rotatebox[origin=c2]{45}{$\blacktriangleright$}$};
\node[inner sep=0pt,right=.4 of a] (x3){};
\fill (x3) circle (1pt);
\node[above right= 1 of x3] (c3){};
\draw[photon] (x3)--(c3) node[pos=0.95]{$\rotatebox[origin=c2]{45}{$\blacktriangleright$}$};
\node[below=1.2 of a] (d2){};
\node[below=1.2 of b2](e2){};
\draw[fermion] (d2)--(e2) node[pos=0.1]{$\blacktriangleright$} node[pos=0.9]{$\blacktriangleleft$};
\node[inner sep=0pt, below= 1.4 of y2](z2){};
\fill (z2) circle(1pt);
\draw[photon](y2)--(z2) node[midway]{$\blacktriangle$};

\node (a) at (4.5,0){};
\node[right=1.4 of a](a2){};
\draw[fermion] (a)--(a2) node[pos=.15]{$\blacktriangleright$} node[pos=.4]{$\blacktriangleleft$}
node[pos=.8]{$\blacktriangleleft$};
\node[above=0.25 of a2](c1){};
\node[below=0.25 of a2](c2){};
\draw[fcut] (c1)--(c2);
  \node[right=1.3 of a2 ] (b2){};
  \draw[fermion] (a2) -- (b2) node[pos=.2]{$\blacktriangleright$} node[pos=0.7] {$\blacktriangleleft$};
 \node[inner sep=0pt,right=.5 of a2] (x2){};
\fill (x2) circle (1pt);
 \node[inner sep=0pt,left=0.5 of a2] (y2){};
\fill (y2) circle (1pt);
\node[above right= 1 of x2] (c2){};
\draw[photon] (x2)--(c2) node[pos=0.95]{$\rotatebox[origin=c2]{45}{$\blacktriangleright$}$};
\node[inner sep=0pt,right=.3 of a] (x3){};
\fill (x3) circle (1pt);
\node[above right= 1 of x3] (c3){};
\draw[photon] (x3)--(c3) node[pos=0.95]{$\rotatebox[origin=c2]{45}{$\blacktriangleright$}$};
\node[below=1.2 of a] (d2){};
\node[below=1.2 of b2](e2){};
\draw[fermion] (d2)--(e2) node[pos=0.1]{$\blacktriangleright$} node[pos=0.9]{$\blacktriangleleft$};
\node[inner sep=0pt, below= 1.4 of y2](z2){};
\fill (z2) circle(1pt);
\draw[photon](y2)--(z2) node[midway]{$\blacktriangle$};
 \node (a3) at (9,0){};
  \node[right=2.4 of a3 ] (b3){};
  \draw[fermion] (a3) -- (b3) node[pos=0.1]{$\blacktriangleright$}
  node[pos=0.26]{$\blacktriangleleft$}
  node[pos=0.6]{$\blacktriangleright$}
  node[pos=0.9] {$\blacktriangleleft$};
 \node[inner sep=0pt,right=1.6 of a3] (x3){};
\fill (x3) circle (1pt);
\node[above right= 1 of x3] (c3){};
\draw[photon] (x3)--(c3) node[pos=0.95]{$\rotatebox[origin=c3]{45}{$\blacktriangleright$}$};
\node[below=1.2 of a3] (d3){};
\node[right=2.4 of d3](e3){};
\draw[fermion] (d3)--(e3) node[pos=0.15]{$\blacktriangleright$} 
node[pos=0.8]{$\blacktriangleleft$};
\node[inner sep=0pt, below= 1.4 of x3](z3){};
\node[below=1.4 of x3](t){};
\draw[photon](x3)--(t) node[pos=0.5]{$\blacktriangle$};
\node[inner sep=0pt,  right= 0.4 of a3](p3){};
\node[above right=1 of p3] (p4){};
\draw[photon] (p3)--(p4) node[pos=0.95]{$\rotatebox[origin=c3]{45}{$\blacktriangleright$}$};
\node[inner sep=0pt,  right= 1 of a3](c){};
\fill[white] (c) circle (5pt);
\node[above=0.25 of c](c1){};
\node[below=0.25 of c](c2){};
\draw[fcut](c1)--(c2);
\node (a3) at (13,-1.2){};
\node (n1) at (4.2+4.5,-0.9) {$+$};
\node (n1) at (4.2,-0.9) {$+$};
\end{tikzpicture}\nonumber\\
    &+\begin{tikzpicture}[baseline=-0.8 cm ,scale=.8]
 \node (a2) at (0,0){};
  \node[right=2.7 of a2 ] (b2){};
  \draw[fermion] (a2) -- (b2) node[pos=0.1]{$\blacktriangleright$} node[pos=0.8] {$\blacktriangleleft$} node[pos=.33]{$\blacktriangleleft$}
  node[pos=.65]{$\blacktriangleright$};
 \node[inner sep=0pt,right=.5 of a2] (x2){};
\fill (x2) circle (1pt);
 \node[inner sep=0pt,right=1.9 of a2] (y2){};
\fill (y2) circle (1pt);
\node[above right= 1 of x2] (c2){};
\draw[photon] (x2)--(c2) node[pos=0.95]{$\rotatebox[origin=c2]{45}{$\blacktriangleright$}$};
\node[inner sep=0pt,right=1.3 of a2] (c){};
\fill[white] (c) circle (5pt);
\node[above=0.25 of c](c1){};
\node[below=0.25 of c](c2){};
\draw[fcut](c1)--(c2);
\node[below=1.2 of a2] (d){};
\node[below=1.2of b2](e2){};
\node[right=1 of d] (d2){};
\draw[fermion] (d)--(e2)node[pos=0.1]{$\blacktriangleright$}
node[pos=0.3]{$\blacktriangleleft$} node[pos=0.8]{$\blacktriangleleft$};
\node[inner sep=0pt, below= 1.4 of y2](z2){};
\fill (z2) circle(1pt);
\draw[photon](y2)--(z2) node[midway]{$\blacktriangledown$};

 \node[inner sep=0pt,right=.4 of d] (x3){};
\fill (x3) circle (1pt);
\node[below right= 1 of x3] (c3){};
\draw[photon] (x3)--(c3)node[pos=0.95]{$\rotatebox[origin=c2]{45}{$\blacktriangledown$}$};
\node (n1) at (4.2,-0.9) {$+$};
 \node (a3) at (4.5,0){};
  \node[right=2.4 of a3 ] (b3){};
  \draw[fermion] (a3) -- (b3) node[pos=0.2]{$\blacktriangleright$} node[pos=0.8] {$\blacktriangleleft$};
 \node[inner sep=0pt,right=1.2 of a3] (x3){};
\fill (x3) circle (1pt);
\node[above right= 1 of x3] (c3){};
\draw[photon] (x3)--(c3) node[pos=0.95]{$\rotatebox[origin=c3]{45}{$\blacktriangleright$}$};
\node[below=1.2 of a3] (d3){};
\node[right=2.4 of d3](e3){};
\draw[fermion] (d3)--(e3) node[pos=0.15]{$\blacktriangleright$} 
node[pos=0.35]{$\blacktriangleleft$}
node[pos=0.8]{$\blacktriangleleft$};
\node[inner sep=0pt, below= 1.4 of x3](z3){};
\node[below=0.6 of x3](t){};
\node[right=.25 of t](t1){};
\node[left=.25 of t](t2){};
\draw[fcut](t1)--(t2);
\fill (z3) circle(1pt);
\draw[photon](x3)--(t) node[pos=0.5]{$\blacktriangle$};
\draw[photon](t)--(z3) node[pos=0.5]{$\blacktriangledown$};
\node[inner sep=0pt,  right= 0.5 of d3](p3){};
\node[below right=1 of p3] (p4){};
\draw[photon] (p3)--(p4) node[pos=0.95]{$\rotatebox[origin=c3]{45}{$\blacktriangledown$}$};
\end{tikzpicture} + \cdots
\end{align}
When the external photons are on-shell, the diagrams with three-particle cuts vanish due to the stability condition. As in the classical limit the exchanged gravitons are in the potential region, and hence off-shell, the diagrams with horizontal cuts do not contribute to the classical results.  Thus, the connected part in Eq.~\eqref{var} at the leading order is entirely quantum. Notice that the configurations in which external photons are emitted from the quartic vertex in QED (or a cubic graviton vertex in gravity) are prohibited by the Feynman rules in the $r/a$ basis. This is because such emissions must originate from quantum vertices, which cannot contribute at tree level as shown by the counting rule in Eq.~\eqref{counting}. In the usual in-out formalism, one must include such diagrams.   
\begin{equation}
    \begin{tikzpicture}
[baseline=-0.8 cm ,scale=.8]  
\node (a) at (0,0){};
  \node[right=2.7 of a ] (b){};
  \draw[fermion] (a) -- (b) node[pos=0.2]{} node[pos=0.8] {} node[pos=.45]{};
 \node[inner sep=0pt,right=.9 of a] (x){};
\fill (x) circle (1pt);
 \node[inner sep=0pt,right=1.7 of a] (y){};
\fill (y) circle (1pt);
\node[above right= 1 of y] (c){};
\draw[photon] (y)--(c) node[pos=0.95]{$\rotatebox[origin=c]{45}{$\blacktriangleright$}$};
\node[above left=1 of y] (c2){};
\draw[photon] (y)--(c2) node[pos=0.95]{$\rotatebox[origin=c]{45}{$\blacktriangle$}$};
\node[below=1.2 of a] (d){};
\node[right=2.7 of d](e){};
\draw[fermion] (d)--(e) node[pos=0.2]{} node[pos=0.8]{};
\node[inner sep=0pt, right= .9 of d](z){};
\fill (z) circle(1pt);
\draw[photon](x)--(z) node[midway]{};
 \node (a2) at (4.5,0){};
  \node[right=2.7 of a2 ] (b2){};
  \draw[fermion] (a2) -- (b2);
 \node[inner sep=0pt,right=.9 of a2] (x2){};
\fill (x2) circle (1pt);
 \node[inner sep=0pt,right=1.7 of a2] (y2){};
\fill (y2) circle (1pt);
\node[inner sep=0pt, above= 0.3 of x2] (c2){};
\fill (c2) circle (1pt);
\draw[photon] (x2)--(c2);
\node[inner sep=0pt,above right=0.5 of c2] (c3){};
\draw[photon] (c2)--(c3) node[pos=0.95]{$\rotatebox[origin=c2]{45}{$\blacktriangleright$}$};
\node[inner sep=0pt,above left=0.5 of c2] (c4){};
\draw[photon] (c2)--(c4) node[pos=0.95]{$\rotatebox[origin=c2]{45}{$\blacktriangle$}$};
\node[below=1.2 of a2] (d2){};
\node[right=2.7 of d2](e2){};
\draw[fermion] (d2)--(e2) node[pos=0.2]{} node[pos=0.8]{};
\node[inner sep=0pt, right= 1.7 of d2](z2){};
\fill (z2) circle(1pt);
\draw[photon](y2)--(z2) node[midway]{};

 \node (a3) at (9,0){};
  \node[right=2.4 of a3 ] (b3){};
  \draw[fermion] (a3) -- (b3);
 \node[inner sep=0pt,right=1.2 of a3] (x3){};
\fill (x3) circle (1pt);
\node[below=1.2 of a3] (d3){};
\node[right=2.4 of d3](e3){};
\draw[fermion] (d3)--(e3) node[pos=0.2]{} node[pos=0.8]{};
\node[inner sep=0pt, right= 1.2 of d3](z3){};
\fill (z3) circle(1pt);
\draw[photon](x3)--(z3);
\node[inner sep= 0 pt, above=0.7 of z3](k1){};
\node[inner sep=0pt, right=0.7 of k1](k2){};
\fill (k2) circle (1pt);
\node[above right= 0.5 of k2](k3){};
\draw[photon] (k2)--(k3) node[pos=0.95]{$\rotatebox[origin=c]{45}{$\blacktriangleright$}$};
\node[below right= 0.5 of k2](k4){};
\draw[photon] (k2)--(k4) node[pos=0.95]{$\rotatebox[origin=c]{45}{$\blacktriangledown$}$};
\draw[photon](k1)--(k2);
\node (n) at (14,-0.8){(not allowed)};
\end{tikzpicture}
\end{equation}
This analysis is rather involved in the $\I/\II$ basis since we need to involve various cut diagrams. For instance, at leading order in the classical expansion, the following diagrams, which include a Compton scattering sub-process, must cancel among themselves 
\begin{align}
    \begin{tikzpicture}[baseline=-0.8 cm,scale=.7]    
\node (a) at (0,0){};
  \node[right=2.7 of a ] (b){};
  \draw[fermion] (a) -- (b);
 \node[inner sep=0pt,right=.7 of a] (x){};
\fill (x) circle (1pt);
 \node[inner sep=0pt,right=1.4 of a] (y){};
\fill (y) circle (1pt);
\node[above right= .8 of y] (c){};
\draw[photon] (y)--(c);
 \node[inner sep=0pt,right=1.9 of a] (y2){};
\fill (y2) circle (1pt);
\node[above right= .8 of y2] (c2){};
\draw[photon] (y2)--(c2);
\node[below=1.2 of a] (d){};
\node[right=2.7 of d](e){};
\draw[fermion] (d)--(e);
\node[inner sep=0pt, right= .7 of d](z){};
\fill (z) circle(1pt);
\draw[photon](x)--(z) node[midway]{};
\node[inner sep=0pt, below=0.1 of y2](ny2){$\textcolor{gray}{\I}$};
\node[inner sep=0pt, below=0.1 of y](ny){$\textcolor{gray}{\I}$};
\node[inner sep=0pt, below=0.1 of z](nz){$\textcolor{gray}{\I}$};
\node[inner sep=0pt, left=0.35 of ny](ny3){$\textcolor{gray}{\I}$};
 \node (a2) at (4.5,0){};
  \node[right=2.7 of a2 ] (b2){};
  \draw[fermion] (a2) -- (b2);
 \node[inner sep=0pt,right=1.1 of a2] (x2){};
\fill (x2) circle (1pt);
 \node[inner sep=0pt,right=1.7 of a2] (y2){};
\fill (y2) circle (1pt);
\node[above right= 0.8 of x2] (c2){};
\draw[photon] (x2)--(c2);
\node[inner sep=0pt,right=.6 of a2] (x3){};
\fill (x3) circle (1pt);
\node[above right= 0.8 of x3] (c3){};
\draw[photon] (x3)--(c3);
\node[below=1.2 of a2] (d2){};
\node[right=2.7 of d2](e2){};
\draw[fermion] (d2)--(e2);
\node[inner sep=0pt, right= 1.7 of d2](z2){};
\fill (z2) circle(1pt);
\draw[photon](y2)--(z2);
\node[inner sep=0pt, below=0.1 of x2](nx2){$\textcolor{gray}{\I}$};
\node[inner sep=0pt, below=0.1 of x3](nx3){$\textcolor{gray}{\I}$};
\node[inner sep=0pt, below=0.1 of z2](nz2){$\textcolor{gray}{\I}$};
\node[inner sep=0pt, right=0.25 of nx2](nx4){$\textcolor{gray}{\I}$};

\node (n1) at (4.2,-1.1) {$+$};

 \node (a2) at (9,0){};
  \node[right=2.7 of a2 ] (b2){};
  \draw[fermion] (a2) -- (b2);
 \node[inner sep=0pt,right=.5 of a2] (x2){};
\fill (x2) circle (1pt);
 \node[inner sep=0pt,right=1 of a2] (x3){};
\fill (x3) circle (1pt);
 \node[inner sep=0pt,right=2.2 of a2] (y2){};
\fill (y2) circle (1pt);
\node[above right= .8 of x2] (c2){};
\draw[photon] (x2)--(c2);
\node[above right= .8 of x3] (c3){};
\draw[photon] (x3)--(c3);
\node[below=1.2 of a2] (d2){};
\node[right=2.7 of d2](e2){};
\draw[fermion] (d2)--(e2);
\node[inner sep=0pt, right= 2.2 of d2](z2){};
\fill (z2) circle(1pt);
\draw[photon](y2)--(z2) node[midway]{};
\draw[fcut,brown](2.7+9,1.2)--(2.7+9,-1.2);
\node (n1) at (4.2+4.5,-1.1) {$+$};
\node[inner sep=0pt, below=0.1 of x2](nx2){$\textcolor{gray}{\I}$};
\node[inner sep=0pt, below=0.1 of x3](nx3){$\textcolor{gray}{\I}$};
\node[inner sep=0pt, below=0.1 of z2](nz2){$\textcolor{gray}{\I}$};
\node[inner sep=0pt, right=1.25 of nx2](nx4){$\textcolor{gray}{\II}$};
\end{tikzpicture}=\mathcal{O}\left(q^{-3}\right).
\end{align}

At one-loop we can either have diagrams with $P_{rr}=2,\; n_a=0$ or $P_{rr}=0,\; n_a=2$. The combination of these diagrams leaves behind factorized six-point diagrams in the classical limit. We discuss one such example and leave the rest as an exercise for the reader. Consider the box topology with two photon emissions. The cut diagram at the leading order is
\begin{align} 
&\begin{tikzpicture}[scale=0.6]
\node[inner sep=0pt] (c) at (-10,-1.1){};
\node[above=0.2 of c](a1){};
\node[left=1 of a1](a){$p_1$};
\draw[fermion] (a)--(a1) node [pos=0.25]{$\blacktriangleright$}; 
\node[below=0.2 of c](b1){};
\node[left=1 of b1](b){$p_1$};
\draw[fermion] (b)--(b1) node [pos=0.25]{$\blacktriangleright$};
\node[right=3.6 of a1](u){$p_1'$};
\node[right=3.6 of b1](v){$p_2'$};
\draw[fermion] (u)--(a1) node [pos=0.09]{$\blacktriangleleft$}  node [pos=0.56]{$\blacktriangleright$} node [pos=0.8]{$\blacktriangleleft$};
\draw[fermion] (v)--(b1) node [pos=0.09]{$\blacktriangleleft$}  node [pos=0.56]{$\blacktriangleright$} node [pos=0.8]{$\blacktriangleleft$};
\node[above right= 1 of a1](k1){$k_1$};
\draw[photon] (a1)--(k1) node[pos=0.99] {$\rotatebox[origin=c]{45}{$\blacktriangleright$}$};
\node[right= 2.2 of a1](a2){};
\node[above right= 1 of a2](k2){$k_2$};
\draw[photon] (a2)--(k2) node[pos=0.99] {$\rotatebox[origin=c]{45}{$\blacktriangleright$}$};
 \node[minimum size=1.5cm, fill = gray!20, draw = black, ultra thick, circle] at (c){};
 \node[right= 2.4 of c] (c2){};
 \node[minimum size=1.5cm, fill = gray!20, draw = black, ultra thick, circle] at (c2){};
\node[right=1 of a1](cx1){};
 \node[minimum size=.05cm, fill = white, draw = white, circle] at (cx1){};
 \node[above=0.1 of cx1] (p1){};
 \node[below=0.15 of cx1] (p2){};
 \draw[fcut](p1)--(p2);
 \node[right=1 of b1](cx2){};
  \node[minimum size=.05cm, fill = white, draw = white, circle] at (cx2){};
  \node[above=0.1 of cx2] (p1){};
 \node[below=0.15 of cx2] (p2){};
 \draw[fcut](p1)--(p2);
\node at (-2,-1.3) (n){$=$};
\node at (-2+9,-1.3) (n){$+\quad \cdots$};
 \node[inner sep=0 pt] (a) at (1,0){};
 \node[inner sep=0 pt, left=0.8 of a] (a1){};
\node[right=0.75 of a] (cx) {};
\draw[fermion] (a1)--(cx) node[pos=0.2] {$\blacktriangleright$}
node[pos=0.42] {$\blacktriangleleft$}
 node[pos=0.8] {$\blacktriangleleft$};
 \node[inner sep=0pt,left=0.3 of a](q1){};
    \node[above right=1 of q1] (q2){$k_1$};
    \draw[photon] (q1) -- (q2) node[pos=0.99] {$\rotatebox[origin=c]{45}{$\blacktriangleright$}$};
 \node[inner sep=0pt,right=1.4 of a](nq1){};
    \node[above right=1 of nq1] (nq2){$k_2$};
    \draw[photon] (nq1) -- (nq2) node[pos=0.99] {$\rotatebox[origin=c]{45}{$\blacktriangleright$}$};
 \node[above=.25 of cx] (x1){};
 \node[below=.25 of cx] (x2){};
 \draw[fcut] (x1)--(x2);
\node[right=2.59 of a] (b){};
 \draw[fermion] (cx)--(b) 
  node[pos=0.18] {$\blacktriangleright$}
   node[pos=0.2] {$\blacktriangleright$}
   node[pos=0.46] {$\blacktriangleleft$}
  node[pos=0.81]  {$\blacktriangleleft$};
 \node[inner sep=0pt, right= 2 of a] (x){};
 \fill (x) circle (1pt);
 \node[inner sep=0pt, right= 0.1 of a] (x2){};
 \fill (x2) circle (1pt);
 \node[inner sep=0pt,below =1.7 of x2] (c){};
 \fill (c) circle (1pt);
 \node[inner sep=0pt,below =1.7 of x] (d){};
 \fill (d) circle (1pt);
 \draw[photon] (x)--(d) node[pos=0.45]{$\blacktriangle$}; 
 \draw[photon] (x2)--(c) node[pos=0.45]{$\blacktriangle$}; 
 \node[right= 0.66 of c](cx){};
 \node[above=.25 of cx] (x1){};
 \node[below=.25 of cx] (x2){};
 \draw[fcut] (x1)--(x2);
 \draw[fermion] (c)--(cx) node [midway]{$\blacktriangleleft$};
 \draw[fermion] (d)--(cx)  node [pos=.6]{$\blacktriangleright$};
 \node[right=1 of c] (cx){};
 \node[inner sep=0pt,right=0.01 of c](c2) {};
 \node[left=1 of c2](c3){};
 \draw[fermion] (c2)--(c3)  node[midway]{$\blacktriangleright$};
  \node[inner sep=0pt,left=0.01 of d](d2) {};
 \node[inner sep=0pt,right=0.6 of d2](d3){};
 \draw[fermion] (d2)--(d3) node[midway]{$\blacktriangleleft$};
\draw[->](-0.2+1,-1.7)--(-0.2+1,-0.5);
  \node (n) at (-.5+1,-1.1){$\ell$};
 \end{tikzpicture}\nonumber\\
&=\int \hat{d}^4\ell\; \mathcal{W}^{h_1}_{k_1}(\ell,k_1) \hat{\d}\left(2p_1\cdot(\ell+k_1)\right)\hat{\d} \left(2p_2\cdot\ell\right) \mathcal{W}^{h_2}_{k_2}(\ell-q,k_1+k_2)\,,
\end{align}
where the outgoing heavy momenta are $p_1+q-k_1-k_2$ and $p_2-q$. The ellipses include various cut diagrams such as cross-boxes and the photon emission from various internal and external lines.
In the classical limit, this becomes a convolution of two leading-order waveforms, thereby demonstrating the classical factorization of such an observable after Fourier transform to impact parameter space and the time domain. Finally, we have two diagrams of the same topology involving  quantum vertices 
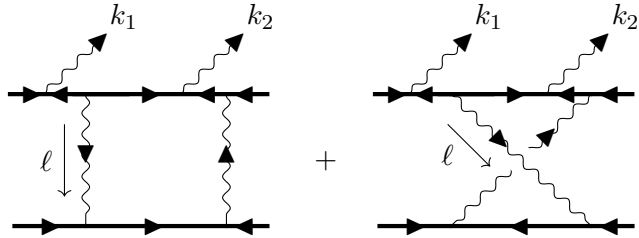
\begin{equation*}
 \begin{tikzpicture}[baseline=0.1 ex,scale=0.8]
\node[inner sep=0 pt] (a) at (6,0){};
 \node[inner sep=0 pt, left=0.8 of a] (a1){};
\node[right=0.75 of a] (cx) {};
\draw[fermion] (a1)--(cx) node[pos=0.2] {$\blacktriangleright$}
node[pos=0.42] {$\blacktriangleleft$};
 \node[inner sep=0pt,left=0.3 of a](q1){};
    \node[above right=1 of q1] (q2){$k_1$};
    \draw[photon] (q1) -- (q2) node[pos=0.99] {$\rotatebox[origin=c]{45}{$\blacktriangleright$}$};
 \node[inner sep=0pt,right=1.4 of a](nq1){};
    \node[above right=1 of nq1] (nq2){$k_2$};
    \draw[photon] (nq1) -- (nq2) node[pos=0.99] {$\rotatebox[origin=c]{45}{$\blacktriangleright$}$};

\node[right=2.59 of a] (b){};
 \draw[fermion] (a)--(b) 
  node[pos=0.4] {$\blacktriangleright$}
   node[pos=0.68] {$\blacktriangleleft$}
   node[pos=0.87]  {$\blacktriangleleft$};
 \node[inner sep=0pt, right= 2 of a] (x){};
 \fill (x) circle (1pt);
 \node[inner sep=0pt, right= 0.15 of a] (x2){};
 \fill (x2) circle (1pt);
 \node[inner sep=0pt,below =1.7 of x2] (c){};
 \fill (c) circle (1pt);
 \node[inner sep=0pt,below =1.7 of x] (d){};
 \fill (d) circle (1pt);
 \draw[photon] (x)--(d) node[pos=0.45]{$\blacktriangle$}; 
 \draw[photon] (x2)--(c) node[pos=0.45]{$\blacktriangledown$}; 
 \node[inner sep=0pt,right=0.01 of c](c2) {};
 \node[left=1 of c2](c3){};
  \node[inner sep=0pt,left=0.01 of d](d2) {};
 \node[inner sep=0pt,right=0.6 of d2](d3){};
 \draw[fermion] (c3)--(d3) 
 node[pos=0.15]{$\blacktriangleright$}
 node[pos=0.55]{$\blacktriangleright$}
 node[pos=0.9]{$\blacktriangleleft$};
 
 \node[inner sep=0 pt] (a) at (12,0){};
 \node[inner sep=0 pt, left=0.8 of a] (a1){};
\node[right=0.75 of a] (cx) {};
\draw[fermion] (a1)--(cx) node[pos=0.2] {$\blacktriangleright$}
node[pos=0.42] {$\blacktriangleleft$};
 \node[inner sep=0pt,left=0.3 of a](q1){};
    \node[above right=1 of q1] (q2){$k_1$};
    \draw[photon] (q1) -- (q2) node[pos=0.99] {$\rotatebox[origin=c]{45}{$\blacktriangleright$}$};
 \node[inner sep=0pt,right=1.4 of a](nq1){};
    \node[above right=1 of nq1] (nq2){$k_2$};
    \draw[photon] (nq1) -- (nq2) node[pos=0.99] {$\rotatebox[origin=c]{45}{$\blacktriangleright$}$};
 
\node[right=2.59 of a] (b){};
 \draw[fermion] (a)--(b) 
  node[pos=0.4] {$\blacktriangleright$}
   node[pos=0.65] {$\blacktriangleleft$}
   node[pos=0.87]  {$\blacktriangleleft$};
 \node[inner sep=0pt, right= 2 of a] (x){};
 \fill (x) circle (1pt);
 \node[inner sep=0pt, right= 0.15 of a] (x2){};
 \fill (x2) circle (1pt);
 \node[inner sep=0pt,below =1.7 of x2] (c){};
 \fill (c) circle (1pt);
 \node[inner sep=0pt,below =1.7 of x] (d){};
 \fill (d) circle (1pt);
 \draw[photon] (x)--(c) node[pos=0.3]{$\rotatebox[origin=c]{45}{$\blacktriangleright$}$} node[midway] (ci) {};
\node[minimum size=0.15cm, fill = white, draw = white, circle] at (ci){};
 \draw[photon] (x2)--(d) node[pos=0.35]{$\rotatebox[origin=c]{45}{$\blacktriangledown$}$}; 
 \node[inner sep=0pt,right=0.01 of c](c2) {};
 \node[left=1 of c2](c3){};
  \node[inner sep=0pt,left=0.01 of d](d2) {};
 \node[inner sep=0pt,right=0.6 of d2](d3){};
 \draw[fermion] (c3)--(d3) 
 node[pos=0.15]{$\blacktriangleright$}
 node[pos=0.55]{$\blacktriangleleft$}
 node[pos=0.9]{$\blacktriangleleft$};
 \draw[<-](-0.3+6.2,-1.7)--(-0.3+6.2,-0.5);
 \node (n) at (-.6+6.2,-1.1){$\ell$};
 \draw[<-](12.9,-1.2)--(12.2,-0.5);
 \node (n) at (12.2,-1){$\ell$};
\node (n2) at (10.2,-1.1){$+$};
 \end{tikzpicture}   
\end{equation*}
The leading order terms in the classical limit cancel due to the causal $i\e$ prescription. To see this, notice the two diagrams only differ by the bottom matter propagator  
\begin{align}
    \frac{1}{(p_2+\ell)^2-m_2^2+i\e}+    \frac{1}{(\ell+q-p_2)^2-m_2^2-i\e}=  {\cal O}(q^0,\ell^0)\,.
\end{align}
Therefore the connected diagrams with quantum vertices are subleading compared to the diagram involving only classical vertices.

\section{Conclusion}
\label{sec:conclusion}
In this work, we have explained how to compute the change in the expectation value of a generic asymptotic observable $\mathbb{O}$ during a scattering process from an amputated causal response function of its local density $\mathcal{O}(x)$, computed using the in-in formalism. For instance, the soft limit of a five-point response function of the stress-energy tensor of a particle $i$ with four amputated legs computes the linear impulse $\Delta \mathbb{P}_i$ of particle $i$ during scattering with an initial two-particle state. We demonstrate that this procedure is exactly equivalent to the KMOC formalism, thus generalizing the findings of Ref.~\cite{CaronHuot2024}.

Although our formalism is general, we have applied it to the computation of classical observables. We find that the soft limit acts as a useful regulator which carefully parameterizes the approach to the asymptotic past and future. This becomes particularly important for certain observables, such as the angular momentum loss, which receive contribution from the long-range interactions present in theories with massless particles. As an example, we computed the angular momentum loss in scalar QED, for which the soft frequency unveils a new region in the loop integration that yields unambiguously the leading contribution to this quantity. It would be interesting to use our method to reproduce the angular momentum loss in gravity at $\mathcal{O}\left(G^3\right)$ from \cite{Manohar:2022dea}, and produce new results beyond this order. Futhermore, we expect that similar considerations will be important to understand other phenomena related to long-range forces, such as violation of peeling \cite{Damour:1985cm, christodoulou2002global} from the perspective of scattering amplitudes.

We also have initiate a study of classical observables from amplitudes with retarded propagators within the Schwinger-Keldysh formalism. This approach ensures that causality is manifest at every step of the calculation, setting it apart from the conventional method of computing classical observables from in-out scattering amplitudes with Feynman propagators. We find that doing the calculations in this way, terms which are singular as $\hbar \to 0$ cancel manifestly for the linear impulse at one-loop, and for waveform and its variance at tree-level. In particular, for the variance, the classical vertices naturally give rise to factorized diagrams as expected for a classical quantity. It would be interesting to extend our analysis beyond one loop. In particular, starting at two loops, both classical and quantum vertices contribute. We expect the contributions from the diagrams with quantum vertices to be $q\sim\hbar$ suppressed compared to those with classical vertices at all loops. We leave the explicit checks of this statement for future work.

\acknowledgments
S.B. would like to acknowledge the hospitality of Institut des Hautes Études Scientifiques (IHES) where this work was initiated, and the support of the University of British Columbia through the Four-Year Fellowship (4YF). J.P.-M. thanks Enrico Herrmann, Michael Ruf, and Zander Moss for many discussions and previous collaboration on related topics.

\appendix

\section{Integrals}
\label{sec:integrals}
The integrals appearing in section \ref{sec: angular-impulse} and Appendix \ref{sec: angular-impulse-scalar} all belong to the integral family
\begin{align}    {I}_{a_1,a_2,a_3,a_4}&=\int \hat{d}^D\ell\frac{1}{\left(p_a\cdot \ell-p_a^0\o\pm i\e\right)^{a_1}\left(p_b\cdot \ell\pm i\e\right)^{a_2}\left(\ell^2\right)\left((q+\ell)^2\right)^{a_4}}\,.
\end{align}
 It is important to note that IBP relations are insensitive to the $i\e$ prescriptions of the integrals.
In particular, the delta function and its derivative appearing in various integrals that we are interested in can be reduced to the above form by using the distributional identity
$$\frac{2\pi i}{(-1)^{n}n!}\d^{(n)}(x)=\frac{1}{(x-i\e)^{n+1}}-\frac{1}{(x+i\e)^{n+1}}.$$
Using \texttt{LiteRed} \cite{Lee:2013mka}, we find  two master integrals 
\begin{align}
    {I}_{1,0,1,0}= {I}_a, \quad {I}_{1,1,1,0}= {I}_{ab}\,.
\end{align}
To evaluate $I_{ab}$, we work in the rest frame of $a$ specified by $p_a^\mu=m_a(1,\bm{0}),\,\;p_b^\mu=\g m_b(1,\bm{v}) $
\begin{align}
    I_{ab}&=\int \hat{d}^4\ell\frac{\hat{\d}_+(\ell^2)\hat{\d}(p_a\cdot\ell-p_a^0\o)}{p_2\cdot\ell\pm i\e}=\int \frac{\hat{d}^3\ell}{2|\bm{\ell}|}\frac{\hat{\d}\left(p_a^0(|\bm{\ell}|-\o)-\bm{p}_1\cdot \bm{\ell}\right)}{\left(p_b^0|\bm{\ell}|-\bm{p}_b\cdot \bm{\ell}\right)}\nonumber\\
    &=\frac{1}{8\pi^2\g m_a m_b} \int_0^\infty d|\bm{\ell}|\d(|\bm{\ell}|-\o)\int\frac{ d\O}{1-v\cos\th}\nonumber\\
    &=\frac{1}{2\pi m_a m_b\g v} \cosh^{-1}\g =\frac{1}{2\pi m_a m_b}\frac{ \cosh^{-1}\g}{\sqrt{\g^2-1}}\,. \label{int_Iab}
\end{align}
$I_{aa}$ is given by the $\g\to 1$ limit of Eq.~\eqref{int_Iab}
\begin{align}
    I_{aa}=\frac{1}{2\pi m_a^2}\,.
\end{align}
Now we discuss the integrals appearing in the calculation of the angular impulse. We start with the evaluation of $\mathcal{I}_{12}$, which upon IBP reduction yields
\begin{align}
    \mathcal{I}_{12}&= 2I_{12}=\frac{1}{\pi m_1m_2}\frac{ \cosh^{-1}\g}{\sqrt{\g^2-1}} \,.
 \end{align}
Next, we have
\begin{align}
        \mathcal{K}_{ab}^{\m}&=\int \hat{d}^4\ell\frac{\hat{\d}_+(\ell^2)\hat{\d}'(p_a\cdot\ell-p_a^0\o)\;q\cdot \ell}{\left(p_b\cdot\ell\pm  i\e\right)^2}\ell^\m
           =Ap_a^\mu+Bp_b^\mu+Cq^\mu\,.
\end{align}
The coefficients $A,B,C$ can be expressed in terms of scalar integrals $p_a\cdot \mathcal{K}_{ab},\; p_b\cdot \mathcal{K}_{ab},\; q\cdot \mathcal{K}_{ab}$. The scalar integrals can be IBP reduced in terms of the master integrals  which gives
\begin{align}
   p_a\cdot \mathcal{K}_{ab}= p_b\cdot \mathcal{K}_{ab} =0.
\end{align}
Taking into account $p_a\cdot q=0=p_b\cdot q$, we obtain $A=B=0$ and
\begin{align}
    C&=\frac{1}{4q^2}\left({I}_{2,2,1,-2}-2q^2{I}_{2,2,1,-1}+q^4 {I}_{2,2,1,0}\right)=\frac{1}{2\pi  m_a^2 m_b^2}\left[\frac{1}{\g^2-1}-\frac{\g \cosh ^{-1}\g}{\left(\g^2-1\right)^{3/2}}\right]
\end{align}
Finally, $\mathcal{K}_{aa}^\mu$ is found by taking $\gamma\to 1$ limit of  ${K}_{ab}^\mu$.

\section{Radiated scalar angular momentum}
\label{sec: angular-impulse-scalar}
In this section, we compute angular momentum loss at one-loop in a simple model of real massive scalars $\F_i$ interacting with a real massless scalar field $\f$ described by the Lagrangian
\begin{align}
    \mathcal{L}&=\frac{1}{2}\pd_\m\f\pd^\m\f+\sum_{i=1,2}\left[\frac{1}{2}\pd_{\m}\F_i\pd^\m\F_i-m_i^2\F^2+\frac{g_i}{2}\F_i^2\f\right]. 
\end{align}
We take the canonical definition of the stress-energy tensor for the massless scalar $\f$. The Feynman rule corresponding to the insertion of the stress-energy tensor of the massless scalar is the same as Eq.~\eqref{p-rule}.  The relevant diagrams are shown in Fig.~\ref{fig:ang-impulse-sc} where we have represented the massless scalar propagator by a dotted line.
\begin{figure}[t]
    \centering
    
\begin{tikzpicture}[scale=1]
\node (a) at (-.8,0){};
\node[right=3 of a](b){};
\node[below=1.5 of a](c){};
\node[right=3 of c](d){};
\draw[fermion] (a)--(b);
\draw[fermion] (c)--(d);
\node[inner sep=0pt,right=0.75 of a](x1){};
\node[inner sep=0pt, right=0.75 of c](x2){};
\draw[scalar](x1)--(x2);
\node[inner sep=0pt,left=0.25 of b](x3){};
\node[inner sep=0pt, left=0.25 of d](x4){};
\draw[scalar](x3)--(x4);
\node[inner sep=0pt, below=1.05 of x1](w){};
\node[left=0.45 of w](w1){};
  \draw [line width=1.5pt, double distance=1pt] (w)-- (w1) ; 
  \node[inner sep=0 pt] (c1) at (0.7-0.8,-.4){};
  \node[inner sep=0 pt, below right=2.2 of c1] (c2){};
  \draw[fcut, brown]  (c1)--(c2);

\node (a) at (3,0){};
\node[right=3 of a](b){};
\node[below=1.5 of a](c){};
\node[right=3 of c](d){};
\draw[fermion] (a)--(b);
\draw[fermion] (c)--(d);
\node[inner sep=0pt,right=0.75 of a](x1){};
\node[inner sep=0pt, right=0.83 of c](x2){};
\node[inner sep=0pt,left=0.27 of b](x3){};
\node[inner sep=0pt, left=0.25 of d](x4){};
\draw[scalar](x3)--(x2) node[midway](m){};
\draw[white,fill=white] (m) circle (0.15);
\draw[scalar](x1)--(x4);
\node[inner sep=0 pt] at (3+2+.75,-1.2)(w){};
\node[left=0.45 of w](w1){};
  \draw [line width=1.5pt, double distance=1pt] (w)-- (w1) ; 
\draw[white,fill=white] (a) circle (.6);
\draw[white,fill=white] (c) circle (.6);
  \node[inner sep=0 pt] (c1) at (1.25+3,-2){};
  \node[inner sep=0 pt, above right=1.8 of c1] (c2){};
  \draw[fcut, brown]  (c1)--(c2);
\node (a) at(7,0){};
\node[right=3 of a](b){ };
\node[below=1.5 of a](c){ };
\node[right=3 of c](d){ };
\draw[fermion] (c)--(d);
\node[inner sep=0pt,right=.6 of a](x1){};
\node[inner sep=0pt, right=.6 of c](x2){};
\node[inner sep=0pt, right=1.5 of a](m1){};
\node[inner sep=0pt, right=1.5 of c](m2){};
\node[inner sep=0pt,left=0.25 of b](x3){};
\node[inner sep=0pt, left=0.25 of d](x4){};
\draw [scalar](x4) arc (15:175:1.15);
\node[above=0.7 of m2](v){}; 
\draw[white,fill=white] (v) circle (0.15);
\node (v2) at (7+.6,-0.13-1.8){};
\draw[white,fill=white] (v2) circle (0.15);
\draw[scalar](m1)--(m2);
\draw[fermion] (a)--(b);
\draw[fermion] (c)--(d);
\node[inner sep=0 pt] at (7+.93,-.7-.53)(w){};
\node[left=0.45 of w](w1){};
  \draw [line width=1.5pt, double distance=1pt] (w)-- (w1) ; 
  \node[inner sep=0 pt] (c1) at (7+1.2,-.5-1.5){};
  \node[inner sep=0 pt, above=1.35 of c1] (c2){};
  \draw[fcut, brown]  (c1)--(c2);
\node (a) at (-.8,3){};
\node[right=3 of a](b){};
\node[below=1.5 of a](c){};
\node[right=3 of c](d){};
\draw[fermion] (a)--(b);
\draw[fermion] (c)--(d);
\node[inner sep=0pt,right=0.75 of a](x1){};
\node[inner sep=0pt, right=0.75 of c](x2){};
\draw[scalar](x1)--(x2);
\node[inner sep=0pt,left=0.25 of b](x3){};
\node[inner sep=0pt, left=0.25 of d](x4){};
\draw[scalar](x3)--(x4);
\node[inner sep=0pt, below=0.47 of x1](w){};
\node[left=0.45 of w](w1){};
  \draw [line width=1.5pt, double distance=1pt] (w)-- (w1) ; 
  \node[inner sep=0 pt] (c1) at (-0.2,-1.4+3){};
  \node[inner sep=0 pt, above right=2.4 of c1] (c2){};
  \draw[fcut, brown]  (c1)--(c2);

\node (a) at (3,3){};
\node[right=3 of a](b){};
\node[below=1.5 of a](c){};
\node[right=3 of c](d){};
\draw[fermion] (a)--(b);
\draw[fermion] (c)--(d);
\node[inner sep=0pt,right=0.75 of a](x1){};
\node[inner sep=0pt, right=0.83 of c](x2){};
\node[inner sep=0pt,left=0.27 of b](x3){};
\node[inner sep=0pt, left=0.25 of d](x4){};
\draw[scalar](x3)--(x2) node[midway](m){};
\draw[white,fill=white] (m) circle (0.15);
\draw[scalar](x1)--(x4);
\node[inner sep=0 pt] at (3+.75+.72,-.52+3)(w){};
\node[left=0.45 of w](w1){};
  \draw [line width=1.5pt, double distance=1pt] (w)-- (w1) ; 
\draw[white,fill=white] (a) circle (.5);
\draw[white,fill=white] (c) circle (.5);
  \node[inner sep=0 pt] (c1) at (1.4+3,-1+3){};
  \node[inner sep=0 pt, above right=1.8 of c1] (c2){};
\node (a) at(7,3){};
\node[right=3 of a](b){ };
\node[below=1.5 of a](c){ };
\node[right=3 of c](d){ };
\draw[fermion] (c)--(d);
\node[inner sep=0pt,right=.6 of a](x1){};
\node[inner sep=0pt, right=.6 of c](x2){};
\node[inner sep=0pt, right=1.5 of a](m1){};
\node[inner sep=0pt, right=1.5 of c](m2){};
\node[inner sep=0pt,left=0.25 of b](x3){};
\node[inner sep=0pt, left=0.25 of d](x4){};
\draw [scalar](x3) arc (345:185:1.15);
\node[above=0.7 of m2](v){}; 
\draw[white,fill=white] (v) circle (0.15);
\node (v2) at (7+.62,0.13+3){};
\draw[white,fill=white] (v2) circle (0.1);
\draw[scalar](m1)--(m2);
\draw[fermion] (a)--(b);
\node[inner sep=0 pt] at (7+.93,-.53+3)(w){};
\node[left=0.45 of w](w1){};
  \draw [line width=1.5pt, double distance=1pt] (w)-- (w1) ; 
  \node[inner sep=0 pt] (c1) at (7+1.2,.3+3){};
  \node[inner sep=0 pt, below=1.35 of c1] (c2){};
  \draw[fcut, brown]  (c1)--(c2);
\end{tikzpicture}
    \caption{The diagrams contributing to the scalar angular momentum loss are shown. We label the momenta as in fig~\ref{fig:ang-impulse-1}.}
    \label{fig:ang-impulse-sc}
\end{figure}\\
 The first diagram gives
\begin{align}
   \mathcal{J}_a&=ig_1^2g_2^2\lim_{\o\rightarrow 0}\lim_{\bm{k}\rightarrow 0}\partial_{\bm{k}}^{[i}\bigg[\int \hat{d}^4q_1 \hat{d}^4q_2\hat{\d}\left(2p_1\cdot q_1+q_1^2\right)\hat{\d}\left(2p_2\cdot q_2+q_2^2\right)e^{-ib_1\cdot q_1-ib_2\cdot q_2}\nonumber\\
    &\times\int \hat{d}^4\ell\frac{2\o \ell^0\hat{\d}_+(\ell^2)\hat{\d}(2p_1\cdot \ell+2k\cdot (p_1-\ell)+k^2)\;\ell^{j]}}{\left(2\ell\cdot k+k^2+i\e\right)\left(2p_2\cdot \ell-i\e\right)\left((q_2-\ell)^2-i\e\right)}\times \hat{\d}^4(k+q_1+q_2)\bigg]
    \,.\label{example-ang-sc}
\end{align}
In evaluating Eq. \eqref{example-ang-sc}, we need to be careful since the $\bm{k}$ derivative hits the delta function and one must integrate by parts before imposing the momentum conservation.
 Taking the soft limits and adding all the diagrams in  Fig.~\ref{fig:ang-impulse-sc}, we obtain the angular impulse of the massless scalar field
\begin{align}
 \D \mathbb{J}_{\phi}^{ij}
&=-\lim_{\o\rightarrow 0}\frac{ig_1g_2}{16}\left[\int \frac{\hat{d}^4q}{q^2}e^{-ib_1\cdot q}\;\hat{\d}\left(p_1\cdot q\right)\hat{\d}\left(p_2\cdot q\right)p_1^{[i}\bigg[g_1g_2\;\mathcal{K}_{12}^{j]}-g_1^2\mathcal{K}_{11}^{j]}\right] \nonumber\\
   & =\frac{1}{8\pi}\left[\frac{g_1^2}{ 3m_1^4}+\frac{g_1g_2}{ m_1^2m_2^2}\left(\frac{1}{\g^2-1}-\frac{\gamma  \cosh ^{-1}\gamma}{\left(\gamma ^2-1\right)^{3/2}}\right)\right]  p_1^{[i}\;\D p_1^{j]}+(1\leftrightarrow 2)\, ,\label{ang_impulse_sc}
\end{align}
where we have used the on-shell conditions $2p_i\cdot q_i+q_i^2=0$. We have also defined the leading order linear impulse of the heavy particle 1:
\begin{align}
   \D p_1^i=-\frac{ig_1g_2}{4}\int \frac{\hat{d}^4q}{q^2}\hat{\d}\left(p_2\cdot q\right)\hat{\d}\left(p_1\cdot q\right)q^{i}e^{-iq\cdot b}. 
\end{align}
This agrees with the small-deflection limit of the result in 
\cite{DiVecchia:2022owy}.


\bibliographystyle{JHEP}
\bibliography{refs.bib}

\end{document}